\UseRawInputEncoding
\documentclass[showpacs,twocolumn,superscriptaddress,nofootinbib,aps,prd,10pt]{revtex4-2} 

\usepackage{hyperref}
\hypersetup{colorlinks=true,linkcolor=blue,citecolor=blue}
\usepackage{tikz}
\usepackage{graphicx}
\usepackage{xcolor,soul,ulem}
\sethlcolor{green}
\usepackage{dcolumn}
\usepackage{bm}
\usepackage{color}
\usepackage{enumitem}
\usepackage{mathrsfs}
\usepackage{multirow} 
\usepackage{amsmath}
\usepackage{amssymb}
\usepackage{cleveref}
\usepackage{orcidlink}

\crefname{equation}{Eqn.}{Eqns.}
\crefname{figure}{Fig.}{Figs.}
\crefname{section}{Sec.}{Sec.}
\crefname{table}{Table}{Tables}
\usepackage{physics}
\providecommand{\dif}{\mathrm{d}} \def\d{\dif}
\setstcolor{red}
\usepackage{verbatim}

\DeclareTextFontCommand{\emph}{\itshape}

\newcommand{\beq}{\begin{equation}}
\newcommand{\eeq}{\end{equation}}
\newcommand{\bea}{\begin{eqnarray}}
\newcommand{\eea}{\end{eqnarray}}
\newcommand{\non}{\nonumber}
\providecommand{\dif}{\mathrm{d}} 
\newcommand{\ce}{{\cal{E}}} 

\newcommand{\cb}{{\cal{B}}}

\def\EE{{\cal E}}
\def\LL{{\cal L}}
\def\BB{{\cal B}}

\def\d{\dif}

\graphicspath{{pic/}}

\definecolor{darkgreen}{rgb}{0.0, 0.75, 0.0}

\begin{document}

\title{Spinning charged test particle dynamics around a Schwarzschild black hole embedded in a homogeneous magnetic field}

\author{Misbah Shahzadi
\orcidlink{0000-0002-3130-1602}}
\email{misbahshahzadi51@gmail.com}
\affiliation{Astronomical Institute of the Czech Academy of Sciences, Bo\v{c}n\'{i} II 1401/1a, CZ-141 00 Prague, Czech Republic}

\author{Martin Kolo{\v s}
\orcidlink{0000-0002-4900-5537}}
\email{martin.kolos@physics.slu.cz}
\affiliation{Research Centre for Theoretical Physics and Astrophysics, Institute of Physics, \\Silesian University in Opava, Bezru\v{c}ovo n\'{a}m\v{s}st\'{i} 13, CZ-74601 Opava, Czech Republic}

\author{Ond\v{r}ej Zelenka
\orcidlink{0000-0003-3639-1587}}
\email{ondrej.zelenka@asu.cas.cz}
\affiliation{Astronomical Institute of the Czech Academy of Sciences, Bo\v{c}n\'{i} II 1401/1a, CZ-141 00 Prague, Czech Republic}

\author{Georgios Lukes-Gerakopoulos
\orcidlink{0000-0002-6333-3094}}
\email{gglukes@gmail.com}
\affiliation{Astronomical Institute of the Czech Academy of Sciences, Bo\v{c}n\'{i} II 1401/1a, CZ-141 00 Prague, Czech Republic}
\affiliation{Department of Physics, University of Thessaly, 35100 Lamia, Greece}

\begin{abstract}
We study the dynamics of spinning charged test particles orbiting a Schwarzschild black hole immersed in a test uniform magnetic field. This setup provides a simple but physically relevant framework for modeling particle motion in magnetized astrophysical environments near compact objects, where both spin-curvature coupling and electromagnetic interactions can play a significant role. The particle trajectories are obtained numerically in both equatorial and off-equatorial configurations, allowing us to examine the influence of spin-curvature and Lorentz forces on the motion. In the equatorial plane, assuming the particle's spin vector is orthogonal to the orbital plane, we derive analytical expressions for the conserved energy and angular momentum, as well as for the radial and orbital frequencies as functions of spin parameter and magnetic parameter. We also construct the corresponding effective potential to determine the allowed regions of particle motion. The equatorial dynamics remain integrable due to the existence of conserved quantities associated with the spacetime symmetries and the alignment of the magnetic field. In contrast, the off-equatorial motion constitutes a non-integrable dynamical system. While limiting subcases of the system, i.e., the spinning neutral and non-spinning charged cases, can be analyzed using two-dimensional Poincar\'{e} surface of sections, the combined system can be reduced only up to three degrees of freedom. Hence, to investigate the resulting complexity, we analyze the phase space using four-dimensional Poincar\'{e} surface of sections along with recurrence analysis, revealing the presence of chaotic behavior for particular choices of parameters and initial conditions. Finally, we compare the dynamics of spinning charged test particles with the limiting cases of non-spinning neutral, spinning neutral, and non-spinning charged particles, thereby distinguishing the respective contributions of spin-curvature and electromagnetic interactions.
\end{abstract}
\keywords{black hole physics, magnetic field, spinning charged particle dynamics}
\maketitle
\setcounter{tocdepth}{2}
\tableofcontents




\section{Introduction} \label{intro}

Deterministic chaos describes systems whose evolution is fully deterministic yet appears random due to sensitive dependence on initial conditions and can be detected by several methods: standard Poincar\'{e} surfaces of section (PS) for 2 degrees of freedom (DoF) systems \cite{Contopoulos:2002ocda}, four-dimensional (4D) PS for 3 DoF systems \cite{Katsanikas:2011IJBC,Moges:2024IJBC}, recurrence analysis \cite{Marwan:2007rps} and the Lyapunov exponents \cite{Skokos:2010LNP}, which are applicable for any number of DoF system. It is well known that geodesic motion in the Kerr or Schwarzschild black hole (BH) spacetime is fully integrable due to the existence of conserved quantities associated with spacetime symmetries, including the Carter constant \cite{Carter:1968PhRv}. As a consequence, only regular motion is allowed in this setting. The emergence of chaotic behavior therefore requires a departure from integrability, which can be achieved either by modifying the background spacetime or by introducing additional interactions that alter the equations of motion \cite{Semerak:2010MNRAS,Sukova:2013MNRAS,GLG2014PhRvD,GLG2016PhRvD,Rad-Kol-Stu:2019:EPJC:,GLG2021hgwa}. In the present work, we follow the latter approach by considering a spinning charged test particle moving in a Schwarzschild background immersed in an external uniform magnetic field.

The dynamics of spinning test particles in general relativity is governed by the Mathisson-Papapetrou-Dixon (MPD) equations \cite{Mathisson:1937:Acta:,Papapetrou:1951:PPSA:,Dixon:1964:NCim:}, which describe how the particle's spin couples to the curvature of spacetime. This spin-curvature interaction gives rise to a force that drives the motion away from geodesic trajectories, leading to qualitatively new dynamical features. The role of spin in relativistic systems has been extensively investigated, revealing phenomena such as orbital precession induced by spin-orbit and spin-spin couplings \cite{Filipe-Jose:2014,Rau-etal:2016:PhRvD:,Drummond2022PhRvD}. In particular, Suzuki and Maeda \cite{Suzuki-Maeda:1997:prd:} demonstrated that the motion of spinning particles in a Schwarzschild spacetime can exhibit chaotic behavior. Since then, the influence of spin on particle dynamics near BHs has been widely studied \cite{Tod-etal:1976:NCimB:,Semerak:1999:MNRAS:,Kyrian-Semerak:2007:MNRAS:,Eva-et-al:2014:PhRvD,Gera-etal:2014:PhRvD:,Gera-etal:2016:PhRvD:,Filipe-etal:2018:PhRvD:,Zelenka-etal:2020:PhRvD:,Ias:Gera:The:2021:PhRvD:,Sha-Gera-Martin:2025PhRvD:}.

Astrophysical BHs are typically surrounded by plasma environments that generate magnetic fields, which can significantly influence particle dynamics near the event horizon. Observations indicate that magnetic field strengths can range from a few Gauss up to $10^8$ Gauss, depending on the astrophysical context \cite{Eatough-etal:2013:Natur:,Gold-etal:2017:ApJ:,EHT:2021:ApJ:}. In this still test-field approximation, the magnetic field does not modify the spacetime geometry but affects the motion of charged particles through electromagnetic forces \cite{Blandford:1977:MNRAS:}. A commonly adopted idealization is the asymptotically uniform magnetic field introduced by Wald \cite{Wald:1974:prd:}, which provides a tractable model capturing essential features of more realistic configurations. The dynamics of charged particles in such magnetized BH spacetimes have been widely explored, yielding important insights into high-energy astrophysical processes \cite{Kol-Stu-Tur:2015:CLAQG:,Kol-Tur-Stu:2017:EPJC:,Sha-Sha:2017:EPJC:,Kol:Tur:Stu:2021:PhRvD:,Martin:2023:EPJC:}.

The combined influence of spin-curvature coupling and electromagnetic forces leads to a rich and complex dynamical system with potential astrophysical relevance. Previous studies have considered various aspects of spinning and charged particle motion, including the existence of energetically bound orbits \cite{Hoj-Hoj:1977:PhRvD:} and applications to quasi-periodic oscillations in accretion systems \cite{Sha-etal:2021:EPJC:}. More generally, the interplay between gravitational and electromagnetic interactions has been investigated in several works \cite{Prasanna:1980:NCimR:,Papadopoulos:2004:GReGr:,Bini-etal:2005:IJMPD:,Filipe-etal:2016:PhRvD:,Zhang-etal:2018:PhRvD:,Zhang-etal:2019:PhLB:}. However, a systematic and comprehensive analysis of spinning charged particle dynamics in magnetized Schwarzschild spacetime, including orbital structure, frequency properties, and chaotic behavior, remains incomplete.

Understanding the dynamics of particles in magnetized BH environments is essential for modeling a wide range of high-energy astrophysical phenomena, including accretion processes, jet formation, and radiation mechanisms near compact objects \cite{Blandford:1977:MNRAS:,McKinney-etal:2012MNRAS:,EHT:2021:ApJ:}. In such settings, electromagnetic interactions play a crucial role due to the presence of magnetic fields generated by surrounding plasma \cite{Wald:1974:prd:,Blandford:1977:MNRAS:}, while spin effects become important for particles with intrinsic angular momentum \cite{Filipe-etal:2016:PhRvD:}. The combined influence of spin-curvature coupling and Lorentz forces can give rise to a rich dynamical structure that cannot be captured by considering either effect in isolation. A detailed investigation of these effects is therefore important for a deeper understanding of particle dynamics in realistic astrophysical environments.

In this work, we address this problem by studying the motion of a spinning charged test particle in a Schwarzschild spacetime immersed in a uniform magnetic field. The dynamics is described by the MPD-Souriau (MPDS) equations supplemented by the Tulczyjew-Dixon (TD) spin supplementary condition (SSC). We first analyze equatorial motion under the assumption that the spin vector is orthogonal to the orbital plane, deriving analytical expressions for the conserved quantities and orbital frequencies, and characterizing the effective potential and stability of circular orbits, including the innermost stable circular orbits (ISCOs). We then consider generic off-equatorial motion and investigate its dynamical properties through numerical integration. In particular, we explore the transition between regular and chaotic regimes using PS and recurrence analysis, providing a detailed characterization of the underlying phase space structure.

The rest of the article is organized as follows. In Sec.~\ref{sec:DS-EQs}, we review the MPDS equations and their formulation. Section~\ref{sec: SchBHMF} introduces the spacetime and its symmetries. Equatorial motion is analyzed in Sec.~\ref{sec: equatorial}, while Sec.~\ref{sec: off-eq} is devoted to the study of generic orbits and their dynamical behavior. Finally, Sec.~\ref{sec: Conc} summarizes our results.

\section{Spinning charged test body dynamics}\label{sec:DS-EQs}

The equations of motion for a spinning test body in a purely gravitational background were first derived by Mathisson \cite{Mathisson:1937:Acta:} and Papapetrou \cite{Papapetrou:1951:PPSA:}, and later reformulated by Dixon \cite{Dixon:1970:RSPSA:}, referred to as MPD equations. By extending Papapetrou's original equations, Dixon and Souriau proposed the MPDS equations governing the motion of a spinning charged test body moving in both gravitational and electromagnetic fields, which contain the spin-curvature and spin-electromagnetism coupling terms. The MPDS equations read \cite{Dixon:1964:NCim:,Dixon:1970:RSPSA:,Dixon:1974:RSPTA:,souriau:1974:,Bini-etal:2000:PhRvD:,Bini-etal:2005:IJMPD:}
\bea \label{eq:spin1}
    \frac{\d x^\alpha}{\d\tau} &=& u^\alpha,\\\label{eq:spin2}
    \frac{\mathrm{D} p^\alpha}{\d\tau} &=& -\frac{1}{2} R^{\alpha}_{\mu\nu\rho} S^{\nu\rho} u^\mu + q {F^{\alpha}}_{\beta} u^{\beta} - \frac{k}{2} S^{\mu\nu} \nabla^{\alpha} F_{\mu\nu} ,\\\label{eq:spin3}
    \frac{\mathrm{D} S^{\alpha\beta}}{\d\tau} &=& p^\alpha u^\beta - p^\beta u^\alpha + k (S^{\alpha \nu} {F_{\nu}}^{\beta} - S^{\beta\nu} {F_{\nu}}^{\alpha}),
\eea
where $\tau$ is the proper time, $\mathrm{D}/\d\tau$ denotes the covariant derivative along the particle trajectory, $R^{\alpha}_{\mu\nu\rho}$ defines the Riemann curvature tensor, $q$ represents the charge of the particle, $S^{\alpha\beta}$ is the antisymmetric spin tensor, $F_{\alpha \beta} = A_{\beta;\alpha} - A_{\alpha; \beta}$ is the electromagnetic field tensor, $A_{\alpha}$ denotes the four-vector potential, $u^\alpha$ is the four-velocity and $p^\alpha$ is the four-momentum of the test particle. The electromagnetic coupling scalar $k$ is defined by
\beq
    k = -\frac{q g}{2 \mu}, 
\eeq
where $g$ is the gyromagnetic ratio and $\mu = \sqrt{- p^{\alpha} p_{\alpha}}$ is the mass of the particle. For a comprehensive analysis of the choice of the factor $g$ and $k$, see \cite{Pra-Vir:1989:PhLA:,Bini-etal:2005:IJMPD:}. The kinematical four-momentum $p^{\alpha}$ is associated with the generalized four-momentum $\pi^{\alpha}$
by the relation
\beq \label{eq:KinCan4mom}
    \pi^{\alpha} = p^{\alpha} + q A^{\alpha}.
\eeq
The term $\frac{1}{2} R^{\alpha}_{\mu\nu\rho} S^{\nu\rho}u^\mu$ on the right-hand side of Eq.~\eqref{eq:spin2} indicates the spin-curvature coupling through a strong gravitational field, while the terms $S^{\mu\nu} \nabla^{\alpha} F_{\mu\nu}$ and $S^{\alpha [\nu} F^{\beta]}{}_{\nu}$ describe the spin-electromagnetism interaction. Note that Eqs.~\eqref{eq:spin1}-\eqref{eq:spin3} are not exactly MPDS pole-dipole equations, but a special case of them, obtained by neglecting the electric dipole of the test body and assuming its magnetic moment dipole tensor to be proportional to the spin tensor \cite{Dixon:1964:NCim:,Dixon:1974:RSPTA:}. By contracting Eq.~\eqref{eq:spin3} with $u_\alpha$, we obtain
\begin{align}
    p^\beta &=m u^\beta+p^\beta_{\rm hid}, \label{eq:hiddenDef}\\
    p^\beta_{\rm hid} &=p^\beta_{\rm hid I}+p^\beta_{\rm hid EM},\\
    p^\beta_{\rm hid I} &=u_\alpha \frac{\mathrm{D} S^{\alpha\beta}}{\d\tau},\\
    p^\beta_{\rm hid EM} &=-u_\alpha k (S^{\alpha \nu} {F_{\nu}}^{\beta} - S^{\beta\nu} {F_{\nu}}^{\alpha}),
\end{align}
where, following the terminology used in Ref.~\cite{Filipe-etal:2016:PhRvD:}, the ``hidden momentum'' $p^\beta_{\rm hid}$ is split into the ``inertial'' part $p^\beta_{\rm hid I}$ and the ``electromagnetic'' part $p^\beta_{\rm hid EM}$, while $m=-p^\alpha u_\alpha$ is the kinetic mass. The spatial dual of the spin-electromagnetism coupling term $(S^{\alpha \nu} {F_{\nu}}^{\beta} - S^{\beta\nu} {F_{\nu}}^{\alpha})$ coincides with the coupling term in Bargmann-Michel-Telegdi equations proposed by Bargmann, Michel, and Telegdi \cite{Bar-Mic-Tel:1959:PhRvL:}. This term is associated with the ``hidden momentum'' induced by the electromagnetic field. In fact, the MPDS Eqs.~\eqref{eq:spin1}-\eqref{eq:spin3} reduce to the Bargmann-Michel-Telegdi equations in the limit of the weak and uniform external field \cite{Cognola-etal:1981:PhLB:}.

A simple scheme for a spinning charged particle can be obtained by setting the electromagnetic coupling scalar $k=0$, or by neglecting the spin-electromagnetism interaction terms $ S^{\nu [\alpha} {F^{\beta]}}_{\nu}$ and $S^{\mu\nu} \nabla^{\alpha} F_{\mu\nu}$ \cite{Hoj-Hoj:1977:PhRvD:}. It is obvious that when the electromagnetic field is switched off ($F^{\alpha \beta} = 0$), the MPDS equations lead to the pole-dipole MPD equations. Furthermore, in the absence of both spin and electromagnetic field ($S^{\alpha \beta} = F^{\alpha \beta} = 0$), we recover the geodesic equation $\mathrm{D} p^\alpha / \d \tau =0$ from the Eqs.~\eqref{eq:spin1}-\eqref{eq:spin3}. In our work, we adopt the aforementioned simple scheme, in which the electromagnetic coupling scalar $k$ is set to zero as done in Refs.~\cite{Pomeranskii2000PhyU,Papadopoulos:2004:GReGr:}. For a detailed discussion on the physical motivation behind the $k=0$ choice, see Sec.~4.2 in Ref.~\cite{Pomeranskii2000PhyU} and the references therein; in our case, the main motivation is to simplify the dynamics. 

The MPDS equations are first-order non-linear ordinary differential equations, but, like the MPD equations, they are not a closed set. Namely, there are fewer equations of motion than the number of variables describing the system's evolution. To address this issue, additional conditions are required. These conditions determine the center of mass of the body, which then serves as a reference point about which the spin and the momentum of the body can be calculated. Various constraints have been proposed to achieve this, and they are known as SSCs \cite{Semerak:1999:MNRAS:}. We use the TD SSC \cite{Dixon:1970:RSPSA:,Dixon:1974:RSPTA:} that reads
\beq\label{eq:TD_spin-cond}
    p_{\mu} S^{\mu \nu} =0,
\eeq
and specifies a unique worldline. The TD SSC is extensively used in numerical calculations, mostly due to the existence of an explicit relation between the four-velocity and the four-momentum of the spinning charged test body\footnote{For the derivation, see Appendix~\ref{Apndx:RelDer}.} \cite{Kunzle:1972:JMP:,Papadopoulos:2004:GReGr:,Gralla-Harte-Wald:2010:PhRvD:}
\beq\label{eq:4v4mrel}
    u^\mu = N \left( p^{\mu} + w^\mu \right),
\eeq
where
\bea
    w^\mu &=&  \frac{2 S^{\mu\nu} p^{\lambda} \left(2 q F_{\nu\lambda} + R_{\nu\lambda\rho\sigma}  S^{\rho\sigma}\right)}{4 \mu^{2} + S^{\alpha\beta} \left( 2 q F_{\alpha\beta} + R_{\alpha\beta\gamma\delta} S^{\gamma\delta} \right )},
\eea
and using the condition $u^\alpha u_\alpha = -1 $, we have
\beq\label{eq:NN-condition}
    N = \frac{1}{\sqrt{\mu^2 - w^\alpha w_\alpha}}.
\eeq
It is clear from Eq.~\eqref{eq:4v4mrel} that the four-velocity $u^{\alpha}$ and the kinematical four-momentum $p^{\alpha}$ are not parallel for spinning charged particles, under TD SCC \eqref{eq:TD_spin-cond}. However, in our setup, both remain timelike due to $u^\alpha u_\alpha = -1 $ and the conservation of the dynamical mass 
\begin{align} \label{eq:DynMassDef}
 \mu^2 = - p^{\alpha} p_{\alpha} > 0,
\end{align}
under the TD SSC. The latter can be proven by contracting Eq.~\eqref{eq:spin3} with $\displaystyle \frac{\mathrm{D} p_\alpha}{\d \tau} p_\beta$ and Eq.~\eqref{eq:spin2} with $u_\alpha$, which leads to $\displaystyle \frac{\mathrm{D} \mu}{\d \tau}=0$.\footnote{For $k\neq 0$, there is a slightly different definition of a conserved mass, see, e.g., Ref.~\cite{HeyFar-etal:2005:PLB:}.} 

It is often useful to use a spin vector $S^\alpha$ instead of the spin tensor, which is defined as
\bea
S_\alpha &=& -\frac{1}{2}\eta_{\alpha\beta\mu\nu} v^{\beta} S^{\mu\nu},\label{eq:spin-vector}
\eea
while we can get the spin tensor from the spin vector as
\bea
 S^{\alpha\beta} &=& -\eta^{\alpha\beta\gamma\delta} S_{\gamma} v_{\delta},\label{eq:spin-tensor}
\eea
where $\eta_{\alpha\beta\mu\nu} = \sqrt{-g} ~ \epsilon_{\alpha\beta\mu\nu}$ is the Levi-Civita tensor, $\epsilon_{\alpha\beta\mu\nu}$ denotes the Levi-Civita symbol, and $v^{\alpha}=p^{\alpha}/\mu$ represents a unit vector parallel to the momentum. Similarly to the mass $\mu$, the measure of the spin
\begin{align} \label{eq:SpinMeasure}
    S^2=\frac{1}{2} S^{\mu\nu} S_{\mu\nu},
\end{align}
is conserved under TD SSC 
\footnote{The proof is provided in Appendix~\ref{Apndx:RelDer}}.

The mass $\mu$ and the spin magnitude $S$ of the particle are constants of motion independently of the symmetry of the background spacetime. However, for every Killing vector field that preserves the electromagnetic field, the following quantity is conserved \cite{Dixon:1970:RSPSA:,Hoj-Hoj:1977:PhRvD:}
\beq\label{eq:Cons_Motion}
    C({\xi}) = p^{\alpha} \xi_{\alpha} -\frac{1}{2} S^{\alpha\beta} \xi_{\alpha ; \beta} + q A^{\alpha} \xi_{\alpha}.
\eeq
For computational convenience, Eq.~\eqref{eq:Cons_Motion} can be rewritten in the form
\beq\label{eq:killing-field2}
C(\xi)=p_{\alpha}\xi^{\alpha} - \frac{1}{2} \left(g_{\rho\sigma} \xi^{\sigma}_{,\alpha} + g_{\mu\rho,\alpha}\xi^{\mu} \right) S^{\rho\alpha} + q A_{\alpha} \xi^{\alpha}.
\eeq

\subsection{Dynamics around a Schwarzschild BH immersed in a uniform magnetic field} \label{sec: SchBHMF}

We consider the motion of spinning charged test bodies in the vicinity of a Schwarzschild BH of mass $M$ immersed in a test uniform magnetic field, and the corresponding geometry is characterized by the line element
\beq
    \d s^2 = -f(r) \d t^2 + f^{-1}(r) \d r^2 + r^2(\d \theta^2 + \sin^2\theta ~ \d \phi^2), \label{eq:SCHmetric}
\eeq
where the function $f(r)$ takes the form
\beq \non
    f(r) = 1 - \frac{2 M}{r},
\eeq
and the associated electromagnetic field with strength $B \in (-\infty, \infty)$, oriented perpendicular to the equatorial plane of the BH, can be described by the electromagnetic four-vector potential $A^{\mu}$ as  
\beq
    A^{\mu} = \frac{B}{2} \xi^{\mu}_{(\phi)}.
\eeq
The commuting Killing vector $\xi_{(\phi)} = \partial / \partial \phi$ generates rotations around the symmetry axis. Thus, the only non-zero covariant component of the four-vector potential of the electromagnetic field takes the form \cite{Wald:1984:book:}
\beq
A_{\phi} = \frac{B}{2} g_{\phi \phi} = \frac{B}{2} r^{2}\sin^{2}\theta .
\eeq

The Killing vector fields $\xi_{(t)} = \partial/\partial t$ and $\xi_{(\phi)}=\partial/\partial \phi$, together with Eq.~\eqref{eq:killing-field2}, give rise to the following quantities
\bea 
    E&=&-C(\xi_{(t)})= -\left(p_{t} + \frac{M}{r^{2}} S^{tr}\right), \label{eq:EnCon} \\\non
    L &=&  C( \xi_{(\phi)} )  = p_{\phi} + r S^{r\phi} \sin^{2}\theta + r^{2} S^{\theta\phi} \sin\theta \cos\theta \\ &+& r^{2} \frac{q B}{2}  \sin^{2}\theta, \label{eq:AngMomCons}    
\eea
where $E$ and $L$ represent the energy and angular momentum of the particle along the symmetry axis, respectively. We introduce for convenience the dimensionless quantities, energy $\ce$, angular momentum $\LL$, spin $\mathcal{S}$, and magnetic parameter $\cb$, by the relations
\beq
    \ce = \frac{E}{\mu}, \quad \LL = \frac{L}{\mu M}, \quad \mathcal{S} = \frac{S}{\mu M}, \quad \cb = \frac{q B M}{2 \mu}.
\eeq
Other dimensionless quantities will be denoted by a hat, e.g., $\hat{r}=r/M$. 

The dynamics of a test body moving on a geodesic orbit around a BH is described by four DoF. This system is integrable, since it possesses four independent and involution integrals of motion, i.e., the energy $\EE$, the axial component of the angular momentum $\LL$, the mass, and the Carter constant \cite{Carter:1968PhRv}, which in the case of a Schwarzschild BH corresponds to the total angular momentum. As shown in Ref.~\cite{Witzany2019CQGra} for the MPD equations ($\BB=0$), the DoF of the respective Hamiltonian system increases from four to six, assuming that the spin measure is conserved. Under the TD SSC, one of the six DoF is redundant. Hence, we are left with five DoF. If there were no magnetic field, the spherical symmetry would allow us due to the aforementioned integrals of motion, to reduce the system to two DoF. The presence of a homogeneous magnetic field reduces the spherical symmetry to axisymmetry, which implies that the total angular momentum is not conserved anymore. In this case, the system can be reduced to up to three DoF. Note that even if the charged body was not spinning ($\mathcal{S}=0$), the presence of the magnetic field would render the system non-integrable with two DoF.

For all the above cases, constraining the system to the equatorial plane renders it integrable, since for three DoF, we have three constants of motion $\LL,~\EE$ and $\mu$. This allows us to tackle the motion analytically. We discuss this case in Sec.~\ref{sec: equatorial}. However, the numerical study of MPDS equations for generic orbits, as presented in Sec.~\ref{sec: off-eq}, entails a bunch of interesting numerical challenges. For example, the efficient integration of the equations of motion over long time intervals requires structure-preserving algorithms, such as symplectic schemes, which have been successfully applied for simulations in various fields of general relativity \cite{Hairer:2006:,Seyrich:2012PhRvD,Gera-etal:2014:PhRvD:,Shahzadi:2023PhRvD}. 

To depict the orbital motion in space, we shall use the Cartesian coordinates, which can be obtained from the Schwarzschild ones by the following coordinate transformation
\begin{align}
   x &= \hat r \cos (\phi)\sin(\theta), \nonumber \\ 
   y &= \hat r \sin(\phi)\sin(\theta), \nonumber \\
   z &= \hat r \cos(\theta).
\end{align}
To show the spin precession, we shall use the spin vector. The explicit relations between the spin tensor $S^{\alpha \beta}$ and spin vector $S^{\alpha}$, employing Eq.~\eqref{eq:spin-vector}, for the case of Schwarzschild BH, can be written as \cite{Suzuki-Maeda:1997:prd:}
\bea \label{eq:SV_expression1}
S_{t} &=& - r^2 \sin\theta \left(v^r S^{\theta \phi} + v^{\theta} S^{\phi r} + v^{\phi} S^{r \theta} \right), \\\label{eq:Sr_component}
S_r &=& r^2 \sin\theta \left(v^t S^{\theta \phi} - v^\theta S^{t \phi} + v^\phi S^{t \theta} \right),\\
S_\theta &=& r^2 \sin\theta \left(v^t S^{\phi r} - v^\phi S^{tr} + v^r S^{t \phi} \right), \\
S_\phi &=& r^2 \sin\theta \left(v^t S^{r \theta} - v^r S^{t \theta} + v^\theta S^{tr} \right). \label{eq:SV_expression4}
\eea

\section{Motion in the equatorial plane} \label{sec: equatorial}

Before we explore the complicated generic off-equatorial motion, let us first consider the motion of a spinning charged test body in the equatorial plane. To achieve this, we assume that the spin vector of the particle is orthogonal to the orbital plane, which is the equatorial plane, and we set
\beq\label{eq:eq_plane}
\theta = \frac{\pi}{2}, \quad S^{\theta \beta}=0, \quad p^{\theta}=0.
\eeq
Substituting $\alpha=\theta$ into Eqs.~\eqref{eq:spin1}-\eqref{eq:spin3}, along with the conditions \eqref{eq:eq_plane}, the Eqs.~\eqref{eq:spin1}-\eqref{eq:spin3}, turns out to be \cite{Hoj-Hoj:1977:PhRvD:}
\beq
\frac{\d \theta}{\d \tau}=0, \quad \frac{\mathrm{D} S^{\theta \beta}}{\d \tau}=0, \quad \frac{\mathrm{D} p^{\theta}}{\d \tau}=0.
\eeq
The orthogonality condition $v^\alpha S_\alpha = 0$, which follows from the definition~\eqref{eq:spin-vector} of the spin vector, implies that, for an arbitrary equatorial orbit with $v^\theta = 0$, all components of the spin vector should be zero except $S^\theta$, i.e.,
\beq\label{eq:spin-vecEq}
  S^{\alpha} = S^{\theta} \delta_{\theta}^{\alpha}.
\eeq
Using Eqs.~\eqref{eq:SpinMeasure} and \eqref{eq:spin-vecEq}, the spin vector can be expressed in terms of spin magnitude $S$ as 
\beq\label{eq:polar-compEq}
    S_{\theta} = - \sqrt{g_{\theta \theta}} ~ S,
\eeq
with $S>0$ ($S<0$) corresponding to a spin vector (anti-) aligned with the total angular momentum, which by convention is always pointing along the positive z-direction. In this work, we assume that $L>0$.

The non-zero components of spin tensor calculated using Eqs.~\eqref{eq:spin-tensor} and \eqref{eq:polar-compEq} read
\bea\label{eq:ST-1}
	S^{tr}&=& -\frac{S}{\mu} \sqrt{-\frac{g_{\theta \theta}}{g}} ~ p_{\phi} = -S^{rt},\\\label{eq:ST-2}
	S^{t \phi} &=& \frac{S}{\mu} \sqrt{-\frac{g_{\theta \theta}}{g}}~p_{r} = - S^{ \phi t},\\\label{eq:ST-3}
	S^{r\phi}&=&- \frac{S}{\mu} \sqrt{-\frac{g_{\theta \theta}}{g}}~p_{t} = -S^{\phi r}.
\eea
For equatorial orbits, Eqs.~\eqref{eq:EnCon} and \eqref{eq:AngMomCons} can be written in the form
\bea
	E & \equiv & - C_{(t)} = - p_{t} + \frac{M S p_\phi}{\mu~r^3},\\
	L & \equiv & C_{(\phi)} = p_{\phi} + r^2 \frac{\mu}{M} \cb -  \frac{S p_t}{\mu}.
\eea
By solving the above equations, we express $p_t$ and $p_{\phi}$ as a function of spin magnitude $S$, angular momentum $L$, and energy $E$
\bea\label{P-1}
	p_{t}&=& \mu \frac{-\hat{r}^{3}\EE + \LL \mathcal{S} - \hat{r}^2 \mathcal{S} \cb } {\hat{r}^{3} - \mathcal{S}^{2}  },\\\label{P-2}
	p_{\phi}&=& \mu M \frac{\hat{r}^3 (\LL - \EE \mathcal{S}) - \hat{r}^5 \cb } {\hat{r}^{3} - \mathcal{S}^{2} }.
\eea

\subsection{An effective potential}

\begin{figure*}[ht]
\includegraphics[width=\hsize]{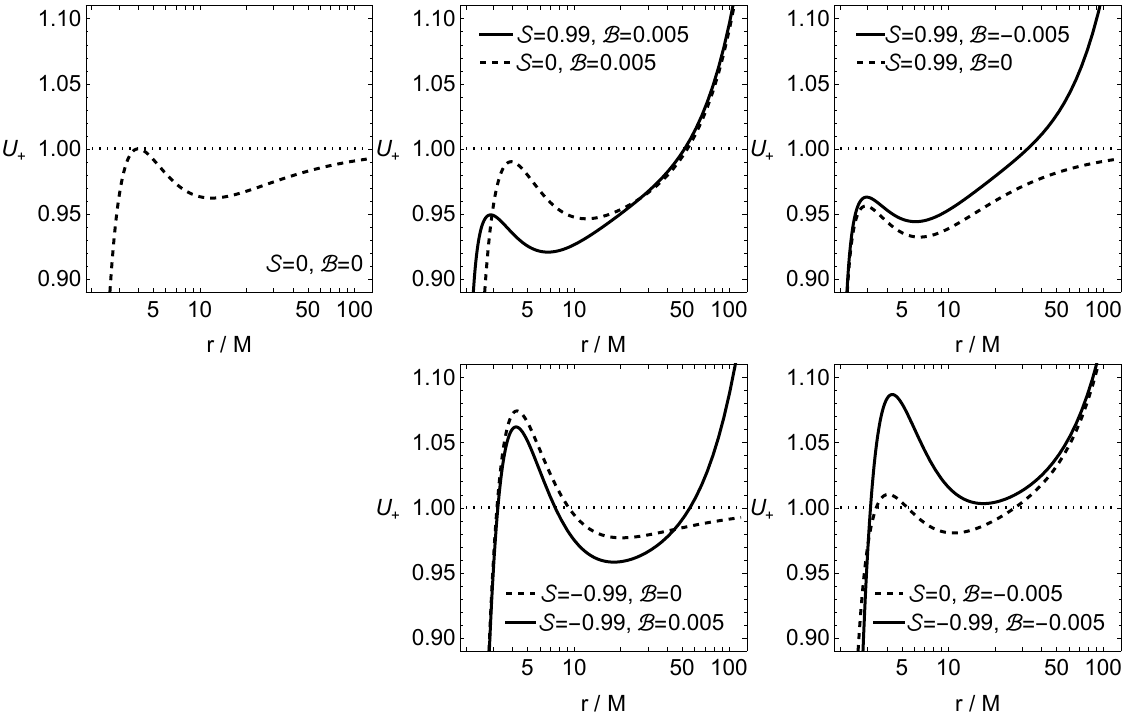}
\caption{Radial profiles of the effective potential $U_{+}$ for a spinning charged test body orbiting a Schwarzschild BH immersed in a uniform magnetic field. We set the magnetic parameter $\cb=\pm 0.005$, the spin parameter $\mathcal{S}=\pm0.99$, and orbital angular momentum $\LL = 4$. Solid curves correspond to the spinning charged case ($\mathcal{S}\neq0$, $\BB\neq0$), while the dashed black curves correspond to subcases. We compare the geodesic $\mathcal{S}=\BB=0$ case (left column) with configurations corresponding to a positive magnetic field ($\BB>0$) for both aligned and anti-aligned spin orientations (middle column), as well as to a negative magnetic field configuration ($\BB<0$) for both aligned and anti-aligned spins (right column). The horizontal dotted line in all plots indicates $\EE=1$.}
\label{fig:Energy}
\end{figure*}

Using the condition $p^\alpha p_\alpha = -\mu^2$, the radial component of four-momentum can be expressed in terms of constants of motion as 
\bea\non 
	  p^{2}_{r} &=&  \frac{\mu^2 \, \hat{r}^2}{(\hat{r}-2)^2}\left[ \frac{ \left(\LL \mathcal{S} -\EE \hat{r}^3  - r^2 \cb \mathcal{S} \right)^2}{\left(\hat{r}^3-\mathcal{S}^2 \right)^2} - \frac{\hat{r}-2}{\hat{r}} \right. \label{eq:pr-component} \\ &-& \left. \left(\hat{r}-2 \right)\hat{r}^3\frac{\left(\LL-\EE \mathcal{S}-\hat{r}^2\mathcal{B} \right)^2}{\left(\hat{r}^3-\mathcal{S}^2 \right)^2} \right]. 
\eea
Setting Eq.~\eqref{eq:pr-component} and its derivative equal to zero, and solving both equations simultaneously, one can find the expressions of energy $\EE$ and angular momentum $\LL$ for a circular equatorial orbit as a function of radius $\hat{r}$, spin magnitude $\mathcal{S}$, and magnetic parameter $\cb$. Since Eq.~\eqref{eq:pr-component} is second order in $\EE$, it admits two solutions, and it is useful to express them in terms of their roots \cite{Suzuki-Maeda:1997:prd:}
\beq\label{eq:U+}
\EE \equiv U_{(\pm)}(r, \mathcal{S}, \cb, \mathcal{L}).
\eeq
The Eq.~\eqref{eq:pr-component} can be rewritten as
\bea
\left(p^{r}\right)^2 &=& \frac{\mu^2}{\eta} \left(\alpha\, \EE^2 - 2\, \EE \frac{\beta\, (\LL-\BB \hat{r}^2)}{\hat{r}} \nonumber \right. \\\label{eq:pr2} &+& \left. \frac{\gamma\, \left(\LL - \BB \hat{r}^2 \right)^2}{\hat{r}^2} - \delta \right),
\eea
where the coefficients read
\bea
\eta &=&  \left(1 - \frac{\mathcal{S}^2}{\hat{r}^3} \right)^2, \\\
\alpha &=& 1 - \frac{(\hat{r}-2) \mathcal{S}^2}{\hat{r}^3},\\\
\beta &=&  - \frac{\mathcal{S}}{\hat{r}^2} \left(\hat{r} - 3 \right),\\\
\gamma &=& - \left(1-\frac{2}{\hat{r}}\right) + \frac{\mathcal{S}^2}{\hat{r}^4}, \\\
\delta &=& \left(1 - \frac{2}{\hat{r}} \right)\, \eta.
\eea
Rearranging Eq.~\eqref{eq:pr2}, we obtain
\beq \label{eq:EffPot2}
\left(p^{r}\right)^2 =  \frac{\alpha\, \mu^2}{\eta} \left(\EE - U_{-} \right) \left(\EE - U_{+} \right),
\eeq
where the functions $U_{\pm}$ take the form
\begin{equation} \label{eq:EffPot3}
        U_{\pm} = \mathcal{F} \pm \sqrt{\frac{\delta}{\alpha} + \mathcal{F}^2 \left(1 -\frac{ \alpha \gamma}{\beta^2} \right)}, 
\end{equation}
with
\begin{align}
 \mathcal{F} = \frac{\beta}{\alpha} \left(\frac{\LL-\BB \hat{r}^2}{\hat{r}} \right) \, .
\end{align}
The roots $U_{\pm}$ define the motion limits on the equatorial plane, for which $p^r$ is a real number. The coefficients  $\eta$ and $\alpha$ are positive\footnote{$\alpha>0$ implies that $1>\left(\frac{\mathcal{S}}{\hat{r}}\right)^2\left(1-\frac{2}{\hat{r}}\right)>\left(\frac{\mathcal{S}}{\hat{r}}\right)^2$, which holds since $\hat{r}>2\mathcal{S}$ for $\mathcal{S}\le 1$.} for motion taking place outside the Schwarzschild's horizon $\hat{r}>2$ with spin $|\mathcal{S}|\leq 1$. For $p^r$ to be a real number, $\EE \leq U_\pm$ or $\EE \geq U_{\pm}$. Moreover, since $\delta>0$ and $\gamma<0$\footnote{$\gamma<0$ implies that $\mathcal{S}^2<\hat{r}^3(\hat{r}-2)$, which holds for $\hat{r} \gtrapprox 2.11$ even if $\mathcal{S}=1$.}, then for $\hat{r}\gtrapprox 2.11$ Eq.~\eqref{eq:EffPot3} implies $U_{+}>0$ and $U_{-}<0$. Hence, for positive energy, the relevant condition is $\EE \geq U_{+}> 0$. We shall refer to $U_{+}$ as the effective potential since the motion is restricted to the region where $\EE \geq U_{+}$, with turning points occurring at $\EE = U_{+}$, where $p^r=0$ implies $u^r=0$, since from Eq.~\eqref{eq:4v4mrel} one can show that $u^r$ is proportional to $p^r$ in the equatorial plane. The minimum of $U_{+}$
\beq\label{eq:minima}
\frac{\partial U_{+}}{\partial \hat{r}}=0,
\eeq
along with
\begin{align} \label{eq:turn_point}
    \EE=U_{+},
\end{align}
defines the circular orbits in the equatorial plane. When $\partial^2 U_{+}/\partial \hat{r}^2>0$, the orbit is stable, while when $\partial^2 U_{+}/\partial \hat{r}^2<0$, it is unstable.

\begin{figure*}[!ht]
\begin{center}
\includegraphics[width=\hsize]{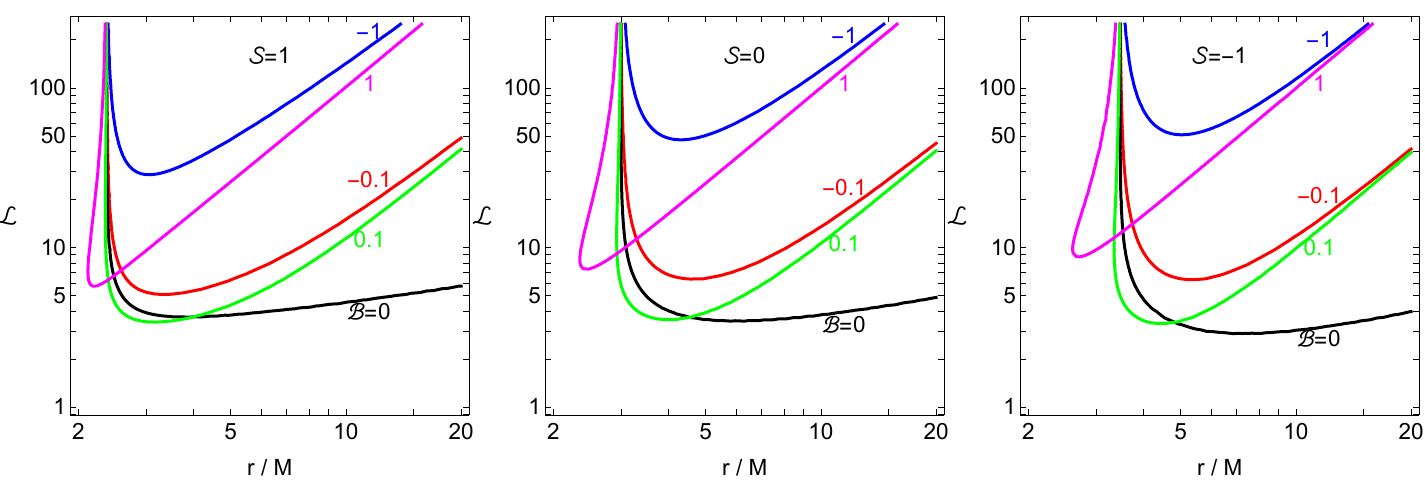}
\end{center}
\caption{Radial profiles of the angular momentum of equatorial circular orbits for a spinning charged test body moving around a Schwarzschild BH immersed in a uniform magnetic field. The middle panel corresponds to the non-spinning case ($\mathcal{S}=0$), while the left and right panels show spin configurations aligned ($\mathcal{S} >0$) and anti-aligned ($\mathcal{S}<0$) with the z-axis, respectively. The black curves correspond to the neutral case ($\cb=0$). 
\label{fig:LL}}
\end{figure*}

Figure \ref{fig:Energy} presents the behavior of the effective potential $U_+$ of a spinning charged test body moving around a Schwarzschild BH immersed in a uniform magnetic field. Since we cannot cover all the parameter space, we shall use as a reference case a geodesic effective potential, in which the spin-curvature force and Lorentz force are absent. We choose an effective potential for $\LL=4$ with a single shallow minimum and a local maximum shown in the top left plot of Fig.~\ref{fig:Energy}. We choose $\EE=1$, which intersects $U_+$ at two points, between which a range of bound orbits lies. The intersection of $\EE=1$ with the local maximum implies the existence of an unstable circular orbit from which a separatrix comes out. When we set the magnetic parameter $\BB=0.005$, but keep $\mathcal{S}=0$ (dashed curve), the effective potential's local maximum drops in value. When the spin is set to $\mathcal{S}=0.99$ (continuous curve), the maximum drops even further, and both local extrema approach the BH, as shown in the top middle plot of Fig.~\ref{fig:Energy}, with respect to the geodesic case. The drop of the local maxima implies that some of the orbits that were bounded in the geodesic case can fall into the BH, while the shift of the local extrema towards the BH implies that bounded orbits can approach closer the BH. A similar behavior can be seen in the top right plot of Fig.~\ref{fig:Energy}, where we set the spin to $\mathcal{S}=0.99$ and keep $\BB=0$ (dashed curve) and then change the magnetic parameter to $\BB=-0.005$. In these examples, the spin aligned with the $z$ axis, $\mathcal{S}=0.99$, results in a lowering of the local extrema of $U_{+}$ and a shift toward the BH. The behavior changes when we set $\mathcal{S}=-0.99$, which is anti-aligned to the $z$-axis, and keep $\BB=0$ (dashed curve in the bottom middle plot of Fig.~\ref{fig:Energy}), then with respect to the geodesic case, the local extrema rise, especially the local maxima, and the distance between the turning points lessens. This implies fewer bound orbits. If we keep $\mathcal{S}=-0.99$ and set $\BB=0.005$ (continuous curve in the bottom middle plot of Fig.~\ref{fig:Energy}), then with respect to the dashed curve, the extrema drops a little bit, but the distance between the turning points decreases even further. This implies even less bounded orbits with smaller eccentricities. In the bottom-right plot of Fig.~\ref{fig:Energy}, the dashed curve represents a case with a non-spinning particle and a magnetic parameter $\BB=-0.005$, showing that the local extrema rise and the distance between the turning points decreases relative to the geodesic case. Setting $\mathcal{S}=-0.99$ along with $\BB=-0.005$ raises the local minimum of the continuous curve (right-bottom plot of Fig.~\ref{fig:Energy}) above $\EE=1$, implying that there are no bounded orbits. In these examples, the anti-align spin to $z$-axis, i.e., $\mathcal{S}=-0.99$, resulted in raising the local extrema of $U_+$. 

 \begin{figure*}[!ht]
\begin{center}
\includegraphics[width=0.9\hsize]{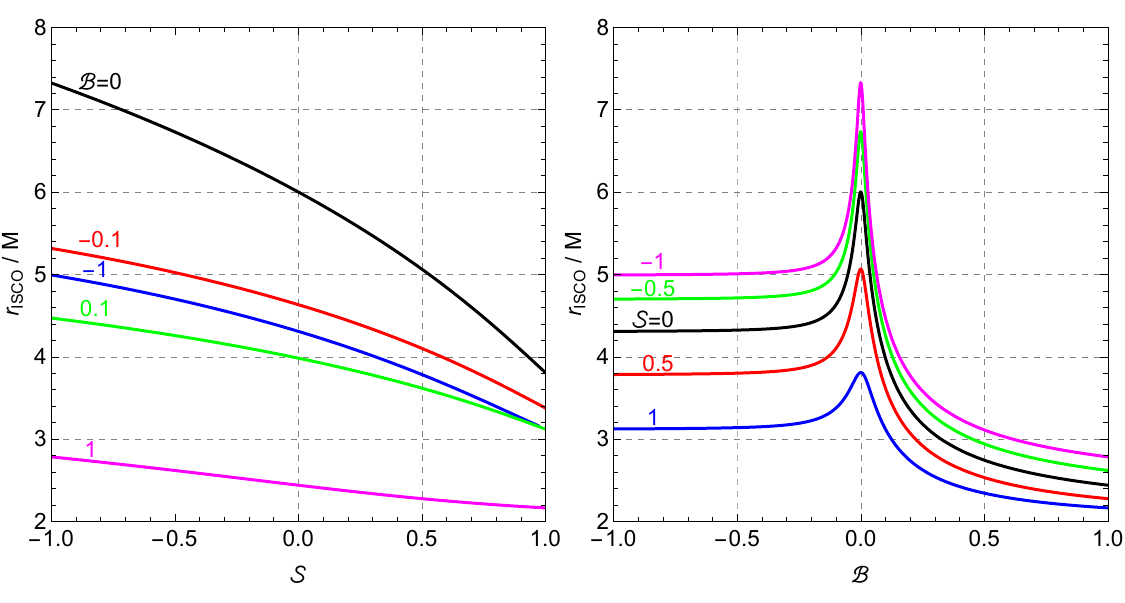}
\end{center}
\caption{The radial position of ISCOs in dependence on the spin $\mathcal{S}$ and magnetic field parameter $\cb$ for spinning charged test particles orbiting around a Schwarzschild BH in an equatorial plane. In the left and right plots, black curves correspond to the spinning neutral and the non-spinning charged cases, respectively.
\label{fig:ISCO}}
\end{figure*}

The local extrema of $U_+$ correspond to circular orbits, and in Fig.~\ref{fig:Energy} we have seen how they shift as the spin-curvature and Lorentz forces change. Hence, in this system, the circular motion is governed by the interplay between gravity, the Lorentz force generated by the magnetic field, and the spin-curvature force produced by the coupling of the test body's spin to the spacetime curvature. Solving the system of Eqs.~\eqref{eq:minima}-\eqref{eq:turn_point} provides circular equatorial orbits and for given $\hat{r}$, we can find the respective $\EE$ and $\LL$ for each radius and plot them. In particular, the angular momentum $\LL$, which is required for circular motion, as a function of the radius is provided in Fig.~\ref{fig:LL}. In the middle panel of Fig.~\ref{fig:LL}, the spin-curvature force is switched off ($\mathcal{S}=0$) and the Lorentz force can act either repulsively or attractively, depending on the direction of the motion, i.e., the sign of $\LL$, and the sign of the charge. Namely, when the Lorentz force is repulsive, the electromagnetic interactions tend to support the orbit against gravity and allow circular motion at smaller radii; this is reflected by a shift of the minimum of $\LL(r)$ towards smaller radii as the magnetic parameter $\BB$ increases. When the Lorentz force is attractive, it acts together with gravity, and the minimum of $\LL(r)$ is displaced towards larger radii.  

Spin changes these behaviours through spin-curvature coupling, and its effect depends on the spin orientation. For spin aligned with the $z$-axis, the spin-curvature force produces a radial shift in the same sense as that induced by the Lorentz force, acting repulsively when the Lorentz force is repulsive and attractively when it is attractive. This reinforces the electromagnetic effect and leads to a stronger shift of the minimum towards smaller radii, as we can see by comparing the left panel and the middle panel of Fig.~\ref{fig:LL}. For anti-aligned spin configurations, the spin-curvature force shifts the minimum in the opposite direction, to larger radii, compared to the aligned case, as we can see from Fig.~\ref{fig:LL}. The minima $\displaystyle \partial \LL/\partial r=0$ seen in Fig.~\ref{fig:LL} are ISCOs. We discuss ISCOs in detail in the next section, where instead of finding them from Fig.~\ref{fig:LL}, we employ a more straightforward method with the help of $U_+$.

\begin{table*}[!ht]
\centering
 \renewcommand{\arraystretch}{1.5}
 \setlength{\tabcolsep}{8pt}
\begin{tabular}{c | c | c c | c}
 Case & Condition & $\LL > 0$ & $\LL < 0$ & Force \\
\hline
\multirow{2}{*}{spinning neutral} & $\mathcal{S} > 0,\,\, \cb = 0$ & repulsive & attractive & \multirow{2}{*}{spin-curvature} \\
& $\mathcal{S} < 0,\,\, \cb = 0$ & attractive & repulsive &  \\[2pt]
\hline
\multirow{2}{*}{non-spinning charged} & $\mathcal{S} = 0, \,\, \cb > 0$ & repulsive & attractive & \multirow{2}{*}{electromagnetic} \\
& $\mathcal{S} = 0,\,\, \cb < 0$ & attractive & repulsive &  \\[2pt]
\hline
\multirow{4}{*}{spinning charged} & $\mathcal{S} > 0,\,\, \cb > 0$ & repulsive & attractive &
  \multirow{4}{*}{\begin{tabular}{@{}c@{}}
  spin-curvature \\[2pt] + \\[2pt] electromagnetic
  \end{tabular}} \\
& $\mathcal{S} > 0,\,\, \cb < 0$ & combined & combined &  \\
& $\mathcal{S} < 0, \,\, \cb > 0$ & combined & combined &  \\
& $\mathcal{S} < 0, \,\, \cb < 0$ & attractive & repulsive &  \\
\hline
\end{tabular}
\caption{Classification of the forces and the resulting motion for spinning charged test particles. The motion can be divided into three cases. The first two rows correspond to spinning neutral particles ($\mathcal{S}\neq0,~\BB=0$), where only the spin-curvature force is present. The next two rows represent non-spinning charged particles ($\mathcal{S}=0,~\BB\neq0$), where only the electromagnetic (Lorentz) force acts. The remaining rows correspond to spinning charged particles ($\mathcal{S}\neq0,~\BB\neq0$) where both spin-curvature and electromagnetic forces contribute.
}
\label{tab1}
\end{table*}

\subsubsection{Innermost stable circular orbits}

ISCOs are important from the astrophysical point of view as they govern thin (Keplerian) accretion disks inner edge. The ISCO, situated at an inflection point of the effective potential, can be determined with the help of the system of equations
\beq\label{ISCO-Eq}
    U_{+} = 0, \quad \frac{\partial U_{+}}{\partial r} = 0, \quad \frac{\partial^2 U_{+}}{\partial r^2} = 0.
\eeq
This inflection point defines the limit between stable and unstable circular orbits. Namely, a small radial perturbation of a stable orbit induces an oscillation around the orbit; while in the case of an unstable orbit, a small radial perturbation will drive the perturbed trajectory exponentially far from its circular origin. In the case of ISCO, a radial perturbation drives the perturbed trajectory linearly with time away from its circular origin.

To see how the radius of ISCOs changes with the change of the parameters $\mathcal{S}$ and $\BB$, we numerically solve the system of Eqs.~\eqref{ISCO-Eq} and plot the radial positions of ISCOs for spinning charged particles in dependence on the spin $\mathcal{S}$ and magnetic parameter $\cb$ in Fig.~\ref{fig:ISCO}. The radius of ISCO decreases with the increase of the magnetic field $\cb$ or spin $\mathcal{S}$ of the spinning charged body. The spin of the body adds an intrinsic contribution to the total angular momentum, allowing the particle to get closer to the region of a compact object. The ISCOs of spinning charged bodies always lie below the ISCOs of spinning neutral bodies. When the spin of the body is directed along the z-axis, the particles have smaller ISCOs as compared to the case when the spin is anti-aligned. In the presence of a repulsive Lorentz force, the ISCOs get closer to the BH with the increase of the magnetic field, while the attractive Lorentz force shifts the ISCOs to constant radii above BH horizon. For two subcases, i.e., spinning neutral bodies ($\cb=0$), and non-spinning charged ($\mathcal{S}=0$) bodies, our results coincide with \cite{Suz-Mae:1998:PhRvD:}, and \cite{Kol-Stu-Tur:2015:CLAQG:}, respectively, while for non-spinning neutralbodies ($\cb=\mathcal{S}=0$), we obtain a well-known radius $r = 6M$.    

\subsubsection{Classification of forces and motion}


As we have seen up to this point, on the basis of the sign of the magnetic field parameter $\cb$ or spin $\mathcal{S}$, the spinning charged body motion can be distinguished into different situations, presented in Tab.~\ref{tab1}. The spin-curvature force is analogous to the electromagnetic Lorentz force, with spin replacing the electric charge as a coupling constant. We compare the motion of two subcases, the purely gravitational case when the electromagnetic field is neglected, and the purely electromagnetic case when the spin-curvature force is neglected. 

For spinning neutral particles, the effective potential shows the symmetry $ (\LL, \mathcal{S}) \leftrightarrow (-\LL, -\mathcal{S}) $, and we have only two situations
\begin{itemize}
\item[+] plus configuration, here $\LL>0, \mathcal{S}>0$ (equivalent to $\LL<0, \mathcal{S}<0$) - the total orbital angular momentum and the spin have the same sign. The spin is aligned with the orbital angular momentum, and the spin-curvature force behaves as a repulsive force.
\item[-] minus configuration, here $\LL>0, \mathcal{S}<0$  (equivalent to $\LL<0, \mathcal{S}>0$) - the total angular momentum and the spin have opposite signs. The spin is anti-aligned with the orbital angular momentum, and the spin-curvature force behaves as an attractive force.
\end{itemize}
For non-spinning charged particles ($\mathcal{S}=0,~\BB\neq0$), the effective potential shows the symmetry $(\LL, \cb) \leftrightarrow (-\LL, -\cb) $, and the following two situations are possible
\begin{itemize}
\item[+] plus configuration, here $\LL>0, \cb>0$ (equivalent to $\LL<0, \cb<0$) - the total angular momentum and the magnetic field have the same sign. The Lorentz force behaves as a repulsive force.
\item[-] minus configuration, here $\LL>0, \cb<0$  (equivalent to $\LL<0, \cb>0$) - the total angular momentum and the magnetic field have opposite sign. The Lorentz force behaves as an attractive force.
\end{itemize}
The situation is more complicated when both electromagnetic and spin-curvature forces are combined. However, the motion can be classified into eight classes depending on the orientation of $\LL$, $\mathcal{S}$, and $\cb$, and the following four situations are possible 
\begin{itemize}
\item[(i)] Case, $\LL>0,~ \mathcal{S}>0,~ \cb>0$ (equivalent to $\LL<0,~ \mathcal{S}<0,~ \cb<0$) - the total angular momentum, the spin, and the magnetic field have the same sign. The magnetic field lines and the spin of the test body are aligned with the total angular momentum. Both spin-curvature and Lorentz forces behave as repulsive forces.
\item[(ii)] Case, $\LL>0,~\mathcal{S}<0,~\cb<0$  (equivalent to $\LL<0,~\mathcal{S}>0,~\cb>0$) - the spin and the magnetic field have the same sign, but the angular momentum $\LL$ has the opposite sign. The magnetic field lines and the spin of the test body are anti-aligned with the total angular momentum. Both spin-curvature and Lorentz forces behave as attractive forces. 
\item[(iii)] Case, $\LL>0,~\mathcal{S}>0,~\cb <0 $ (equivalent to $\LL<0,~\mathcal{S}<0,~\cb>0)$ - the spin and the total angular momentum have the same sign, while the magnetic field $\cb$ has the opposite sign. The spin of the body is aligned with the total angular momentum, and the magnetic field lines are oriented in the opposite direction with respect to the total angular momentum. Here, the spin-curvature force is repulsive, and the Lorentz force acts as an attractive force.
\item[(iv)] Case, $\LL>0,~\mathcal{S}<0,~\cb>0 $ (equivalent to $\LL<0,~\mathcal{S}>0,~\cb<0)$ - the angular momentum and the magnetic field parameters have the same sign, while the spin of the body is with opposite sign. The spin of the body is anti-aligned with its total angular momentum, while the magnetic field lines are oriented in the same direction as the total angular momentum. Here, the spin-curvature force is attractive, and the Lorentz force is repulsive.   
\end{itemize}

\subsubsection{Analytical expressions for circular orbits up to $\mathcal{O}(\mathcal{S}^2)$}

Having closed-form analytical expressions at hand has several advantages, such as enabling faster calculations and providing deeper insight into the dynamics. Following this line of thought, in this section, we provide analytical expressions for $\LL$ and $\EE$ up to $\mathcal{O}(\mathcal{S}^2)$ as shown below.

The time and radial component of Eq.~\eqref{eq:4v4mrel} on the equatorial plane can be written as a function of constants of motion, in the form
\begin{widetext}
\bea
\label{time-Eq}
\frac{\d \hat{t}}{\d \hat{\tau}} &=& \frac{\ce}{f(\hat{r})} - \left( \frac{\LL}{\hat{r}^2} - \BB  \right)  \left( \frac{1}{ \hat{r}-2} + 2 \BB \left( \LL - \BB \hat{r}^2\right)\right) \mathcal{S} 
+ \left(\frac{\EE \mathcal{F}_0}{\hat r^4 (\hat r-2)} \right) \mathcal{S}^2
+ \mathcal{O}(\mathcal{S}^3),\\\label{radial-Eq}
\frac{\d \hat{r}}{\d \hat{\tau}} &=& \mathcal{F}_{1} - \EE \mathcal{F}_{1} \left(2 \BB +  \frac{\left(\hat{r}-3 \right) \left( \LL - \BB \hat{r}^2 \right)}{\left(\hat{r} - 2 \right) \left(\hat{r}^2 + \left( \LL - \BB \hat{r}^2 \right)^2  \right) - \EE^2 \hat{r}^3 } \right) \mathcal{S} + \left(\frac{\mathcal{F}_1 \mathcal{F}_2}{2} \right) \mathcal{S}^2 + \mathcal{O}(\mathcal{S}^3), 
\eea
\end{widetext}
where the function $\mathcal{F}_{1}$ takes the form 
\beq
\mathcal{F}_{1} =  \frac{\sqrt{\hat{r}^2 \left[2 + \left(\EE^2-1\right) \hat{r} \right]  - \left(\hat{r} - 2 \right) \left(\LL - \BB \hat{r}^2 \right)^2}}{\sqrt{\hat{r}^3}},
\eeq
while the expressions for $\mathcal{F}_0$ and $\mathcal{F}_2$ are provided in the Appendix \ref{sec:anal_exp}. Following \cite{Wald:1984:book:}, we take the square of Eq.~\eqref{radial-Eq} and rearrange the terms up to $\mathcal{O}(\mathcal{S}^2)$ such that we arrive at 
\beq \label{eq:VeffApp}
\frac{1}{2} \left(\frac{\d \hat{r}}{\d \hat{\tau}}\right)^2 + V_{\rm eff}(\hat{r}, \EE, \LL, \mathcal{S}, \cb) = 0,
\eeq
where an appropriately approximate effective potential $V_{\rm eff}$ takes the form 
\bea\non
V_{\rm eff} &=& -\frac{\EE^2}{2}+\frac{1}{2} f(\hat{r}) \left[1 + \left(\frac{\LL}{\hat{r}} - \BB \hat{r} \right)^2 \right] \\\non &+& \EE \left[ \frac{\chi}{\hat{\hat{r}}^3} - 2 \cb\, f(\hat{r}) \left(\frac{\LL}{\hat{r}} - \cb \hat{r} \right)^2 \right] \mathcal{S} \\ &+& \frac{1}{2 \hat r^{12}}\mathcal{H} \mathcal{S}^2 + \mathcal{O}(\mathcal{S}^3),\label{eq:Veffap}
\eea
with
\beq
\chi  = \cb \hat{r}^2 \left(1 + \left( 2\, \ce^2 - 1 \right) \hat{r} \right) - \LL \left(\hat{r}-3 \right), 
\eeq
while the expression for $\mathcal{H}$ is provided in the Appendix~\ref{sec:anal_exp}. For vanishing spin $\mathcal{S}$, Eq.~\eqref{eq:Veffap} reduces to the effective potential for non-spinning charged particles \cite{Kol-Stu-Tur:2015:CQGra:}. For $\BB = 0$, one can obtain the effective potential for spinning neutral particles in Schwarzschild \cite{Suzuki-Maeda:1997:prd:} up to $\mathcal{O}(\mathcal{S}^2)$, and $\mathcal{S}=\cb=0$ leads to the non-spinning neutralparticle case \cite{Wald:1984:book:}. For a given spin $\mathcal{S}$, magnetic field $\cb$, and radial distance $\hat{r}$, the solution of a system of equations
\begin{align}
  V_{\rm eff} &=0,\label{eq:Veff0}\\
  \frac{\partial V_{\rm eff}}{\partial \hat{r}} &=0, \label{eq:derVeff0}  
\end{align}
can give the energy $\ce$ and the angular momentum $\LL$. In order to find the analytical expressions for the energy $\EE$ and the orbital angular momentum $\LL$ as a function of $\hat{r}$, magnetic parameter $\BB$, and spin parameter $\mathcal{S}$, we apply the power series expansion in terms of spin $\mathcal{S}$. Specifically, we expand $\EE$ and $\LL$ as a series in $\mathcal{S}$, representing their dependence on spin, and substitute the series expansions  
\bea\label{eq:EE}
\EE &=& \EE_{0} + \EE_{1} \mathcal{S} + \EE_{2}\, \mathcal{S}^2 + \mathcal{O} (\mathcal{S}^3),\\
\LL &=& \LL_{0} +  \LL_{1} \mathcal{S} + \LL_{2} \mathcal{S}^2 + \mathcal{O} (\mathcal{S}^3),\label{eq:LL}
\eea
in Eqs.~\eqref{eq:Veff0} and \eqref{eq:derVeff0}, where $\EE_{n}(\hat r,\BB)$ and $\LL_{\rm n}(\hat r,\BB)$ for $n=0,1,2,$ are the expansion coefficients to be determined.

The resulting expressions for the coefficients $\EE_{0}(\hat r, \BB)$, $\LL_{0}(\hat r, \BB)$, $\EE_{1}(\hat r, \BB)$, and $\LL_{1}(\hat r,\BB)$ for a spinning charged test body take the form
\bea
\EE_{0} &=& \left(\hat{r}-2 \right) \frac{\sqrt{ \hat{r}-3 + 2 \BB^2 \left(\hat{r}-2 \right) \hat{r}^2 - 2\BB \mathcal{F}}}{ \left(\hat{r}-3 \right) \sqrt{\hat{r}}},\\
\LL_{0} &=& \frac{-\BB \hat{r}^2 \pm \hat{r} \mathcal{F}}{\hat{r}-3},\\\label{eq:E0}
\EE_{1} &=& -\frac{\left(\mathcal{F} - \BB \left(\hat{r} - 2 \right) \hat{r} \right)^2}{2 \mathcal{F} \left(\hat{r}-3 \right)^2 \hat{r}}, \\
\LL_{1} &=& \frac{ \left(\hat{r}-2 \right) \left(\BB \hat{r}^3 + 2 \mathcal{F} \hat{r} - 9 \mathcal{F} \right)}{2 \mathcal{F} \left(\hat{r}-3 \right)^2 \sqrt{\hat{r}}} \mathcal{X}_1, 
\eea
where we have defined the functions $\mathcal{F}$ and $\mathcal{X}_1$ as
\bea
 \mathcal{F} &=& \sqrt{\hat{r} \left(1+\BB^2  (\hat{r}-2)^2 \, \hat{r}  \right)-3}\,,\\
 \mathcal{X}_1 &=& \sqrt{2 \BB^2 (\hat{r}-2) \hat{r}^2-2 \BB \mathcal{F} + \hat{r}-3}\,.
\eea
To find the analytical expressions for $\LL_1(\hat r, \BB)$ and $\LL_2(\hat r, \BB)$, we use the positive branch of the function $\LL_0(\hat r,\BB)$ following Ref.~\cite{Kol-Stu-Tur:2015:CLAQG:}. Obviously, the coefficients $\EE_{0}(\hat r, \BB)$ and $\LL_{0}(\hat r, \BB)$ correspond to the energy and the orbital angular momentum of a non-spinning charged test body orbiting a Schwarzschild BH \cite{Kol-Stu-Tur:2015:CQGra:}, while $\EE_{1}(\hat r, \BB)$ and $\LL_{1}(\hat r, \BB)$ represent the linear spin corrections to the energy and the orbital angular momentum of a spinning charged test body. The expressions for coefficients $\EE_{2}(\hat r, \BB)$ and $\LL_{2}(\hat r, \BB)$, which account for the quadratic spin corrections, are provided in the Appendix~\ref{sec:anal_exp}. 

\begin{figure*}
\begin{center}
\includegraphics[width=\hsize]{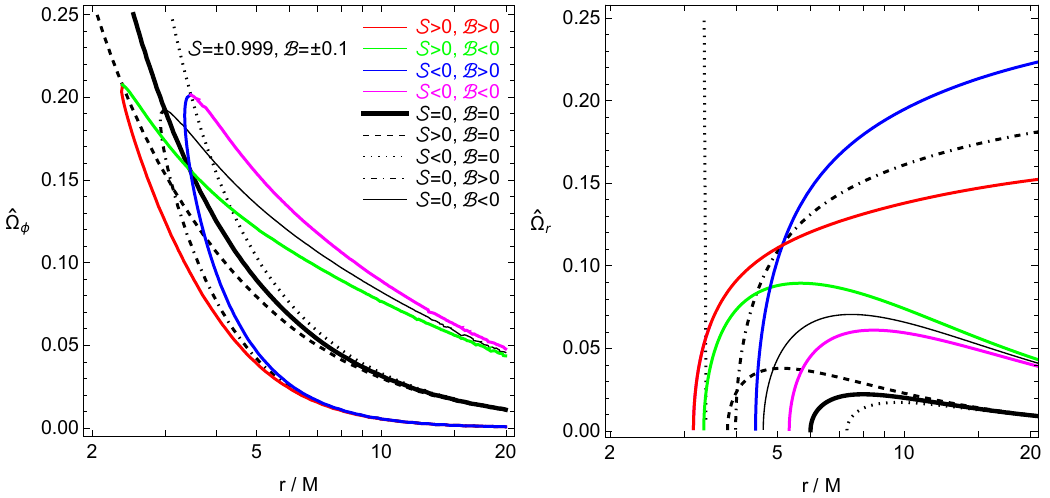}
\end{center}
\caption{
Radial profiles of orbital frequency $\hat \Omega_\phi$ (left panel) and radial frequency $\hat \Omega_{r}$ (right panel) for spinning charged test body orbiting Schwarzschid BH immersed in uniform magnetic field. We choose the parameters $\BB = \pm 0.1$, and $\mathcal{S} = \pm 0.999$. The black thick solid curves correspond to the geodesic case of a non-spinning neutral particle $(\mathcal{S}=0,~ \BB=0)$.
The black dashed, dotted curves represent spinning but neutral configurations with the particle spin aligned $(\mathcal{S}>0,~ \BB=0)$, and anti-aligned $(\mathcal{S} < 0,~ \BB=0)$ with the $z$-axis, respectively.
The black dot-dashed and thin solid curves correspond to the non-spinning charged case with attractive $(\mathcal{S}=0,~ \BB > 0)$, and repulsive $(\mathcal{S}=0,~ \BB < 0)$ Lorentz forces. 
The colored curves show combined configurations with different sign combinations of the test body spin $\mathcal{S}$ and external uniform magnetic field $\BB$. We see that the deviations from the geodesic behaviour become significant toward smaller radii, reflecting the growing influence of spin-curvature coupling and electromagnetic interactions.
}\label{fig_omega_phi}
 \end{figure*}

\subsection{Frequencies of equatorial orbits} \label{sec: Frequencies}

In this section, we focus on the radial epicyclic and azimuthal frequencies. We derive them in closed analytical form up to $\mathcal{O}(\mathcal{S}^2)$.

\subsubsection{Radial epicyclic frequency}

As we mentioned earlier in the text, if a test body is slightly displaced from the equilibrium position situated at a minimum of the effective potential $V_{\rm eff}(\hat{r}, \mathcal{S}, \BB)$ at $r_0$ and $\theta_{0} = \pi/2$, corresponding to a stable circular orbit, then the particle will start to oscillate around the minimum realizing the epicyclic motion, which can be approximated by linear harmonic oscillations. In particular, having at hand Eq.~\eqref{eq:VeffApp} of the approximative effective potential allows us to follow \cite{Wald:1984:book:} to find the epicyclic frequency around a circular equatorial orbit at radius $R_0$. Thus, we expand the approximative effective potential around $R_0$ up to quadratic terms in $r$ as
\bea\non
    V_\textrm{eff} &=&V_\textrm{eff}|_{\hat{r}=\hat{R}_0} 
    +\frac{\partial V_\textrm{eff}}{\partial \hat{r}}|_{\hat{r}=\hat{R}_0}(\hat{r}-\hat{R}_0) \\\label{eq:Veff} &+& \frac{1}{2}\frac{\partial^2 V_\textrm{eff}}{\partial \hat{r}^2}|_{\hat{r}=\hat{R}_0}(\hat{r}-\hat{R}_0)^2.
\eea   
Substituting Eq.~\eqref{eq:Veff} into Eq.~\eqref{eq:VeffApp}, and using Eqs.~\eqref{eq:Veff0} and \eqref{eq:derVeff0} at $\hat r = R_{0}$, we obtain
\begin{align}
    \left(\frac{\d \hat{r}}{\d \hat{\tau}}\right)^2 + \frac{\partial^2 V_\textrm{eff}}{\partial \hat{r}^2}|_{\hat{r}=\hat{R}_0}(\hat{r}-\hat{R}_0)^2  = 0,
\end{align}
and thus arrive at a harmonic oscillator's equation, yielding
\beq \label{eq:rad_freq}
 \hat{\omega}^{2}_r = \frac{\partial^2 V_\textrm{eff}}{\partial \hat{r}^2}.
\eeq

For a spinning charged test particle subject to the combined influence of gravitational and spin-curvature couplings, as well as the electromagnetic forces, the radial frequency  of the harmonic oscillations takes the form
\bea\non
\hat{\omega}^{2}_{r} &=& \frac{1}{\hat{r}^5} \Bigl[3 \LL^2 \left(\hat{r}-4 \right) + \hat{r}^2 \left(\cb^2 \hat{r}^3 + 4 \cb \LL- 2 \right) \Bigl] \\\non &+&
\frac{2 \ce}{\hat{r}^5} \Bigl[\cb r^2 \left(1 - 2 \cb^2 \hat{r}^3 \right) - 3 \LL \left(\hat{r} - 6 \right) - 8 \BB^2 \LL \hat{r}^2 \\&-& 6 \LL^2 \cb \left(\hat{r} - 4 \right) \Bigl] \mathcal{S} +  \mathcal{Y} \mathcal{S}^2 +\mathcal{O}(\mathcal{S}^3),\label{Omega_R} 
\eea
where the specific angular momentum $\LL$ is given by Eq.~\eqref{eq:LL}, and the coefficient $\mathcal{Y}$ of the quadratic spin term is provided in the Appendix~\ref{sec:anal_exp}. The first term in Eq.~\eqref{Omega_R} corresponds to the non-spinning charged case \cite{Kol-Stu-Tur:2015:CLAQG:}. For $\cb= \mathcal S=0$, Eq.~\eqref{Omega_R} reduces to the radial angular frequency of the harmonic oscillation for non-spinning neutral particles \cite{Wald:1984:book:}.

The angular frequencies $\omega_{\alpha}$ measured with respect to the particle's proper time $\tau$, given by
\beq
	\omega_\alpha = \frac{\d x^\alpha}{\d \tau},
\eeq
where $\alpha \in {r, \theta, \phi}$, are associated with the angular frequencies measured by the static distant observers $\Omega$, given by
\beq
\Omega_{\alpha} = \omega_{\alpha} \frac{\d \tau}{\d t}.
\eeq
The radial $\Omega_{r}$ angular frequency associated with the distant observer for the spinning charged test particle harmonic oscillations around a Schwarzschild BH is given by
\bea\non
\hat{\Omega}^{2}_{r}  &=& \frac{(\hat{r}-2)^2}{\ce^2 \hat{r}^7} \rho_{1} + \frac{(\hat{r}-2)^2}{\ce^3 \hat{r}^8} \Bigg[2 \BB \hat{r}^2 \left(\BB^2 \hat{r}^3 - 3 \right) \\\non &+& 2 \EE^2 \hat{r} \Big \{6 \BB J_{z}^2 \left( \hat{r} - 4 \right) + \LL \hat{r} \left( 8 \BB^2 \hat{r} - 3  \right) \Big \} \\ &+& 36 \EE^2 \LL\, \hat{r} + \rho_{2} \Bigg] \mathcal{S} + \frac{\mathcal{Z}}{\EE^4 \hat{r}^{13}} \mathcal{S}^2 + \mathcal{O}(\mathcal{S}^3). \label{distant_Omega_R}
\eea
The coefficients $\rho_{1}$ and $\rho_{2}$ take the form
\bea
\rho_{1} &=& 3 \LL^2 \left( \hat{r} - 4 \right) + 4 \BB \LL r^2 + \hat{r}^2 \left( \BB^2 \hat{r}^3 - 2 \right),\\ \rho_{2} &=& \frac{2 \LL}{\hat{r}^2} \Bigl[1 + 2 \BB \left(\hat{r}-2 \right) \left( \BB \hat{r}^2 - \LL \right) \Bigl] \rho_{1},
\eea
and the specific energy $\EE$ and specific angular momentum $\LL$ are given by Eqs.~\eqref{eq:EE} and \eqref{eq:LL}, respectively, while the coefficient $\mathcal{Z}$ of the quadratic spin term is given in the Appendix~\ref{sec:anal_exp}.

The graphical behavior of the radial frequency $\Omega_{r}$ as a function of radial distance $\hat{r}$ for a spinning charged test body is shown in the right panel of Fig.~\ref{fig_omega_phi}, with nine possible combinations of spin $\mathcal{S}$ and magnetic field $\BB$. In the non-spinning neutral case, i.e., geodesic case ($\BB=0,~ \mathcal{S}=0$), the radial frequency $\Omega_{r}$ decreases as the particle approaches the BH and vanishes at a radius $r=6M$. When particle spin $\mathcal{S}$ is included while the magnetic field $\BB$ is set to zero ($\BB=0,~ \mathcal{S}\neq0$), the behavior of $\Omega_{r}$ changes depending on the orientation of the particle's spin. For aligned spin with the z-axis, the radial frequency profiles shift toward smaller radii compared to the geodesics case, indicating that radial oscillations persist closer to the BH. In contrast, when spin is anti-aligned with the z-axis, the radial profiles shift towards larger radii, indicating a reduction in the range over which stable radial motion is possible. These differences are more prominent at the smaller radii, while at large radii, both cases gradually approach the geodesic behavior.

For the non-spinning charged case ($\BB\neq0,~ \mathcal{S}=0$), the presence of the magnetic field $\BB$ also modifies the radial frequency through the Lorentz force. An attractive Lorentz force increases the radial frequency $\hat \Omega_r$ and allows stable radial oscillations over a wider range of radii, whereas a repulsive Lorentz force tends to decrease the radial frequency. When both spin-curvature and electromagnetic effects are included ($\BB \neq 0, ~\mathcal{S}\neq0$), the radial frequency exhibits different behavior depending on whether the spin-curvature and Lorentz forces act in the same direction or in opposite directions. These differences are more prominent in the strong-field region close to the BH, where both spin-curvature coupling and electromagnetic effects are more significant.

\subsubsection{Azimuthal frequency}

For circular equatorial orbits, the radial and vertical components of the four-velocity vanish ($u^r = u^\theta = 0$). In this case, the system of MPDS Eqs.~\eqref{eq:spin2} and \eqref{eq:spin3} results in trivial identities, except for the components $\mathrm{D} p^r / \d \tau$ and $\mathrm{D} S^{t \phi} / \d \tau$. These components satisfy the following relation
\beq
\frac{\d p^r}{\d \tau} = \frac{\d S^{t \phi}}{\d \tau}=0.
\eeq
After further calculations, we obtain
\bea\non
&& p^t M \left(S u^\phi - u^t \mu \right)  + 2\, \frac{\mu^2}{M} \BB\, r^3 u^\phi \\ &+& p^\phi \left(\mu\, r^3 u^\phi + 2 M S u^t \right) = 0,\label{eq:pr}
\eea
\beq\label{eq:Stphi}
 p^t \left(\mu u^\phi -\frac{M S u^t}{r^3}\right) = p^\phi \left(\mu u^t - S u^\phi\right).
\eeq
For vanishing magnetic field $\BB=0$, Eqs.~\eqref{eq:pr} and \eqref{eq:Stphi} reduce to the spinning neutral case \cite{Jafar-etal:2020PhRvD:}. 

To solve the system of Eqs.~\eqref{eq:pr} and \eqref{eq:Stphi}, it is useful to introduce the quantity $W=p^{\phi}/p^t$, which through the definition of dynamical rest mass $\mu = \sqrt{-p^\alpha p_\alpha}$, establishes a relationship between the $p^t$ component and $W$, given by
\beq \label{eq:sqrtPt}
p^t = \pm \frac{\mu}{\sqrt{-g_{tt} - g_{\phi \phi} W^2}}.
\eeq
The orbital frequency of the spinning charged test body reads
\beq
\Omega_\phi = \frac{u^\phi}{u^t}.
\eeq
Using the normalization condition of four-velocity, i.e., $u^\alpha u_\alpha = -1$, we can write
\beq
u^t = \pm \frac{1}{\sqrt{-g_{tt}-g_{\phi \phi}\, \Omega_{\phi}^2}}.
\eeq
We choose the positive root in the above expression of $u^{t}$ to ensure that the flow of the proper time and the coordinate time have the same orientation, i.e., both increase along the particle's worldline. A similar argument applies to the Eq.~\eqref{eq:sqrtPt}.

Substituting $u^{\phi} = \Omega_{\phi} u^t$ and $p^{\phi} = W p^{t}$ into Eq.~\eqref{eq:Stphi}, we obtain the expression for $W$ in terms of spin parameter $\mathcal{S}$ and orbital frequency $\hat \Omega_{\phi}$, given by
\beq\label{eq:W}
W = \frac{\mathcal{S} - \hat r^3 \,  \hat \Omega_{\phi}}{\hat r^3 \left( \mathcal{S}\,  \hat \Omega_\phi - 1 \right)}.
\eeq
For vanishing spin $\mathcal{S}=0$, Eq.~\eqref{eq:W} reduces to $W=\Omega_{\phi}$. Inserting Eqs.~\eqref{eq:sqrtPt}-\eqref{eq:W} into Eq.~\eqref{eq:pr}, and after simplication, we obtain
\bea\non
-1 + \mathcal{S} \hat{\Omega}_\phi + \frac{\hat{\Omega}_\phi  \left( \mathcal{S}-\hat{r}^3 \hat{\Omega}_\phi \right)}{\mathcal{S} \hat{\Omega}_\phi -1}+\frac{2 \mathcal{S} \left(\mathcal{S}-\hat{r}^3 \hat{\Omega}_\phi \right)}{\hat{r}^3 (\mathcal{S} \hat{\Omega}_\phi -1)} \\ + 2 \BB \hat{r}^{5/2} \hat{\Omega}_\phi  \sqrt{\hat{r}-2-\frac{\left(\mathcal{S}-\hat{r}^3 \hat{\Omega}_\phi \right)^2}{\hat{r}^3 (\mathcal{S} \hat{\Omega}_\phi -1)^2}}=0. \label{eq:Omega_Phi}
\eea
To solve the polynomial Eq.~\eqref{eq:Omega_Phi} for $\hat{\Omega}_{\phi}$, we employ a power series expansion in terms of the spin $\mathcal{S}$ of the spinning charged test body. Specifically, we expand $\hat{\Omega}_\phi$ as a series in $\mathcal{S}$, reflecting its dependence on the spin. Substituting the series expansion
\beq\label{eq:Anyl_Omega_Phi}
\hat{\Omega}_{\phi} = \hat{\Omega}_{0} + \hat{\Omega}_{1} \mathcal{S} + \hat{\Omega}_{2} \mathcal{S}^2 + \mathcal{O} (\mathcal{S}^3), 
\eeq
into the Eq.~\eqref{eq:Omega_Phi}, where $\hat{\Omega}_{n}$ ($n=0,1,2$), are the expansion coefficients to be determined, and $\mathcal{O} (\mathcal{S}^3)$ represents the higher terms which are neglected for this approximation. The resulting expressions for the coefficients $\hat{\Omega}_0(r,\BB)$, and $\hat{\Omega}_1(r,\BB)$ for spinning a charged test body around a Schwarzschild BH take the form 
\begin{widetext}
\bea
\hat{\Omega}_{0} &=& \frac{1}{\sqrt{\hat{r}^3 \left(1 + 2 \BB \hat{r} \left(\BB (\hat{r}-2) \hat{r} + \sqrt{\hat{r} \left(\BB^2 \hat{r} (\hat{r}-2)^2+1\right)-3}\right)\right)}},\\
\hat{\Omega}_{1} &=& \frac{\hat{\Omega} _0 \left(\sqrt{\hat{r}-2 - \hat{r}^3 \hat{\Omega} _0^2 } \left(2 + \hat{r}^3 \hat{\Omega} _0^2\right)-2 \BB \hat{r}^{5/2} \hat{\Omega} _0 \left(\hat{r}^3 \hat{\Omega} _0^2-1\right)\right)}{2 \hat{r}^{5/2} \left( \BB \left(2-\hat{r}+2\BB \hat{r}^3 \hat{\Omega} _0^2 \right) - \hat{\Omega} _0 \sqrt{\hat{r} \left(\hat{r}-2 - \hat{r}^3 \hat{\Omega} _0^2 \right)}\right)},
\eea
\end{widetext}
where $\hat \Omega_{0}$ corresponds to the orbital frequency for a non-spinning charged test body, whereas $\hat \Omega_{1}$ and $\hat \Omega_{2}$ show the linear and quadratic corrections due to the spin of the spinning charged test body, respectively. The expression for $\hat \Omega_{2}$ is given in the Appendix~\ref{sec:anal_exp}. For vanishing spin $\mathcal{S}=0$, Eq.~\eqref{eq:Anyl_Omega_Phi} reduces to the orbital frequency for a non-spinning charged test body around Schwarzschild BH \cite{Kol-Stu-Tur:2015:CQGra:}. 

The behaviour of the orbital frequency $\hat \Omega_{\phi}$ for a spinning charged test body around the Schwarzschlild BH has been presented in the left panel of Fig.~\ref{fig_omega_phi}. We provide a comparison of all nine possible configurations corresponding to different combinations of particle spin $\mathcal{S}$ and magnetic field $\BB$. The geodesics reference case $(\BB=0,~ \mathcal{S}=0)$ shows the monotonic decrease of the orbital frequency $\hat \Omega_{\phi}$ with increasing radius. When spin is included in the absence of a magnetic field ($\BB=0,~ \mathcal{S}\ne0$), the orbital frequency is modified depending on the spin orientation: spin aligned with the z-axis shifts the frequency profile toward smaller radii, while anti-aligned spin shifts it toward larger radii relative to the geodesic case.

For non-spinning charged particles ($\BB\ne0,~ \mathcal{S}=0$), the magnetic field also alters the orbital frequency, with an attractive Lorentz force enhancing $\hat \Omega_{\phi}$ and a repulsive Lorentz force suppressing it. When both spin and electromagnetic effects are included ($\BB \ne 0,~ \mathcal{S}\ne0$), their combined effects yield distinct behaviors, depending on whether the spin-curvature and Lorentz forces act along the same direction or oppose each other. These deviations from the geodesic behavior become more significant as the orbit approaches the BH, where spin-curvature coupling and electromagnetic effects are stronger, while at large radii, all configurations gradually approach the geodesic limit.

\begin{figure*}
\includegraphics[width=\hsize]{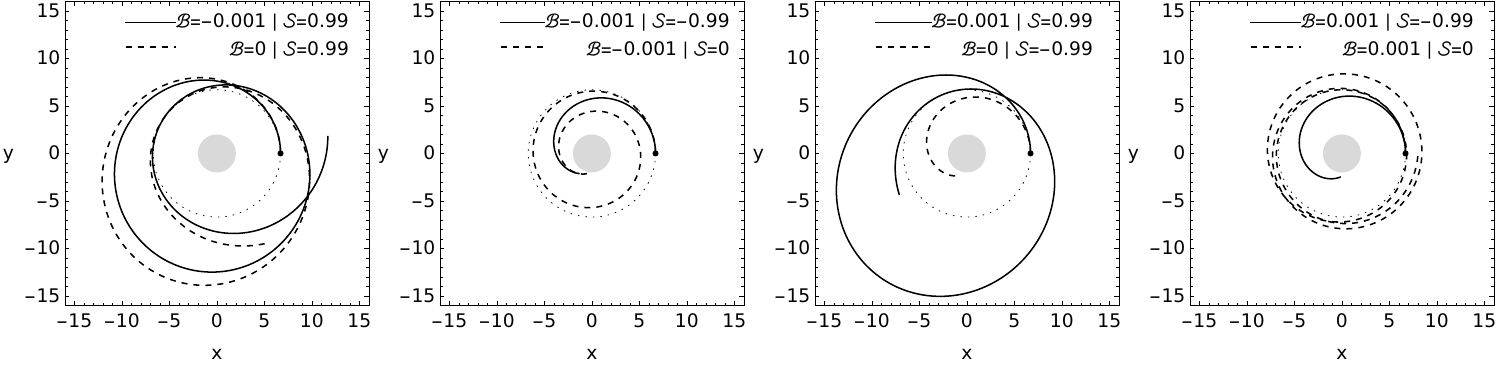}
\caption{
Spinning charged equatorial trajectories demonstrating the effects of Lorentz and spin-curvature forces on the test body.
The particle initial position ($\hat{r}_0=6.7$), orbital velocity ($\hat p^\phi=0.0822342$), and total integration time ($\hat \tau_{\rm end}=300$) remain the same for all the trajectories. These initial conditions correspond to a geodesic ($\mathcal{S}=\cb=0$) circular orbit (dotted circle).
The dashed curves in the first and third columns correspond to spinning neutral subcase ($\cb=0, \, \mathcal{S} = \pm 0.99$), while the dashed curves in the second and fourth columns correspond to non-spinning charged subcase ($\cb = \pm 0.001, \, \mathcal{S} = 0$). The solid curves correspond to spinning charged particles by employing the four combinations of $\cb=\pm0.001$, and $\mathcal{S}=\pm0.99$.
\label{fig:trajectories1}}
\end{figure*}

\begin{figure*}
\includegraphics[width=\hsize]{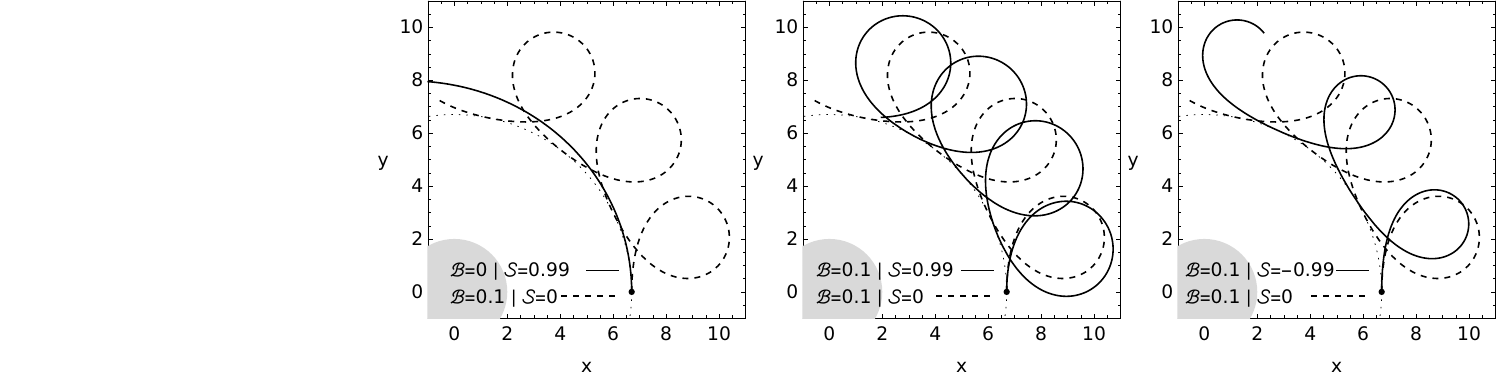}
\caption{
Spinning charged particle equatorial trajectories demonstrating the effects of Lorentz and spin-curvature forces on charged particle motion curls. The particle initial position ($\hat{r}_0=6.7$), orbital velocity ($\hat p^\phi=0.0822342$), and total integration time ($\hat \tau_{\rm end}=90$) are the same for all trajectories and correspond to a geodesic ($\mathcal{S}=\cb=0$) circular orbit (dotted curves). The dashed curves show the non-spinning charged case ($\mathcal{S}=0,~\BB \neq 0$) with magnetic parameter $\BB=0.1$, and we use them as a reference case to explore how the spin-curvature force affects the curl motion.
In the first column, the solid curve corresponds to the neutral aligned spin case with spin parameter $ \mathcal{S} = 0.99$, while the solid curves in the second and third columns correspond to the spinning charged particles case with aligned and anti-aligned spin configurations, respectively, with parameters $\BB=0.1,~\mathcal{S}= \pm 0.99$.
}
\label{fig:trajectories2}
\end{figure*}

\subsection{Equatorial trajectories}\label{subsec:eq-trajec}

To illustrate the geometric properties of particle motion and the resulting orbital structure, we numerically integrate the motion of a spinning charged test body in the background of a non-rotating Schwarzschild BH immersed in a uniform magnetic field, employing the Gauss Runge-Kutta scheme. The initial conditions cannot be chosen arbitrarily to evolve the system; instead, the initial data must satisfy the constraints~\eqref{eq:TD_spin-cond}, \eqref{eq:DynMassDef}, \eqref{eq:SpinMeasure}, \eqref{eq:EnCon}, and \eqref{eq:AngMomCons}. We initialize the motion in the equatorial plane and impose conditions to ensure that it remains confined to this plane. In particular, we set $\theta = \pi/2$, $\hat r = \hat r_{0}$, $t = \phi = 0$, and impose vanishing radial and polar momentum components, $p^r =0$, $ p^{\theta}=0$. Moreover, we choose the constants of motion $\ce$, $\LL$, $\mathcal{S}$ along with the magnetic parameter $\cb$, such that the particle motion remains bounded. The initial components of the spin tensor are determined from Eqs.~\eqref{eq:ST-1}-\eqref{eq:ST-3}, while the time and azimuthal components of the four-momentum, $p^t$ and $p^{\phi}$ are obtained from Eqs.~\eqref{P-1} and \eqref{P-2}. To check the accuracy of our numerical integration, we compute the numerical relative errors of the conserved quantities, namely the energy $\EE$, angular momentum $\LL$, mass $\mu$, and spin $\mathcal{S}$ throughout the evolution, using the constraint Eqs.~\eqref{eq:TD_spin-cond}, \eqref{eq:DynMassDef}, \eqref{eq:SpinMeasure}, \eqref{eq:EnCon}, and \eqref{eq:AngMomCons}. In addition to the conserved quantities, we also examine the preservation of the TD SSC \eqref{eq:TD_spin-cond} throughout the evolution of the system. More details are provided in the Appendix \ref{Apndx:IntAcc}. 

As already explained in Sec.~\ref{sec: SchBHMF}, the motion in an equatorial plane is integrable, and therefore regular. To demonstrate the effect of the spin-curvature and Lorentz forces on the particle dynamics, we plot representative examples of equatorial trajectories in Figs.~\ref{fig:trajectories1} and \ref{fig:trajectories2}. We show the trajectories for all nine possible combinations of spin orientation ($\mathcal{S}<0$, $ \mathcal{S}=0$, $ \mathcal{S}>0$) and magnetic field ($\BB<0$, $\BB=0$, $\BB>0$) by setting $\cb=\pm0.001$, and $\mathcal{S}=\pm 0.99$, while keeping the same initial position and orbital velocity corresponding to a reference geodesic. Depending on the sign of the magnetic field parameter $\BB$ and spin parameter $\mathcal{S}$, the particle experiences either an attractive or repulsive contribution to the radial force, which modifies its orbital motion relative to the geodesic case.

In the non-spinning case ($\mathcal{S}=0$), the Lorentz force alone determines the deviation from the geodesic motion. A repulsive Lorentz force $(\cb>0)$ pushes the particle outward, producing wider trajectories and slowing the azimuthal motion compared to the reference orbit (dashed curve in the fourth column of Fig.~\ref{fig:trajectories1}). In contrast, an attractive Lorentz force $(\cb<0)$, strengthens the inward tendency, leading to tighter motion and a higher orbital speed (dashed curve in the second column of Fig.~\ref{fig:trajectories1}).

When spin effects are included ($\mathcal{S} \neq 0$), the spin-curvature force modifies the motion depending on the orientation of the particle's spin relative to the orbital angular momentum. For aligned spin ($\mathcal{S} > 0$), the spin-curvature force is repulsive, pushing the orbit outward (dashed curve in the third column of Fig.~\ref{fig:trajectories1}). The repulsive Lorentz force ($\cb > 0$), combined with the repulsive spin-curvature force, produces the strongest outward deviation (solid curve in the third column of Fig.~\ref{fig:trajectories1}). In contrast, for $\cb < 0$, the attractive Lorentz force partially offsets the spin-induced widening (solid curve in the first column of Fig.~\ref{fig:trajectories1}).

Conversely, for anti-aligned spin ($\mathcal{S} < 0$), the spin-curvature force becomes attractive, driving the motion inward (dashed curve in the third column of Fig.~\ref{fig:trajectories1}). In combination with $\cb < 0$, this can lead to rapid capture by the central BH (solid curve in the second column of Fig.~\ref{fig:trajectories1}). For $\cb > 0$, however, the repulsive Lorentz force counteracts the inward drift and reduces the radial infall of the particle (solid curve in the fourth column of Fig.~\ref{fig:trajectories1}).

In the presence of a repulsive Lorentz force ($\cb>0$), the particle motion exhibits orbital curls as it evolves along its trajectory, see Fig.~\ref{fig:trajectories2}. When the test body's spin is aligned $(\mathcal{S}>0)$ as shown in the second column of Fig.~\ref{fig:trajectories2}, the repulsive spin-curvature force reduces the orbital frequency (solid curve), leading to a slower azimuthal motion and more closely spaced curls compared to the non-spinning case ($\mathcal{S}=0$; dashed curve). In contrast, for anti-aligned spin $(\mathcal{S}<0)$ as shown in the third column of Fig.~\ref{fig:trajectories2}, the attractive spin-curvature force increases the orbital frequency, resulting in faster orbital motion and more widely separated curls (solid curve) compared to the non-spinning case (dashed curve). This behavior demonstrates that the spacing of the curls directly reflects the change in orbital frequency: positive spin decreases the angular velocity, whereas negative spin enhances it. The combined influence of the Lorentz and spin-curvature forces therefore modifies the local structure of the orbit even when the motion remains bounded. 

\begin{figure*}
\includegraphics[width=0.85\hsize]{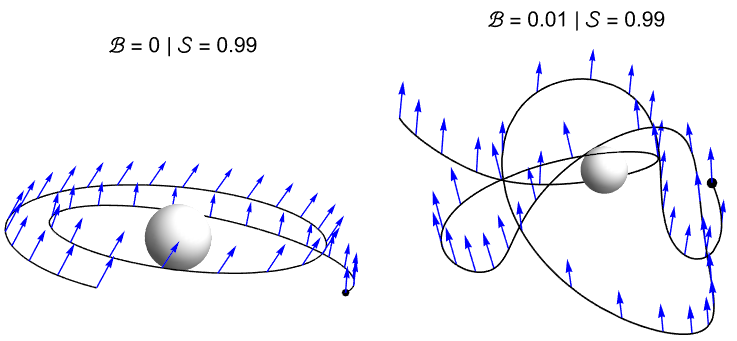}
\caption{Off-equatorial trajectories for spinning neutral ($\BB=0$, $\mathcal{S} \neq 0$; left panel) and spinning charged particles ($\BB\neq0$, $\mathcal{S} \neq 0$; right panel) with the same initial conditions ($\hat r_0 = 10.75,~ \EE = 0.95,~ \LL = 4,~ \BB = 0.01,~ \mathcal{S} = 0.99$). Blue arrows show the direction of the spin of the spinning charged test body along the trajectory, and the black dot represents the initial position of the test body. The neutral case exhibits regular, quasi-periodic motion confined near a plane. In contrast, the inclusion of the magnetic field leads to significant deviations from planar motion and the emergence of chaotic trajectories. \label{fig:off_trajectories1}}
\end{figure*}

\section{Off-equatorial orbital dynamics}\label{sec: off-eq}

In this section, we investigate the off-equatorial motion of the spinning charged test body, where the spin vector, orbital angular momentum, and magnetic field of the central BH are not generally aligned, and explore the dynamical features of the underlying system.

\subsection{Off-equatorial trajectories}

We now turn to the off-equatorial motion, allowing the particle to deviate from the plane $\theta = \pi/2$. Unlike the equatorial case, where the dynamics remain integrable, the inclusion of polar motion introduces additional DoF that couple nonlinearly to the radial motion through both spin-curvature and Lorentz forces. As a result, the orbital dynamics becomes significantly richer, admitting precessing and, in certain parameter regimes, chaotic trajectories.

To illustrate the influence of the spin-curvature force and the Lorentz force on particle dynamics beyond the equatorial plane, we plot representative off-equatorial trajectories in Fig.~\ref{fig:off_trajectories1}. We use the same procedure for computing initial data as described in Subsec. \ref{subsec:eq-trajec} for equatorial orbits, except for $p^{\theta}$, which is calculated using the condition $p^\alpha p_\alpha = - \mu^2$. In addition, we visualize the evolution of the particle's spin vector along the trajectory (blue arrows), computed using the corresponding expressions for the spin vector components (Eqs.~\eqref{eq:Sr_component}-\eqref{eq:SV_expression4}). 

The qualitative impact of the magnetic field is clearly demonstrated by comparing the two panels in Fig.~\ref{fig:off_trajectories1}. In the absence of a magnetic field ($\BB=0$), the spinning neutral particle exhibits, for the particular choice of parameters considered here, regular, quasi-periodic motion, with the trajectory remaining confined to a narrow region around a deformed orbital plane. In contrast, when the magnetic field is present ($\BB \neq0$), the combined action of spin-curvature coupling and the Lorentz force leads to a substantial deformation of the orbit. The motion is no longer confined to a quasi-planar region and instead explores a larger portion of the available phase space, displaying irregular and aperiodic behavior characteristic of chaos. This transition highlights the role of electromagnetic interactions in enhancing the complexity of the dynamics.

\subsection{Chaotic dynamics}\label{sec:ChaoDyn}

When an unperturbed dynamical system is subjected to a perturbation that leads to non-integrability, the non-integrable system exhibits distinct behaviors compared to its unperturbed counterpart. One prominent effect is the appearance of chaotic motion around the resonances, triggered by the perturbation as stated by the Poincar\'{e}-Birkhoff theorem \cite{Birkhoff1913}. In particular, for two DoF, after the perturbation from a resonant torus of the integrable system, only an even number of periodic orbits survive; half of them are stable and the other half unstable, creating the Birkhoff chain. In this chain, islands of stability form around the stable orbits, and they interchange with the unstable orbits, from which asymptotic manifolds (stable and unstable branches) emanate. These manifolds twist and fold in a very peculiar manner; motion on such a manifold is called \emph{chaotic}. A stable asymptotic manifold and an unstable asymptotic manifold cross: if the manifolds emanate from periodic orbits of a single resonance, they are called \emph{homoclinic}; if they emanate from different resonances, they are called \emph{heteroclinic}. Depending on the strength of the perturbation, deterministic chaos can even dominate the system's dynamics. Nevertheless, even in a weakly perturbed system as defined by the Kolmogorov-Arnold-Moser (KAM) theorem \cite{Arnold1963}, where the chaotic behaviour is minimal, the system can still exhibit non-integrable features in the vicinity of the resonances, while most of the other regular structures survive deformed. These structures are called \emph{KAM tori} and in the vicinity of the resonances, homoclinic chaos appears. As the perturbation increases, the resonances start to overlap and heteroclinic chaos appears.

To gain insight into orbital dynamics and the emergence of chaos, it is necessary to explore the structure of the phase space. This can be achieved by well-established methods. In this work, we utilize such two methods, i.e., the PS and the recurrence analysis, which are described in the following subsections.

\begin{figure*}
\begin{center}
\includegraphics[width=\hsize]{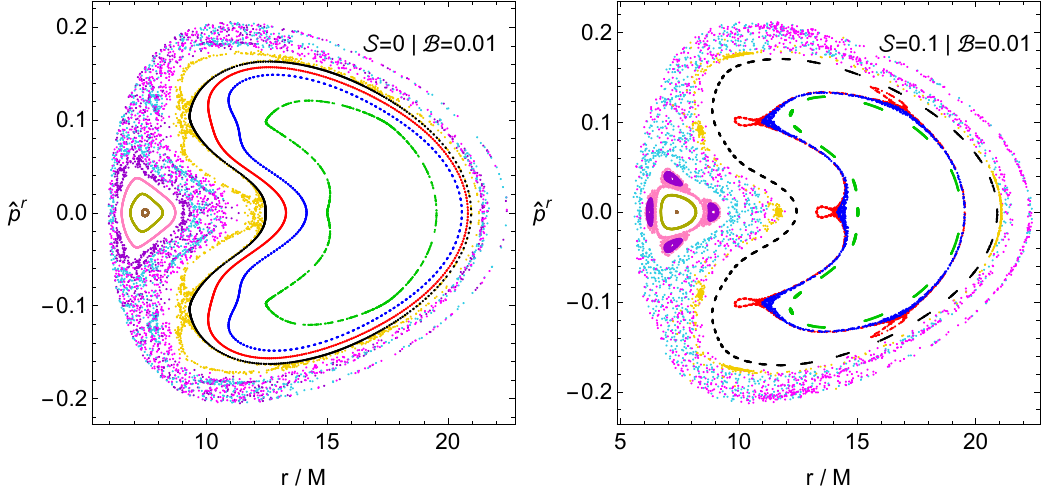}
\end{center}
\caption{The PSs computed for $\theta=\pi/2$, $\hat p^{r}=0$, and $\hat p^\theta>0$, employing eleven initial conditions with $\hat{r}_0$ varying from $6.5 $ to $15 $, with orbital parameters $\EE = 0.9555,\, \LL=4,\, \BB=0.01$, and $\mathcal{S}=0.1$. Each trajectory is represented by 1500 intersection points. The left plot corresponds to the non-spinning charged ($\mathcal{S}=0$, $\BB \neq 0$) case, while the right plot represents the spinning charged ($\mathcal{S} \neq 0$, $\BB \neq 0$) case. Note that only the non-spinning charged case (left plot) can provide a proper 2D PS, while the spinning charged case cannot.
}\label{fig_PS}
 \end{figure*}

\subsubsection{Poincar\'e sections}\label{sec: PS}

The PS is one of the most powerful and widely used techniques for visualizing the qualitative behaviour of dynamical systems in Hamiltonian systems that are reducible to two Dof. A PS exploits the symplectic structure of a Hamiltonian system and allows us to study its flow on a 2D hypersurface of the phase space. In a two DoF Hamiltonian system, each regular trajectory moves on a $\mathbb{T}^2$ torus, and these tori nest to form a tori foliation through which the Hamiltonian flow takes place. Taking a 2D cut through this foliation, which is not tangent to the Hamiltonian flow, provides the 2D hypersurface of a PS. Practically, a PS is constructed by considering the intersections of trajectories sharing the same integrals of motion with the aforementioned hypersurface. When a trajectory pierces this hypersurface, the corresponding phase space points are recorded, and the pattern constructed on the hypersurface forms the PS. Originally introduced by Poincar\'{e}, this method provides a simplified but insightful view of systems with two DoF; the PS provides a complete 2D graphical representation of a 4D phase space manifold, revealing different structure and stability features associated with resonances, regular, and chaotic motion. For example, a smooth zero-width closed curve, usually in the shape of a deformed ellipse, suggests a regular (KAM) orbit, while scattered points filling a non-zero-width region suggest chaotic motion.

As discussed in Sec.~\ref{sec: SchBHMF}, a spinning neutral body ($\mathcal{S}\neq0,\, \cb=0$), or a non-spinning charged ($\mathcal{S}=0,\, \cb\neq0$) moving around a Schwarzschild BH can be reduced to two DoF. This is why standard 2D PSs have been used in dynamical studies of the aforementioned cases, see, e.g., \cite{Suzuki-Maeda:1997:prd:, Zelenka-etal:2020:PhRvD:, Pan-Kol:2019EPJC:}. However, the combination of the Lorentz and the spin-curvature force allows us to reduce the number of DoF to only three. In three DoF, the tori foliation consists in general of $\mathbb{T}^3$ nested tori, and we need a 4D hypersurface for a PS. Hence, the standard 2D PS is not adequate to provide accurate information for the phase space structures, since a 2D hypersurface is merely a projection plane of a 4D hypersurface. To depict a 4D PS, Patsis and Zachilas in \cite{Patsis-Zachilas:1994IJBC:} suggested using 3D plots enhanced with a color spectrum that suggests the 4th dimension. Regular toroidal patterns with smooth color transition on such a 4D PS suggest a regular (KAM) orbit. Mixing of colors or the appearance of irregular patterns on the 4D PS signals chaotic behavior. This approach has been employed in studies of diverse systems \cite{Katsanikas:2011IJBC,Moges:2024IJBC}, including an approximation of the MPD equations \cite{Gera-etal:2016:PhRvD:}.

The phase space of bound geodesic orbits in the Schwarzschild spacetime contains only nested tori of regular orbits. A cut through this nesting is portrayed on a PS by zero-width ellipsoids nested around a fixed point lying in the center. Such a formation is called the \emph{main island of stability}. When additional interactions are introduced, such as spin-curvature coupling or electric charge in the presence of an external magnetic field, the integrability is broken, and the phase space develops a richer and more complex structure. An illustration of this behaviour is shown in the left panel of Fig.~\ref{fig_PS} for non-spinning charged ($\mathcal{S}=0$, $\BB=0.01$) test particles.  We observe that the KAM curves of the main island of stability are surrounded by a chaotic layer, within which smaller islands of stability associated with various resonances are embedded.

If we switch on the spin-curvature coupling by setting $\mathcal{S}=0.01$ and depict the trajectories with the same initial conditions as those in the left plot of Fig.~\ref{fig_PS}, we obtain the right panel of Fig.~\ref{fig_PS}. The latter plot is no longer a PS, but a projection of a 4D PS on the $(\hat r, \hat p^r)$ plane. The right panel looks like a distorted version of the left panel. Some structures seem to be similar in both panels, but one cannot tell for sure what is chaotic and what is not. To address this issue, we present the same trajectories using both 2D projections and 4D PSs in Fig.~\ref{fig_2D+4D}, employing the same color coding as in Fig.~\ref{fig_PS} for direct comparison. While distinct trajectories may give rise to similar patterns in the 2D projections, the corresponding 4D PSs clearly distinguish between smooth invariant structures and those that are broken or distorted. To gain further insight into the underlying dynamics, we further employ recurrence analysis discussed in the following section.

\subsubsection{Recurrence analysis}\label{sssec:recurrence_analysis}

Recurrence analysis is a method to study the properties of a dynamical system by observing its recurrences, i.e., the temporal correlation of its states. For a generic phase space of dimension $d$, let us denote the $i$-th phase space vector in a time series $\vec{x}_i$, where $i = 1,\dots,l$, and $l$ is the total number of sampled points in the trajectory. Then we define the \emph{recurrence matrix}~\cite{Marwan:2007rps}
\begin{equation}
    \mathbf{R}_{i,j} = \begin{cases}
			1, & \left|\vec{x}_i - \vec{x}_j\right|\leq \epsilon ~,\\
            0, & \left|\vec{x}_i - \vec{x}_j\right|>\epsilon ~,
		 \end{cases}\quad i,j=1,\dots,l ~,
\end{equation}
where $\epsilon$ is a parameter called the \emph{recurrence threshold}. In addition, typically, all elements on the main diagonal are set to zero. A visualization of the recurrence matrix is called the \emph{recurrence plot} (RP): a 2D plot wherein points with $\mathbf{R}_{i,j} = 1$ are plotted in black, and the others in white.

The above definition relies on the full knowledge of phase vectors in the time series. When limited data are available, such as observations of radiation from a complex source, the full phase space can be reconstructed using \emph{time delay embedding}. Given a time delay $T$ and embedding dimension $n$, the reconstructed phase space vectors are given by the expression
\begin{equation}
    \vec{X}_i = \left(\vec{x}_i, \vec{x}_{i+T}, \dots, \vec{x}_{i+\left(n-1\right)T}\right).
\end{equation}
According to Takens' theorem, there exists a diffeomorphism between the original and reconstructed phase spaces~\cite{Takens:1981emb}, and embedding faithfully reproduces the dynamics of the original system. The optimal time delay $T$ is typically estimated as the first distinct minimum of the time series' mutual information, and the embedding dimension $n$ is determined by the false nearest neighbors algorithm~\cite{Sukova:2015naa}.

A RP contains characteristic features based on which one can discern order from chaos. Primarily, a quasi-periodic orbit will yield clear lines parallel to the main diagonal stretching through the entire plot, while a chaotic orbit will only contain shorter diagonal lines within square blocks corresponding to periods of sticky motion (a chaotic orbit remaining very close to regular tori for an extended period of time, reproducing its characteristics). This holds for systems where the 2D PS method is not feasible due to, e.g., higher number of dimensions, data from only one channel, or a non-conservative nature of the system, and is thus an invaluable tool in a vast amount of applications~\cite{glg18,kopacek10,Sukova:2015naa,Sukova:2016zlf}.

Furthermore, a thorough study of the plot's features can be used to extract numerical characteristics that do not rely on a visual inspection of the RPs and can also be employed to estimate various dynamical invariants. This is called \emph{recurrence quantification analysis} (RQA). Most recurrence quantifiers involve the distribution of diagonal and vertical lines in the RP. We make use of several of these to aid in the classification of orbits. For more details on RQA, we refer the interested reader to~\cite{Marwan:2007rps}.

The recurrence rate (RR) is the average density of points in the RP,
\begin{equation}
    \mathrm{RR} = \frac{1}{l^2}\sum_{i=1}^l\sum_{j=1}^l \mathbf{R}_{i,j} ~.
\end{equation}
Naturally, when RR is very low or very high, the corresponding plots carry little information and are difficult to read. Typically, recurrence thresholds are chosen such that $\mathrm{RR}\in\left[0.05,0.2\right]$. In this work, we set recurrence thresholds such that in all RPs it holds $\mathrm{RR} = 0.1$.

Furthermore, we define the following recurrence RQA quantifiers:
\begin{itemize}
    \item determinism ($\mathrm{DET}$), the fraction of recurrence points in the RP that form diagonal lines
    \item longest diagonal line ($L_{\max}$)
    \item laminarity ($\mathrm{LAM}$), the fraction of recurrence points in the RP that form vertical lines
    \item longest white vertical line ($W_{\max}$), defined as the longest vertical line containing no recurrence points
\end{itemize}
These quantities can be used to extract information about the system without relying on an ambiguous visual interpretation. Most significantly, for a regular orbit, the $L_\mathrm{max}$ indicator is close to the length of the trajectory $l$, provided that a sufficient $l$ value is used, and $\mathrm{DET}$ is close to one, as the majority of RPs form parallel diagonal lines at a constant spacing. At the same time, since a few recurrence points are present in the spaces between the diagonal lines, both $\mathrm{LAM}$ and $W_\mathrm{max}$ remain low as both black and white vertical lines are interrupted at the level of the diagonal line spacing by this structure.

In a chaotic orbit, on the other hand, many more vertical structures are present, both white and black. This is connected to transitions between sticky periods and general chaos. Thus, both $\mathrm{LAM}$ and $W_\mathrm{max}$ then rise to much higher values than for a regular orbit. The $\mathrm{DET}$ indicator may still produce high values in the case of sticky chaos, since fairly short sticky periods will also contribute all of their recurrence points in its computation; in strongly chaotic orbits, it typically decreases to much lower values. The $L_\mathrm{max}$ will, again, respond to stickiness and produce a value corresponding to the length of the longest sticky period.

We therefore consider $L_\mathrm{max}$ the primary indicator of chaos: values close to $l$ point towards a regular orbit, very low values towards a strongly chaotic orbit, and intermediate values towards weakly chaotic sticky orbits. We also compute the other indicators mentioned above as supporting characteristics.


\begin{table}[ht]
\centering
\renewcommand{\arraystretch}{1.4}
\begin{tabular}{|c|c|c|c|c|c|c|c|c|}
\hline
$\hat{r}_0$ & $T$ & $n$ & $\epsilon$ & $\mathrm{DET}$ & $L_\mathrm{max}$ & $\mathrm{LAM}$ & $W_\mathrm{max}$ \\ \hline
 $6.50$ & $1$ & $2$ &  $0.315$ & $0.880$ & $1468$ &      $0$ &   $30$ \\ \hline
 $7.35$ & $1$ & $2$ & $0.0392$ & $0.996$ & $1495$ &      $0$ &   $66$ \\ \hline
 $8.20$ & $1$ & $2$ &  $0.197$ & $0.938$ & $1495$ &      $0$ &   $32$ \\ \hline
 $9.05$ & $2$ & $4$ &   $2.73$ & $0.740$ &  $377$ & $0.736$ & $1093$ \\ \hline
 $9.90$ & $4$ & $4$ &   $3.35$ & $0.559$ &  $180$ & $0.593$ & $1328$ \\ \hline
$10.75$ & $4$ & $5$ &   $5.15$ & $0.465$ &  $121$ & $0.595$ & $1335$ \\ \hline
$11.60$ & $2$ & $4$ &   $3.85$ & $0.765$ &  $199$ & $0.0139$ &  $283$ \\ \hline
$12.45$ & $3$ & $4$ &   $3.16$ & $0.907$ & $1462$ &      $0$ &   $33$ \\ \hline
$13.30$ & $3$ & $3$ &   $2.82$ & $0.783$ & $1477$ &      $0$ &   $36$ \\ \hline
$14.15$ & $3$ & $2$ &   $1.34$ & $0.554$ & $1483$ &      $0$ &   $78$ \\ \hline
$15.00$ & $2$ & $2$ &   $1.12$ & $0.886$ & $1425$ &      $0$ &   $46$ \\ \hline
\end{tabular}
\caption{RQA values for the non-spinning charged particle case (left panel of Fig.~\ref{fig_PS}). The recurrence thresholds $\epsilon$ and the $\mathrm{DET}$ and $\mathrm{LAM}$ indicators are rounded to three significant digits. 
}
\label{tab:RQA_spinless}
\end{table}


\begin{table}[ht]
\centering
\renewcommand{\arraystretch}{1.4}
\begin{tabular}{|c|c|c|c|c|c|c|c|c|}
\hline
$\hat{r}_0$ & $T$ & $n$ & $\epsilon$ & $\mathrm{DET}$ & $L_\mathrm{max}$ & $\mathrm{LAM}$ & $W_\mathrm{max}$ \\ \hline
 $6.50$ & $1$ & $3$ &   $0.490$ & $0.921$ & $1030$ &      $0$ & $194$ \\ \hline
 $7.35$ & $1$ & $3$ & $0.00963$ & $0.923$ & $1491$ &      $0$ &  $24$ \\ \hline
 $8.20$ & $1$ & $3$ &   $0.283$ & $0.909$ & $1487$ &      $0$ &  $25$ \\ \hline
 $9.05$ & $1$ & $3$ &   $0.270$ & $0.915$ & $1346$ &      $0$ &  $83$ \\ \hline
 $9.90$ & $2$ & $5$ &    $4.45$ & $0.734$ &  $413$ & $0.801$ & $974$ \\ \hline
$10.75$ & $2$ & $5$ &    $7.47$ & $0.514$ &  $153$ & $0.324$ & $801$ \\ \hline
$11.60$ & $2$ & $4$ &   $0.770$ & $0.857$ &  $992$ &      $0$ & $503$ \\ \hline
$12.45$ & $2$ & $4$ &    $3.47$ & $0.895$ & $1473$ &      $0$ &  $20$ \\ \hline
$13.30$ & $2$ & $3$ &    $1.61$ & $0.918$ &  $148$ &      $0$ & $425$ \\ \hline
$14.15$ & $2$ & $3$ &    $1.96$ & $0.920$ &  $305$ &      $0$ &  $75$ \\ \hline
$15.00$ & $2$ & $3$ &    $1.79$ & $0.669$ & $1479$ &      $0$ &  $16$ \\ \hline
\end{tabular}
\caption{RQA values for the spinning charged particle case (right panel of Fig.~\ref{fig_PS}). The recurrence thresholds $\epsilon$ and the $\mathrm{DET}$ and $\mathrm{LAM}$ indicators are rounded to three significant digits.
}
\label{tab:RQA_spinning}
\end{table}




\begin{figure*}
\begin{center}
\includegraphics[width=0.77\hsize]{ 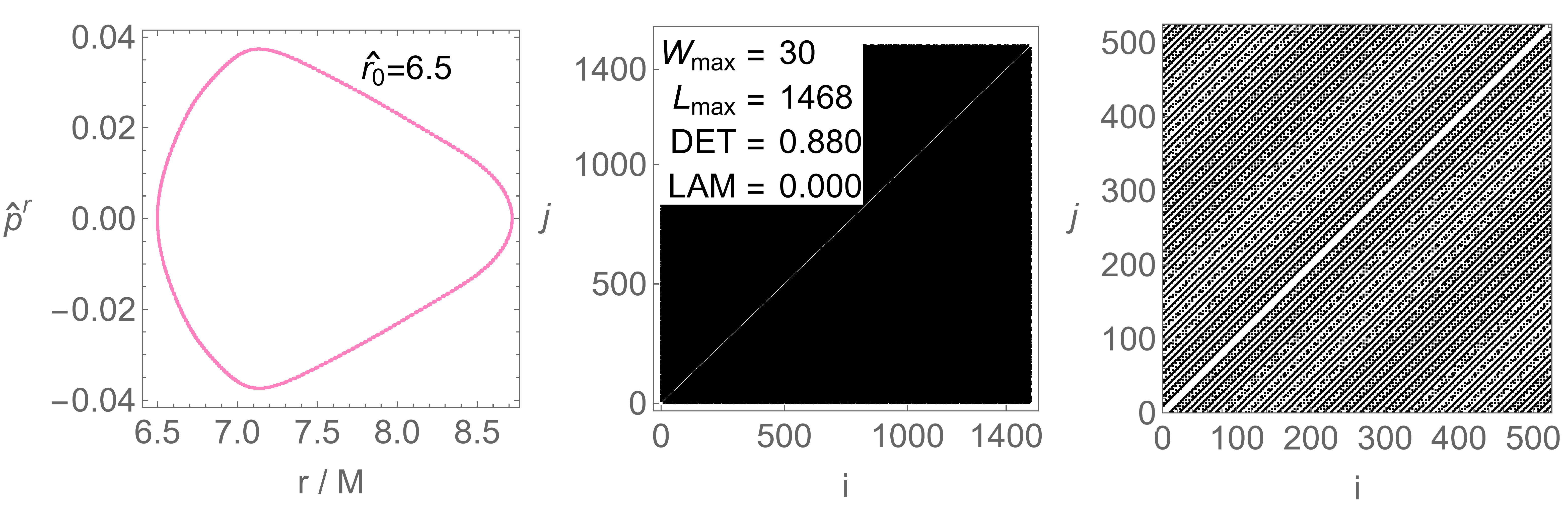}
\includegraphics[width=0.77\hsize]{ 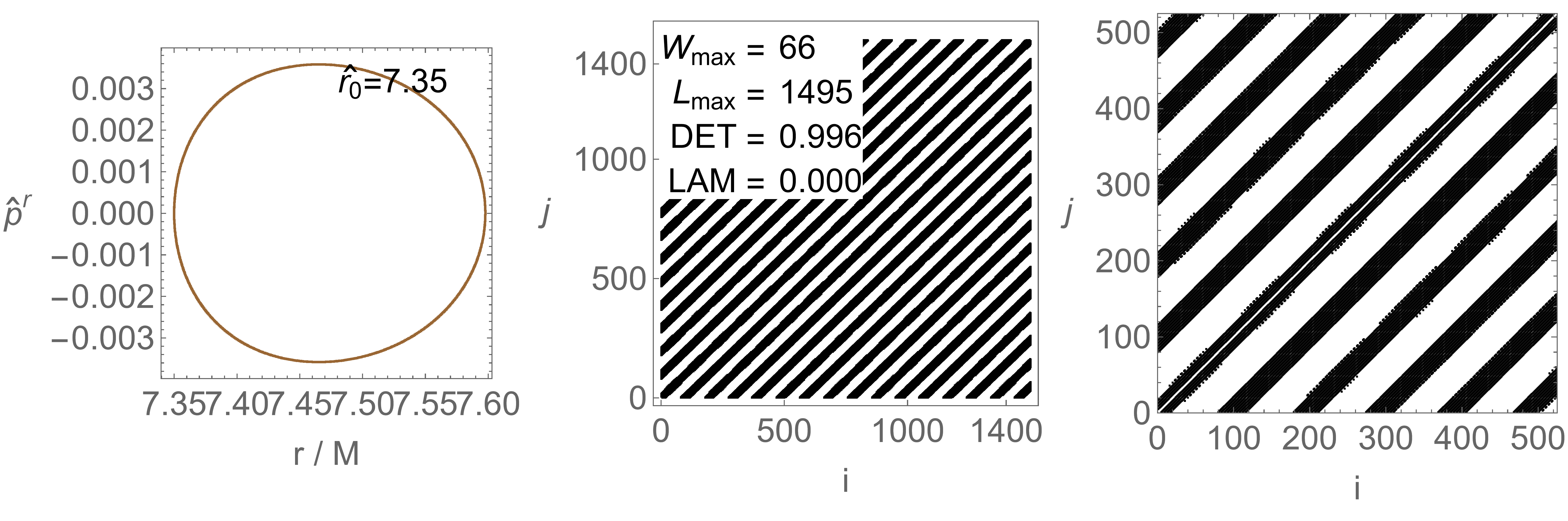}
\includegraphics[width=0.77\hsize]{ 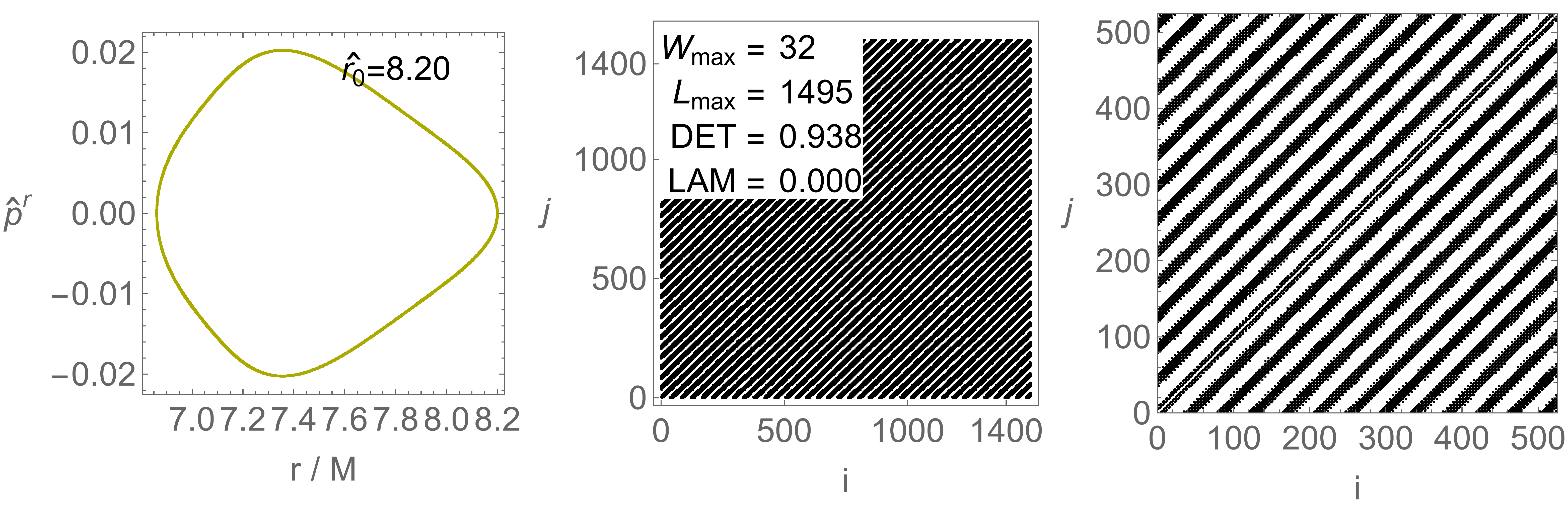}
\includegraphics[width=0.77\hsize]{ 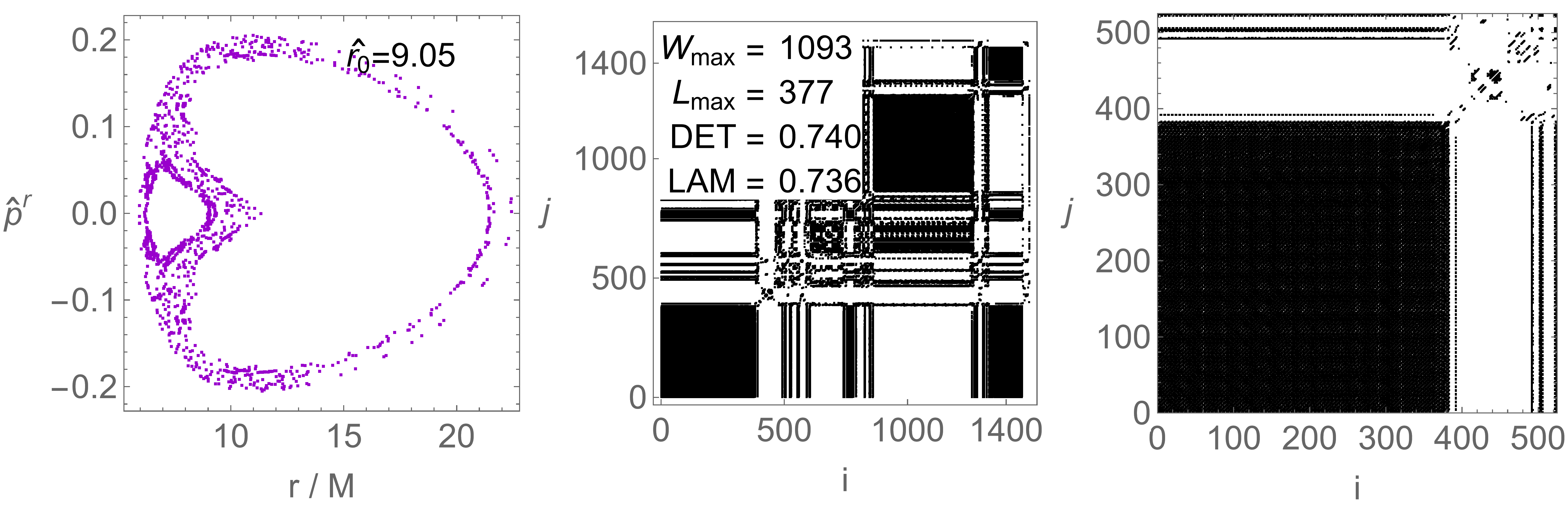}
\includegraphics[width=0.77\hsize]{ 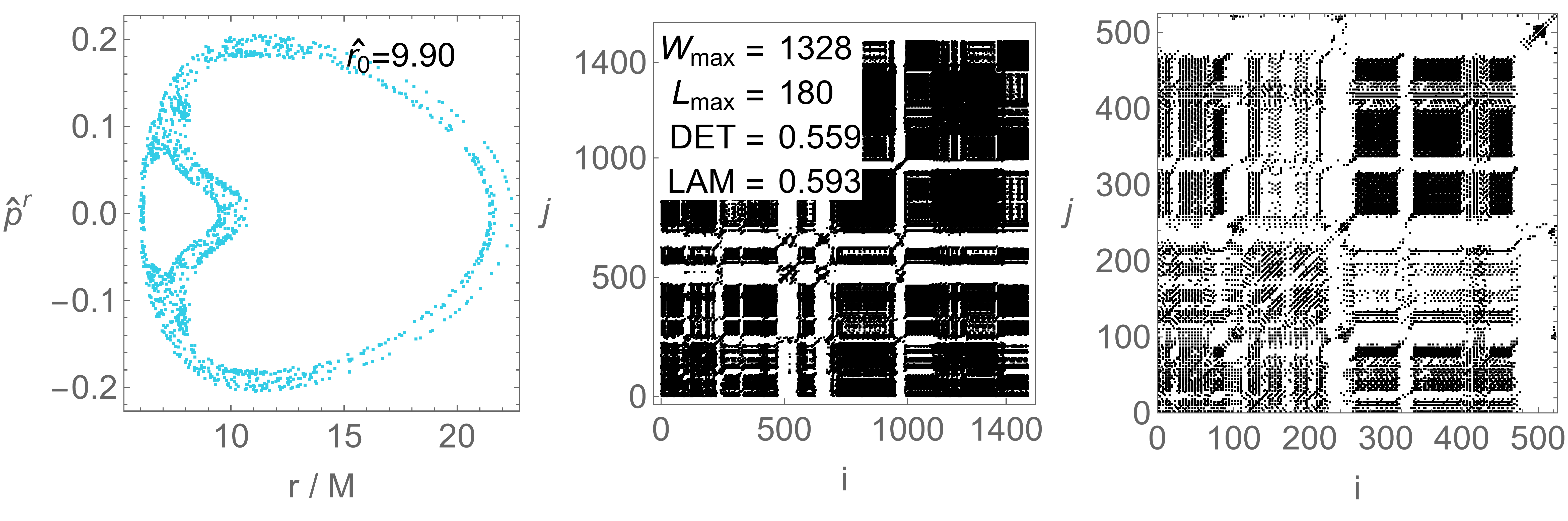}
\end{center}
\caption{
The 2D PSs (first column), RPs (second column), and zoomed-in RPs (third column) corresponding to the trajectories for a non-spinning charged particle ($\mathcal{S}=0,~\cb\neq 0$) presented in Fig.~\ref{fig_PS}, with the same color coding. Each row corresponds to a different initial radius $\hat{r}_0$, shown in the first column. The associated RQA indicators ($W_{\rm max},~ L_{\rm max}$, LAM, and DAT) are presented in the second column and Tab.~\ref{tab:RQA_spinless}.
}\label{fig2_2D+4D}
\end{figure*}

\addtocounter{figure}{-1}
\begin{figure*}
\begin{center}
\includegraphics[width=0.67\hsize]{ 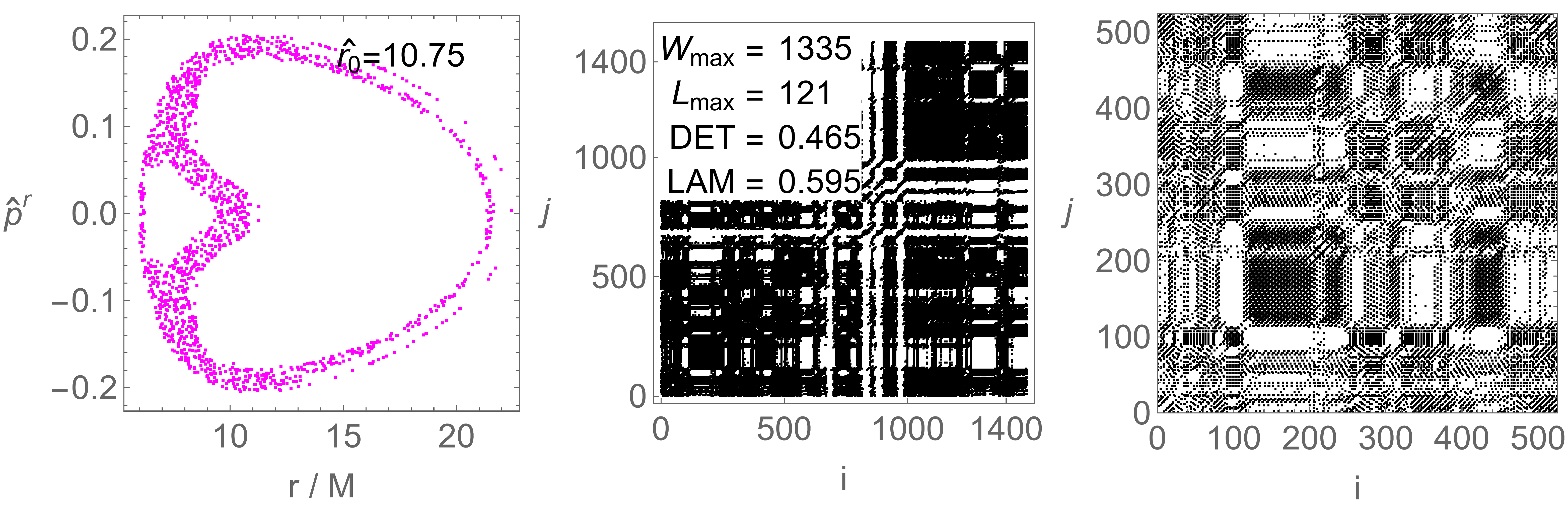}
\includegraphics[width=0.67\hsize]{ 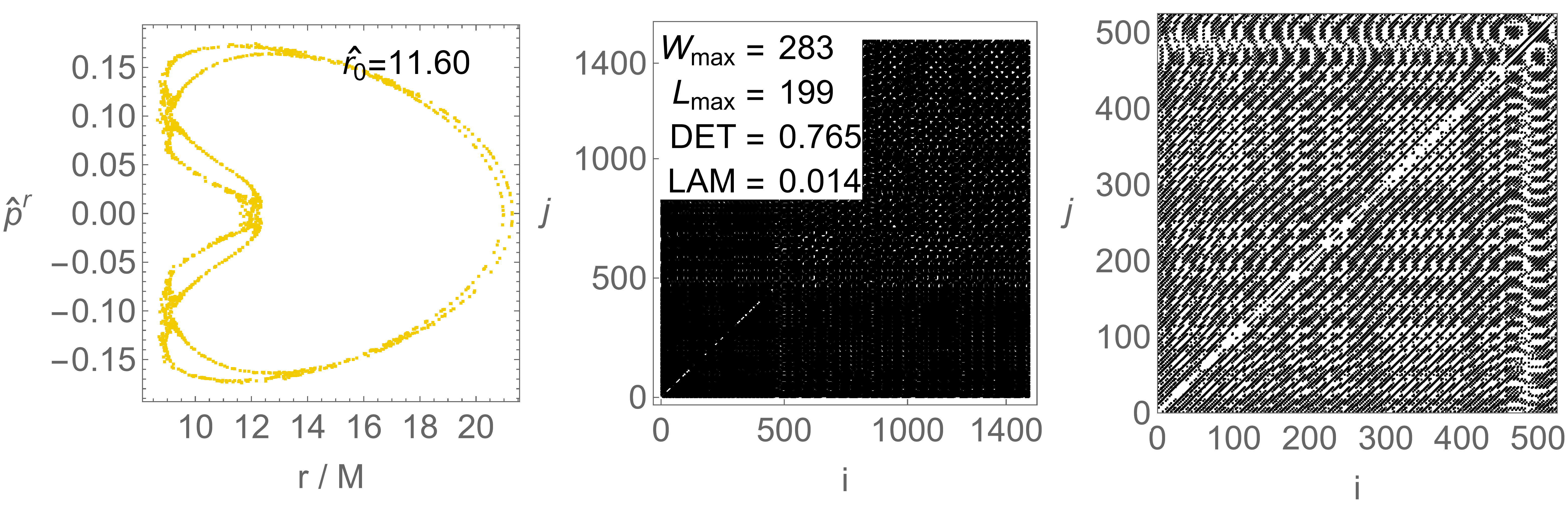}
\includegraphics[width=0.67\hsize]{ 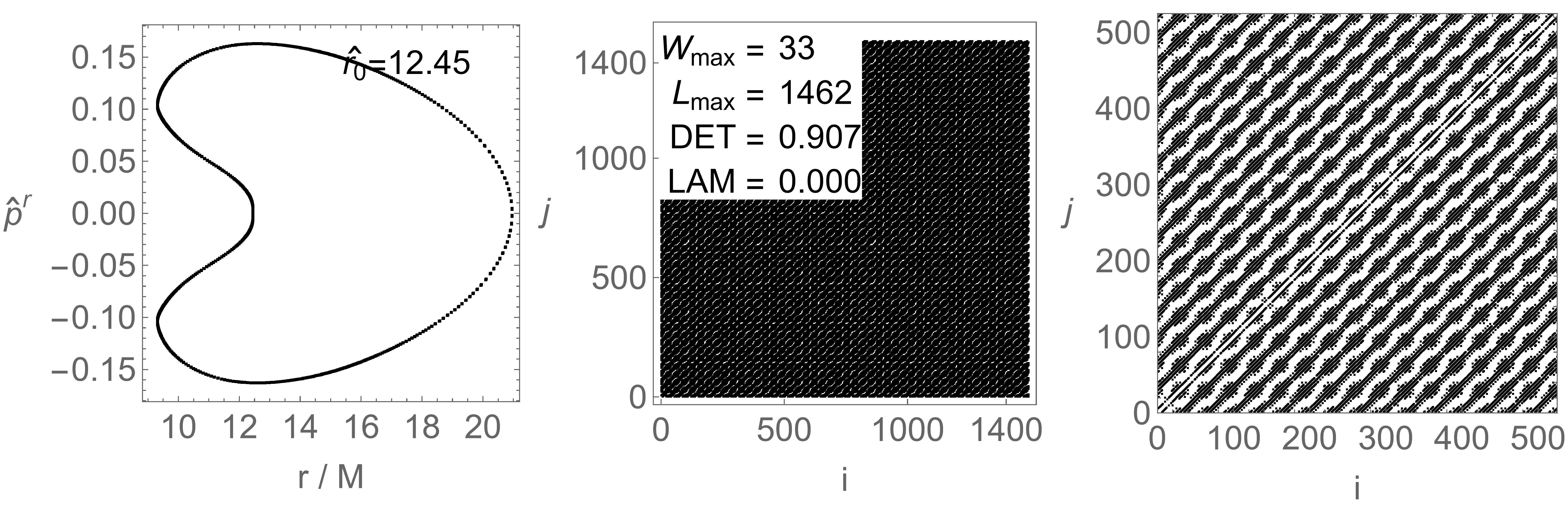}
\includegraphics[width=0.67\hsize]{ 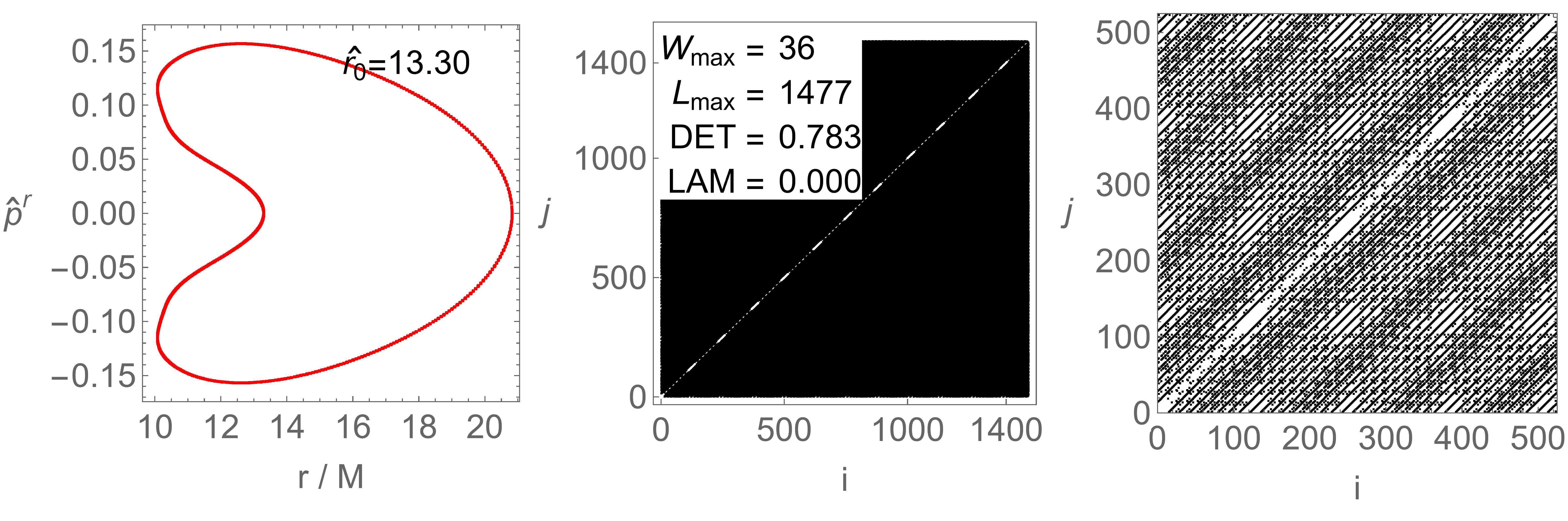}
\includegraphics[width=0.67\hsize]{ 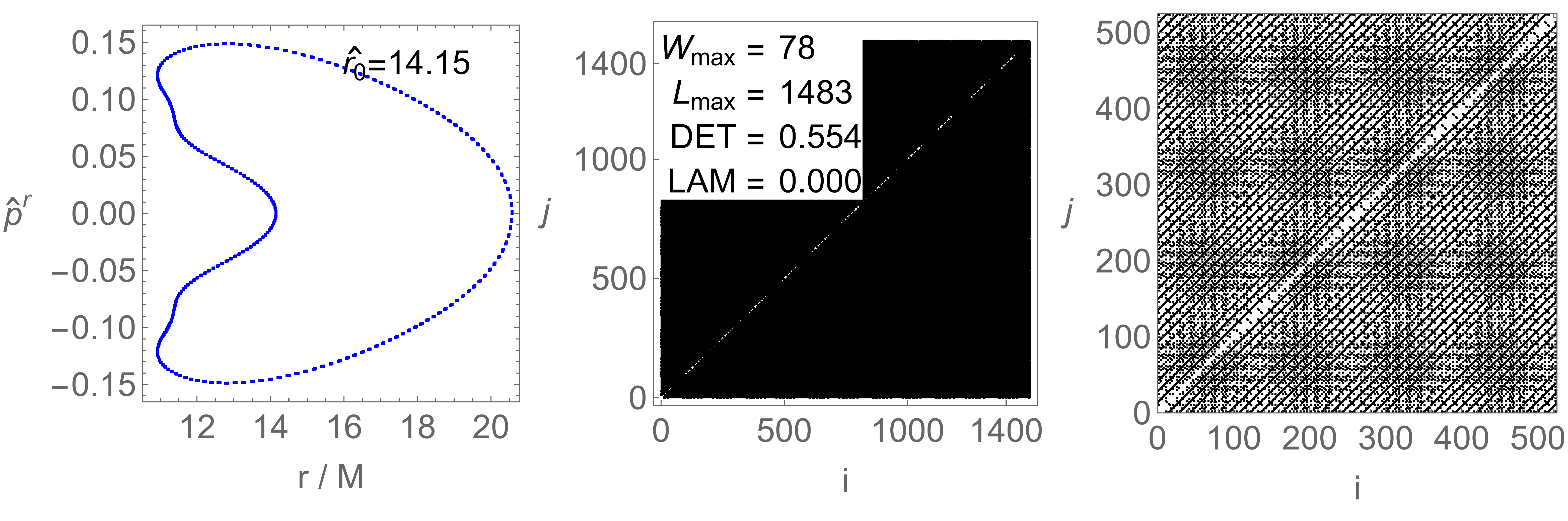}
\includegraphics[width=0.67\hsize]{ 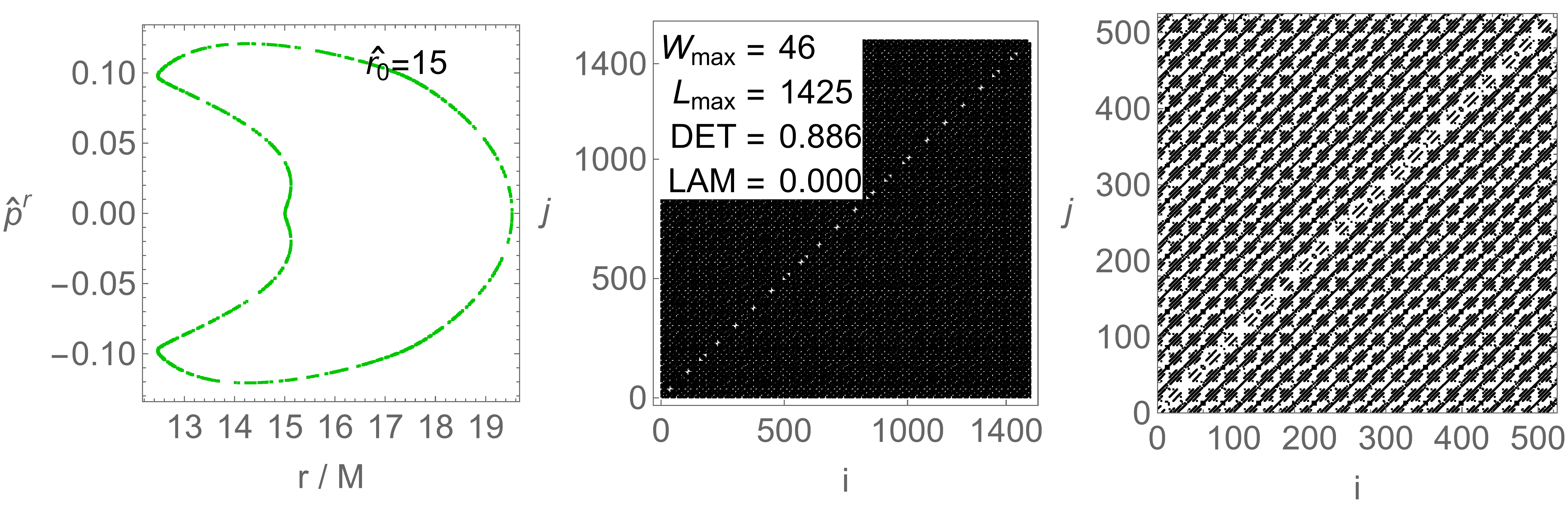}
\end{center}
\caption{Continued...} 
\end{figure*}

\begin{figure*}
\begin{center}
\includegraphics[width=\hsize]{ 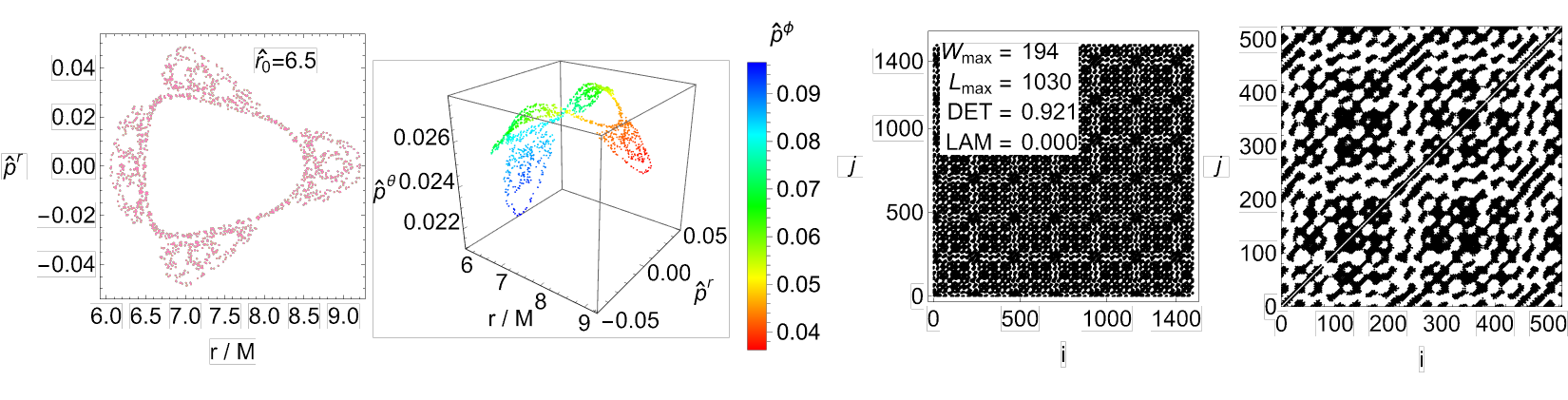}
\includegraphics[width=\hsize]{ 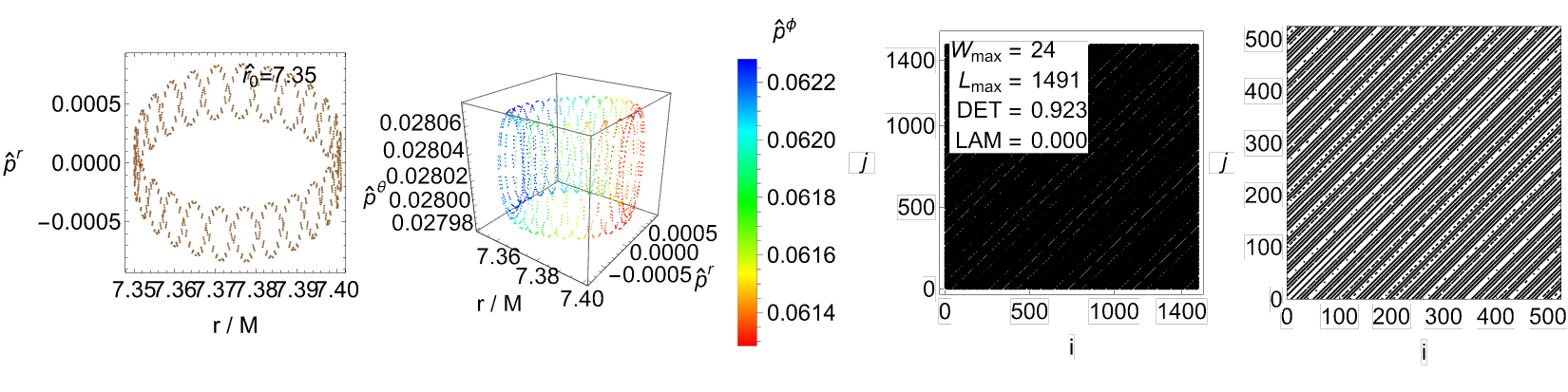}
\includegraphics[width=\hsize]{ 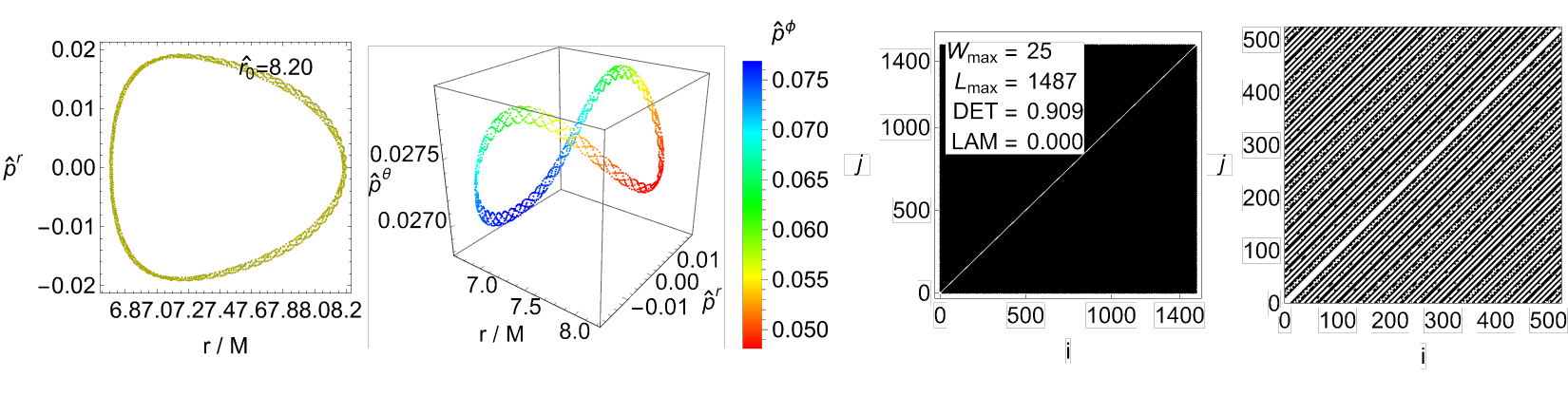}
\includegraphics[width=\hsize]{ 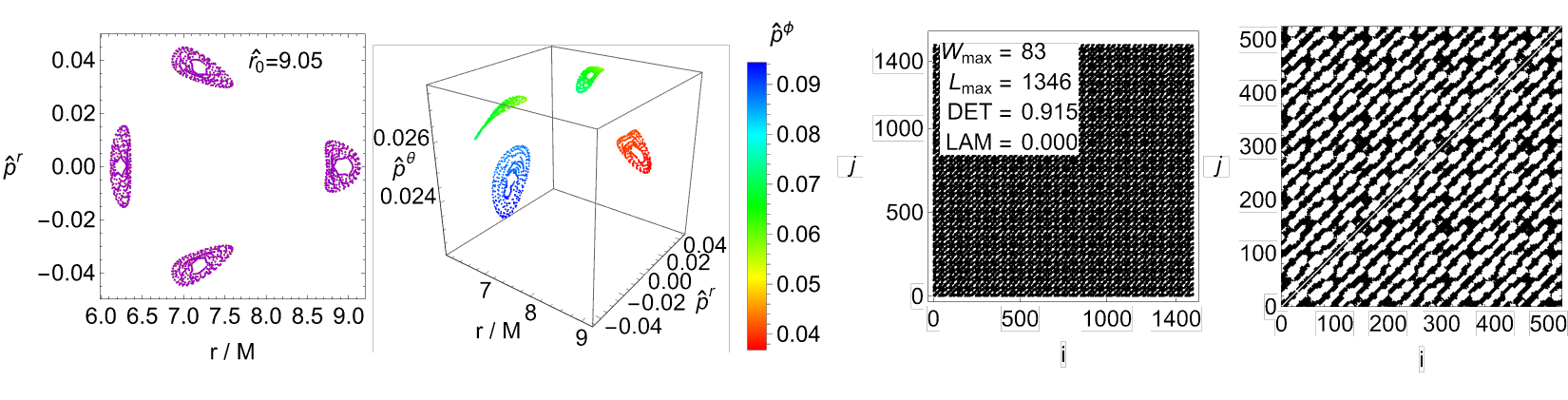}
\includegraphics[width=\hsize]{ 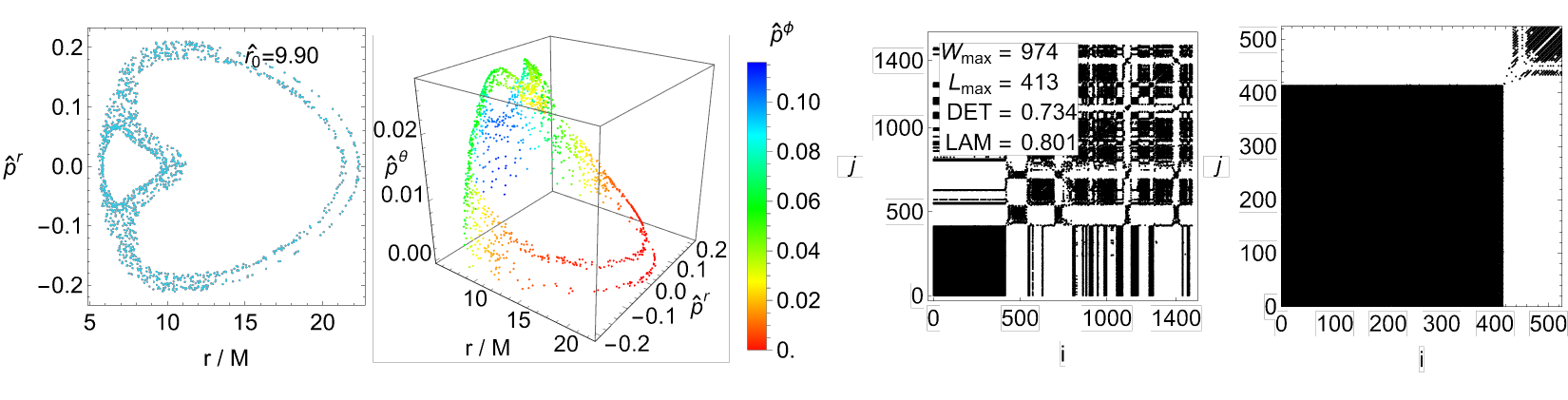}
\end{center}
\caption{The 2D projections on the ($\hat r\,, \hat p^r$) plane  (first column), 4D PSs (second column), RPs (third column), and zoomed-in RPs (fourth column) corresponding to the trajectories for spinning charged test body presented in Fig.~\ref{fig_PS}, with the same color coding. The colors in 4D plots correspond to the momentum associated with the coordinate $\phi$. Each row corresponds to a different initial radius $\hat{r}_0$, shown in the first column. The associated RQA quantifiers are displayed in the third column and Tab.~\ref{tab:RQA_spinning}.
}\label{fig_2D+4D}
\end{figure*}

\addtocounter{figure}{-1}
\begin{figure*}
\begin{center}
\includegraphics[width=0.9\hsize]{ 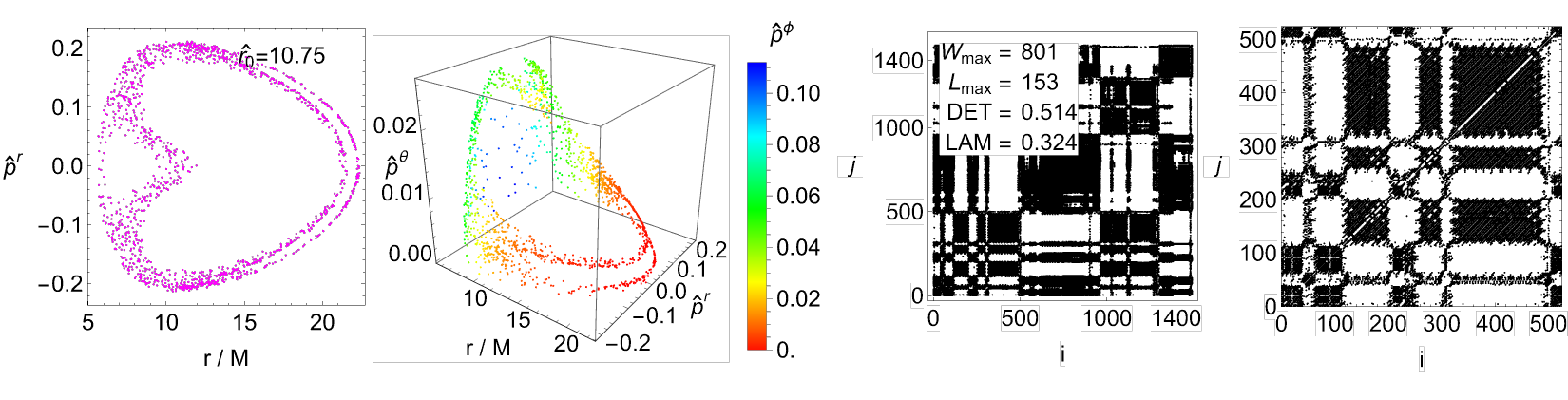}
\includegraphics[width=0.9\hsize]{ 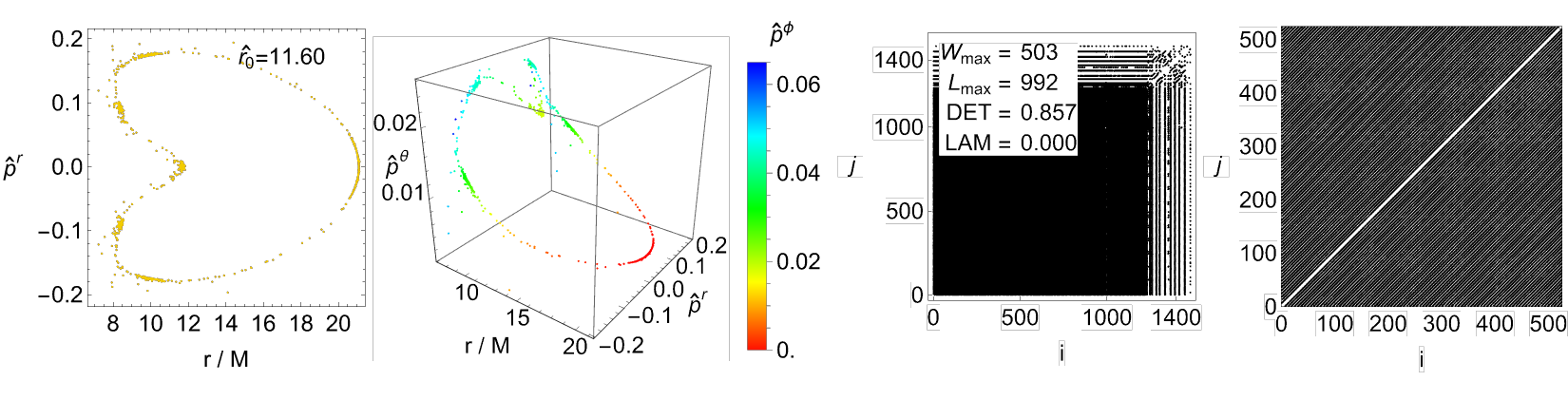}
\includegraphics[width=0.9\hsize]{ 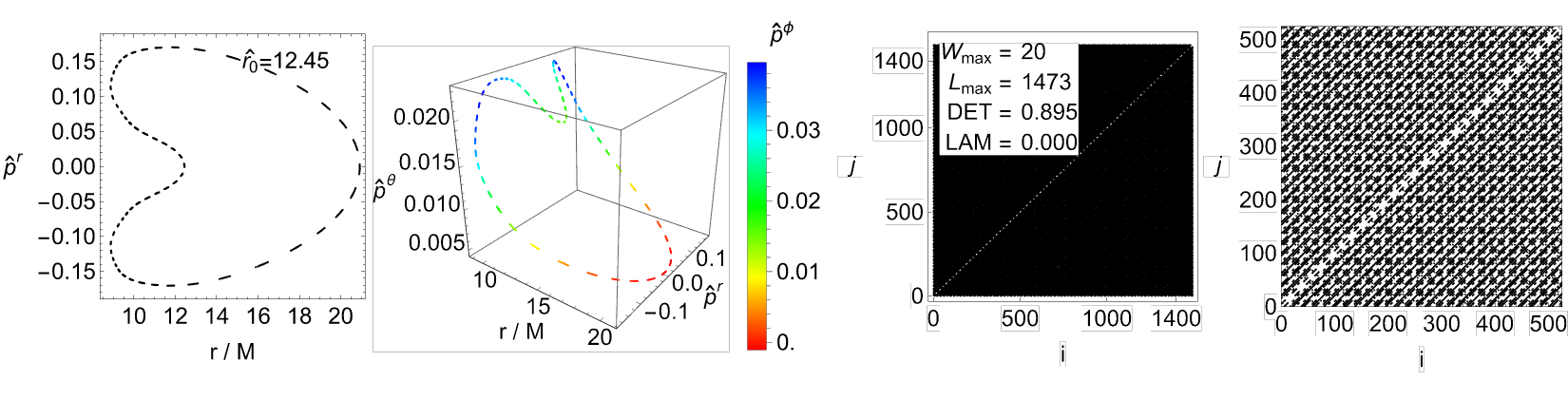}
\includegraphics[width=0.9\hsize]{ 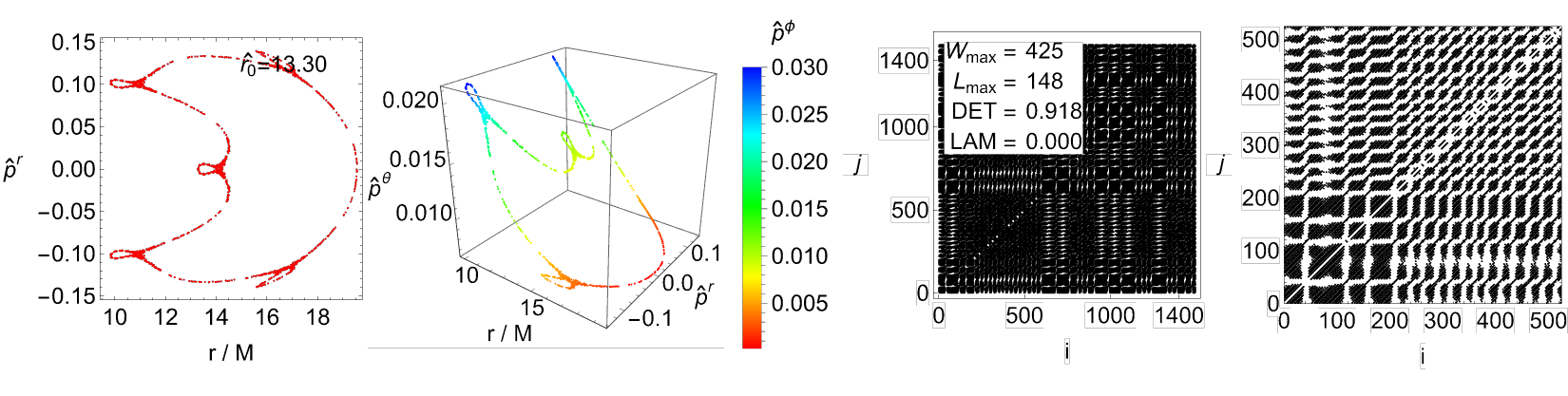}
\includegraphics[width=0.9\hsize]{ 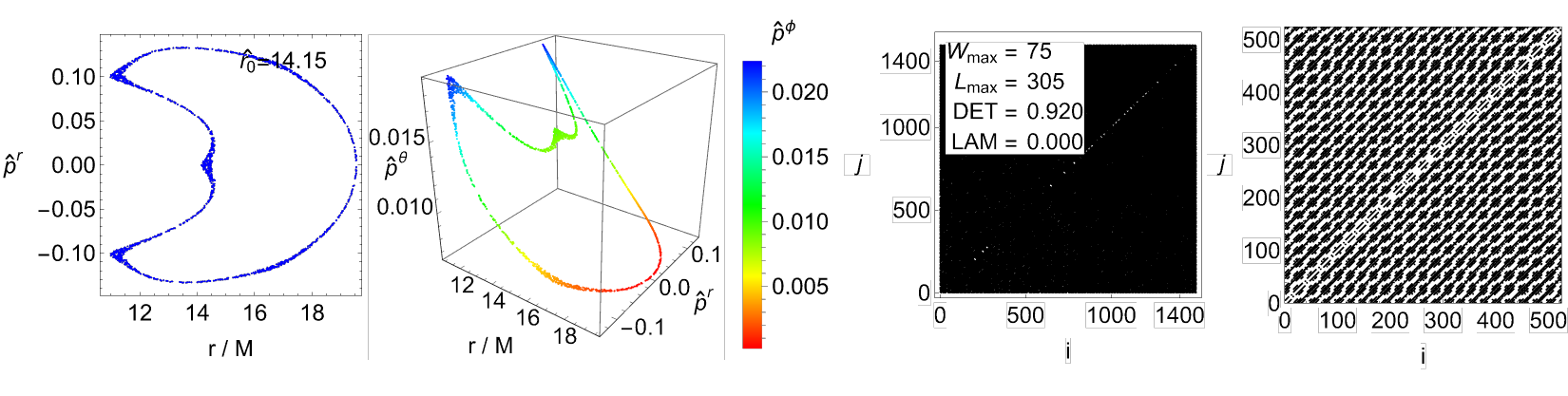}
\includegraphics[width=0.9\hsize]{ 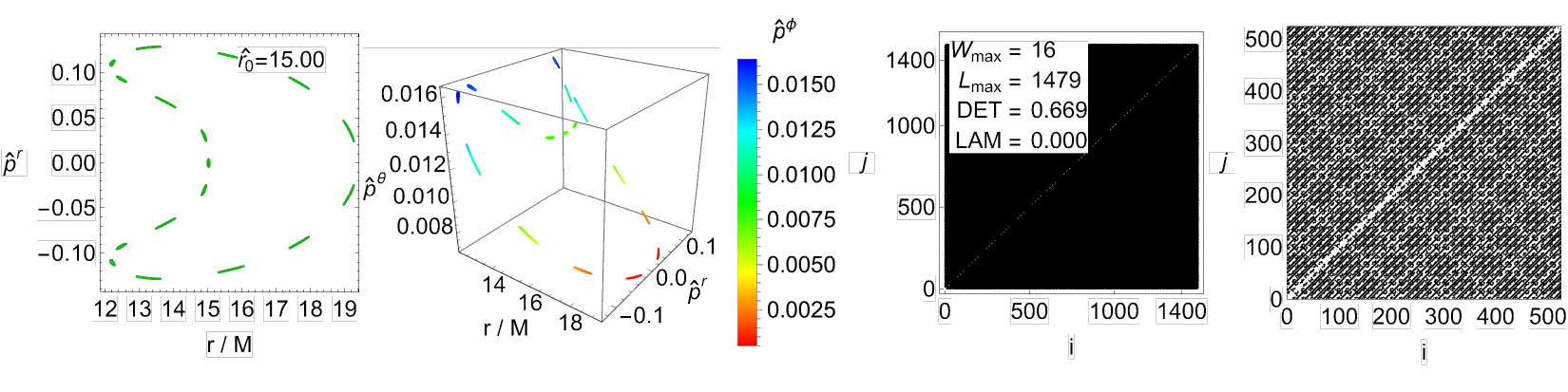}
\end{center}
\caption{Continued...}
\end{figure*}

\subsection{Recognition of regular and chaotic behavior}

To distinguish regular from chaotic orbits, we use RQA described in Sec.~\ref{sssec:recurrence_analysis}. Following~\cite{PhysRevD.111.084003}, we use PS data for RQA, specifically, we use only the $r$ coordinate of the PS data with embedding. The parameters used to produce all RPs are given in the corresponding figure captions, alongside some RQA quantifiers.


\subsubsection{Non-spinning charged case}

In the non-spinning charged case ($\mathcal{S}=0$, $\cb \neq 0$), the PS is 2D as discussed in Sec.~\ref{sec: PS}. It can thus be interpreted unambiguously and cross-referenced with the RPs and quantifiers. The corresponding PSs and RPs are shown in Fig.~\ref{fig2_2D+4D}, and the respective RQA values are listed in Tab.~\ref{tab:RQA_spinless}.

The first three and final four are regular orbits, demonstrated by smooth closed curves in the PS. The remaining orbits with $\hat{r}_0=9.05,\, 9.90,\, 10.75,\, 11.60$ are chaotic, the latter being part of a resonance inside the island of stability, and the former three forming the chaotic sea.

In the first three regular orbits with $\hat{r}_0=6.5,\, 7.35,\, 8.20$, there is a clear correspondence between the PSs and the RPs, which display long diagonal lines in the RP. The following four chaotic, shown as dispersed points in the PS, display characteristic square-like structures in the RP. However, in the remaining 4 regular orbits with initial position $\hat{r}_0=12.45,\, 13.30,\, 14.15,\, 15$, the RPs are not as clear: all display faint square-like structures.

RQA clearly reveals the regularity of these orbits: $L_\mathrm{max}$ can be seen to have values above 1400 (limited by the total number of data points, 1500) for regular orbits, while below 400 for all chaotic orbits. This demonstrates that the $L_\mathrm{max}$ is more reliable in the identification of order and chaos than a visual inspection of RPs. It should be noted that its inverse is related to the maximum finite Lyapunov exponent~\cite{Marwan:2007rps}.

The $W_\mathrm{max}$ and $\mathrm{LAM}$ can also be seen to respond well: both clearly separate chaotic and regular behaviors, marginally distinguishing weaker homoclinic chaos in the orbit with $\hat{r}_0=11.60$ as well. The $\mathrm{DET}$ indicator proves unreliable, as evidenced by e.g. the regular orbit with $\hat{r}_0=14.15$ producing $\mathrm{DET} = 0.554$.

\subsubsection{Spinning charged case}

In the case of spinning charged particles ($\mathcal{S}\neq0,\, \cb\neq0$), we first study the 2D projections of the PSs. Therein, we expect a regular trajectory to display a closed curve with epicyclic oscillations superposed due to the coupling between the radial and spin DoF. A chaotic orbit, on the other hand, is expected to manifest again as a set of dispersed points.

The 4D PSs show a similar behavior with the epicyclic oscillations expanded into the third dimension and the color space: thus, regular orbits display the shape of a torus with smooth color variations. Chaotic orbits are again expected to show dispersed points. Let us interpret the shapes of the PSs and compare them with RQA indicators.

The first orbit, with $\hat{r}_0 = 6.5$, produces a set of 4 islands as well as a thinner curve, without visible ordered epicyclic structures. This shape of an invariant manifold cannot correspond to a torus; therefore, we expect this orbit to be weakly chaotic. In accordance, $L_\mathrm{max} = 1030$ is not close to the upper bound of 1500, but also above the lower hundreds expected in strongly chaotic orbits.

The two following orbits with $\hat{r}_0=7.35,\, 8.20$ both follow the shape expected for regular orbits. Accordingly, $L_\mathrm{max}$ values close to 1500 confirm their regularity. Similarly, the orbit with $\hat{r}_0=9.05$ appears as four islands without irregular structures; we therefore classify it as a resonant regular orbit. Its $L_\mathrm{max} = 1346$ again suggests that this assessment is correct, despite being slightly lower than in the case of KAM orbits.

The following three orbits with $\hat{r}_0=9.90,\, 10.75,\, 11.60$ all present diffuse points with no apparent structure and are therefore expected to lie in the chaotic sea. In the former two cases, very low values of $L_\mathrm{max}=413,\, 153$, respectively, confirm this interpretation; in the latter, the seemingly contradicting $L_\mathrm{max} = 992$ is obviously due to a very sticky period lasting beyond 1000 points in the PS, as can be seen in the corresponding RP. After this time, the orbit can be seen to diffuse into the chaotic sea as well.

Of the remaining points, the first and last, with $\hat{r}_0=12.45,\, 15.00$ present as high-order resonant islands in the PS. Oscillations in the spin DoF are small enough not to be visible in the plots. The recurrence quantifier $L_\mathrm{max}$ is again in agreement with our interpretation, producing values greater than 1400 in both cases.

Finally, the orbits with $\hat{r}_0=13.30,\, 14.15$ demonstrate signs of homoclinic chaos: while apparently within a resonance (the former encapsulating its islands, the latter merely thickening in the corresponding points), the islands are not separated. The corresponding values $L_\mathrm{max} = 148,\, 305$, respectively, confirm this interpretation.

It is thus clear that a visual interpretation of 4D PSs as well as their 2D projections can distinguish dynamical properties of the orbits, supported by an independent classification scheme based on RQA. Further study reveals that the remaining indicators $W_\mathrm{max}$, $\mathrm{DET}$, and $\mathrm{LAM}$ are less reliable than $L_\mathrm{max}$, which agrees with our interpretation in all cases.

\section{Discussion and Conclusions} \label{sec: Conc}

We have explored the dynamics of a spinning charged test body around a Schwarzschild BH immersed in an external uniform magnetic field, examining the combined effects of gravitational, spin-curvature, and Lorentz forces on the particle's motion within the framework of MPDS equations under TD SSC. We analyzed both in- and off-equatorial motion, and compared our results with three limiting cases: non-spinning neutral case ($\mathcal{S}=0$, $\BB=0$), spinning neutral case ($\mathcal{S}\neq0$, $\BB=0$), and non-spinning charged case ($\mathcal{S}=0$, $\BB\neq0$).

By restricting the motion to the equatorial plane and assuming that the spin vector is orthogonal to the orbital plane, we derived analytical expressions for the energy and angular momentum of the spinning charged test body. In the absence of spin-curvature force ($\mathcal{S}=0$, $\BB\neq0$), the Lorentz force can act either repulsively or attractively, depending on the sign of the magnetic parameter $\BB$. When the Lorentz force is repulsive, electromagnetic interactions help counteract gravity and allow the circular motion at smaller radii. Conversely, when the Lorentz force is attractive, it reinforces the gravity, causing the minimum of angular momentum $\LL$ to move to larger radii. 

Spin modifies these behaviours through spin-curvature coupling, and its impact is determined by the spin orientation. When the spin is aligned with the z-axis, the spin-curvature force induces a radial shift in the same direction as the Lorentz force - repulsive when the Lorentz force is repulsive and attractive when it is attractive. As a result, it enhances the electromagnetic effect, producing a more pronounced shift of the minimum toward smaller radii. In contrast, for anti-aligned spin configurations, the spin-curvature force drives the minimum in the opposite direction, shifting it toward larger radii relative to the aligned case.   

We analyzed the characteristics of the effective potential due to the combined gravitational, spin-curvature, as well as Lorentz forces acting on a spinning charged particle. We numerically compute the radial positions of the ISCOs in dependence on the spin parameter $\mathcal{S}$ and magnetic parameter $\BB$. The ISCO radius decreases as either the magnetic field or the spin of the spinning charged body increases. The particle's spin contributes intrinsically to the total angular momentum, enabling it to orbit closer to the compact object. Consequently, the ISCOs of spinning charged bodies are located at smaller radii than those of spinning neutral bodies. Furthermore, when the spin is aligned with the z-axis, the ISCO radius is smaller than in the anti-aligned case. In the presence of a repulsive Lorentz force, increasing the magnetic field draws the ISCO closer to the BH, whereas an attractive Lorentz force pushes the ISCO outward, away from the event horizon.

We studied the perturbative oscillatory motion of a spinning charged test body around its circular equatorial orbits, and obtained analytical expressions for the radial and the orbital frequencies in terms of the spin parameter $\mathcal{S}$ and the magnetic parameter $\BB$. In the geodesic limit, where both charge and spin are absent ($\mathcal{S}=0,~ \BB=0$), the radial epicyclic frequency $\Omega_r$ decreases as the particle approaches the BH and vanishes at $r = 6M$. When spin is introduced while keeping the magnetic field zero ($\mathcal{S}\neq0,~\BB=0$), the behaviour of orbital frequency depends on the spin orientation. For aligned spin with the z-axis, the radial frequency profiles shift toward smaller radii relative to the geodesic case, indicating that stable radial oscillations extend closer to the BH. In contrast, for anti-aligned spin, the profiles shift toward larger radii, reducing the region where stable radial motion is possible. These deviations are most pronounced at small radii, while at large distances, both cases asymptotically approach the geodesic behavior. 

When both spin and charge are included ($\mathcal{S}\neq0,~\BB\neq 0$), the radial frequency reflects the interplay between spin-curvature and electromagnetic effects. Its behavior depends on whether these two forces act in the same direction or oppose each other. These combined effects become especially significant in the strong-field region near the BH, where both spin-curvature coupling and electromagnetic interactions are most pronounced.

In the non-spinning but charged case ($\mathcal{S}=0, ~ \BB \neq0$), the magnetic field modifies the radial frequency through the Lorentz force. An attractive Lorentz force enhances the radial frequency $\Omega_r$ and permits stable oscillations over a wider radial range, whereas a repulsive Lorentz force tends to suppress $\Omega_r$. 

The behaviour of the orbital frequency $\Omega_\phi$ follows trends broadly consistent with those observed for the radial frequency $\Omega_r$, while exhibiting distinct quantitative differences. In the geodesic limit, $\Omega_\phi$ decreases monotonically with radius. The inclusion of spin or charge shifts the frequency profile depending on spin orientation and the nature of the Lorentz force: aligned spin and attractive electromagnetic interaction enhance the orbital motion and shift it inward, while anti-aligned spin or repulsive interaction produces the opposite effect. When both effects are present, their interplay determines the overall behavior. These deviations are most significant near the BH and diminish at large radii.

To examine the geometric properties of the motion of the spinning charged test body and the resulting orbital structure, we numerically integrate the equations of motion, employing the Gauss Runge-Kutta scheme, and plot the trajectories for both in- and off-equatorial planes. We compare the trajectories for all nine possible combinations of spin orientation ($\mathcal{S}<0$, $ \mathcal{S}=0$, $ \mathcal{S}>0$) and magnetic field ($\BB<0$, $\BB=0$, $\BB>0$), and show that the Lorentz and spin-curvature forces can either widen or tighten the orbit relative to the reference geodesic, depending on the sign of the magnetic parameter $\BB$ and spin parameter $\mathcal{S}$.

In the non-spinning charged case ($\mathcal{S}=0,~\BB \neq0$), the deviation is governed entirely by the Lorentz force, with repulsive interaction shifting the motion outward and attractive interaction drawing it inward. When spin is included ($\mathcal{S} \neq 0,~\BB \neq0$), the spin-curvature force further modifies the trajectories: aligned spin tends to enhance outward motion, whereas anti-aligned spin favors inward motion and can even lead to rapid capture by BH. In the repulsive Lorentz case, the appearance and spacing of orbital curls additionally reflect the change in orbital frequency $\Omega_\phi$, showing that the interplay between electromagnetic and spin-curvature effects alters not only the overall size of the orbit but also its local structure.

We have also examined the off-equatorial trajectories, both in the absence and the presence of the magnetic field. Although spinning neutral particles are known to exhibit chaotic dynamics, the inclusion of the magnetic field, however, leads to a substantial increase in dynamical complexity and the appearance of more prominent chaotic behavior, demonstrating the combined effect of spin-curvature and electromagnetic interactions. To further investigate the underlying phase space structure, we employ the PSs. While standard 2D PSs are sufficient for systems that can be reduced to two DoF, they become insufficient when both spin-curvature and electromagnetic interactions are present, as the dynamics extend to three DoFs. To capture the full phase space structure, we therefore also employed 4D PSs, visualized through 3D projections with color coding. 

Our results show that these additional interactions break the integrability of geodesic motion, leading to a richer phase space in which regular and chaotic trajectories coexist. In particular, for suitable choices of parameters and initial conditions, the motion of spinning charged particles exhibits clear signatures of chaotic behavior. To support these conclusions, we have also applied RQA. While a visual inspection of RPs is insufficient to distinguish regular orbits from weakly chaotic ones, we have used RQA (namely, the longest diagonal line ($L_{\rm max}$) and longest white vertical line ($W_{\rm max}$) indicators) to demonstrate that our interpretation of 4D PSs is correct.

A natural extension of this work is to explore more general and astrophysically relevant scenarios. In particular, it would be interesting to investigate the dynamics of a spinning charged test body in rotating Kerr BH spacetime, where frame-dragging effects are expected to further enrich the interplay between spin-curvature and electromagnetic interactions. Additionally, a more systematic exploration of the parameter space, including stronger magnetic fields and varying spin magnitudes, could provide deeper insight into the onset of chaotic behavior. Finally, incorporating radiation reaction effects and examining the implications for gravitational wave emission may help bridge the gap between theoretical predictions and observable astrophysical phenomena.

\section*{Acknowledgments}

We sincerely thank L. Filipe O. Costa for his insightful discussions and valuable guidance on the dynamics of spinning charged particles. M.S. has been supported by GA\v{C}R-25-15272I, M.K. by the Institute of Physics, Silesian University in Opava, and O.Z. by the PPLZ fellowship of the Czech Academy of Sciences.

\appendix 

\section{Derivation of the four-velocity four-momentum relation and spin conservation}\label{Apndx:RelDer}


To derive the four-velocity four-momentum relation~\eqref{eq:4v4mrel} for the MPDS equations, one can follow the steps provided in Ref.~\cite{Semerak:1999:MNRAS:} for the MPD equations. First, we plug Eq.~\eqref{eq:spin2} into the equation produced by applying the covariant derivative on the TD SSC~\eqref{eq:TD_spin-cond}. Then, we eliminate $u^\mu$ using the relation obtained by contracting Eq.~\eqref{eq:spin3} with $p_\alpha$, in which the covariant derivative of the TD SSC is reused. For the final step, one needs to take into account that because of the antisymmetry of $S^{\mu\nu}$, it holds that 
\beq
R_{\alpha\mu\nu\rho}S^{\beta\mu}S^{\alpha \kappa}=R_{\alpha\mu\nu\rho}S^{\beta[\mu}S^{\alpha] \kappa}=\frac{1}{2}R_{\mu\alpha\nu\rho}S^{\alpha\mu}S^{\beta \kappa}, 
\eeq
and that a similar relation holds for the term containing $F_{\alpha\beta}$, since the latter is antisymmetric as well.

To prove the conservation of test body spin $S$, Eq.~\eqref{eq:spin3} can be contracted with $S_{\alpha\beta}$ leading to 
\beq
\displaystyle S\frac{\mathrm{D} S}{\d \tau}= 2 k S_{\alpha\beta} S^{\alpha\nu} {F_\nu}^{\beta}. 
\eeq
Using Eq.~\eqref{eq:spin-tensor}, one can write 
\beq
S_{\alpha\beta} S^{\alpha\nu}=S_\beta S^\nu-S^2 (\delta_\beta^\nu +v_\beta v^\nu), 
\eeq
which implies that $S_{\alpha\beta} S^{\alpha\nu}$ is a symmetric tensor, and where $\delta_\alpha^\nu$ is the Kronecker's delta. Since the contraction of a symmetric and an antisymmetric tensors vanishes, thus $2 k S_{\alpha\beta} S^{\alpha\nu} F^{~\beta}_\nu = 0$, which implies $\displaystyle \frac{\mathrm{D} S}{\d \tau}= 0,~\forall~k$, see also Ref.~\cite{Holten:1991:NPB:,HeyFar-etal:2005:PLB:}.

\section{Integration scheme and accuracy}\label{Apndx:IntAcc}

\begin{figure*}
\begin{center}
\includegraphics[width=\hsize]{ 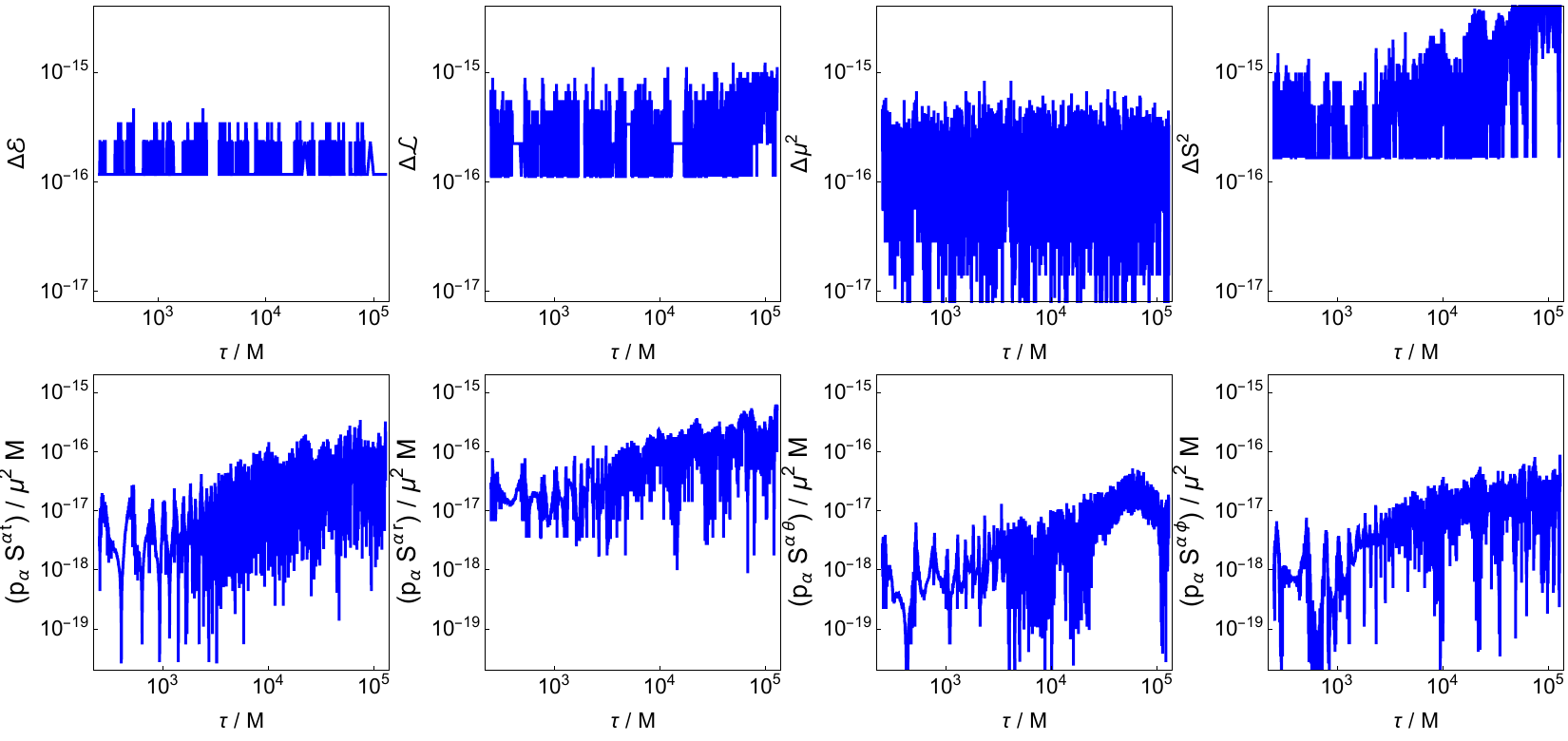}
\end{center}
\caption{Logarithmic plots showing the relative errors in the energy $\EE$, angular momentum $\LL$, mass $\mu$, and spin magnitude $\mathcal{S}$ of the spinning charged test body (first row), corresponding to regular off-equatorial orbit ($\hat r_{0}= 8.2$) presented in Fig.~\ref{fig_PS}. The bottom row represents the numerical evolution for the TD SSC \eqref{eq:TD_spin-cond}, throughout the integration. 
\label{fig_Err_regular}}
\end{figure*}

\begin{figure*}
\begin{center}
\includegraphics[width=\hsize]{ 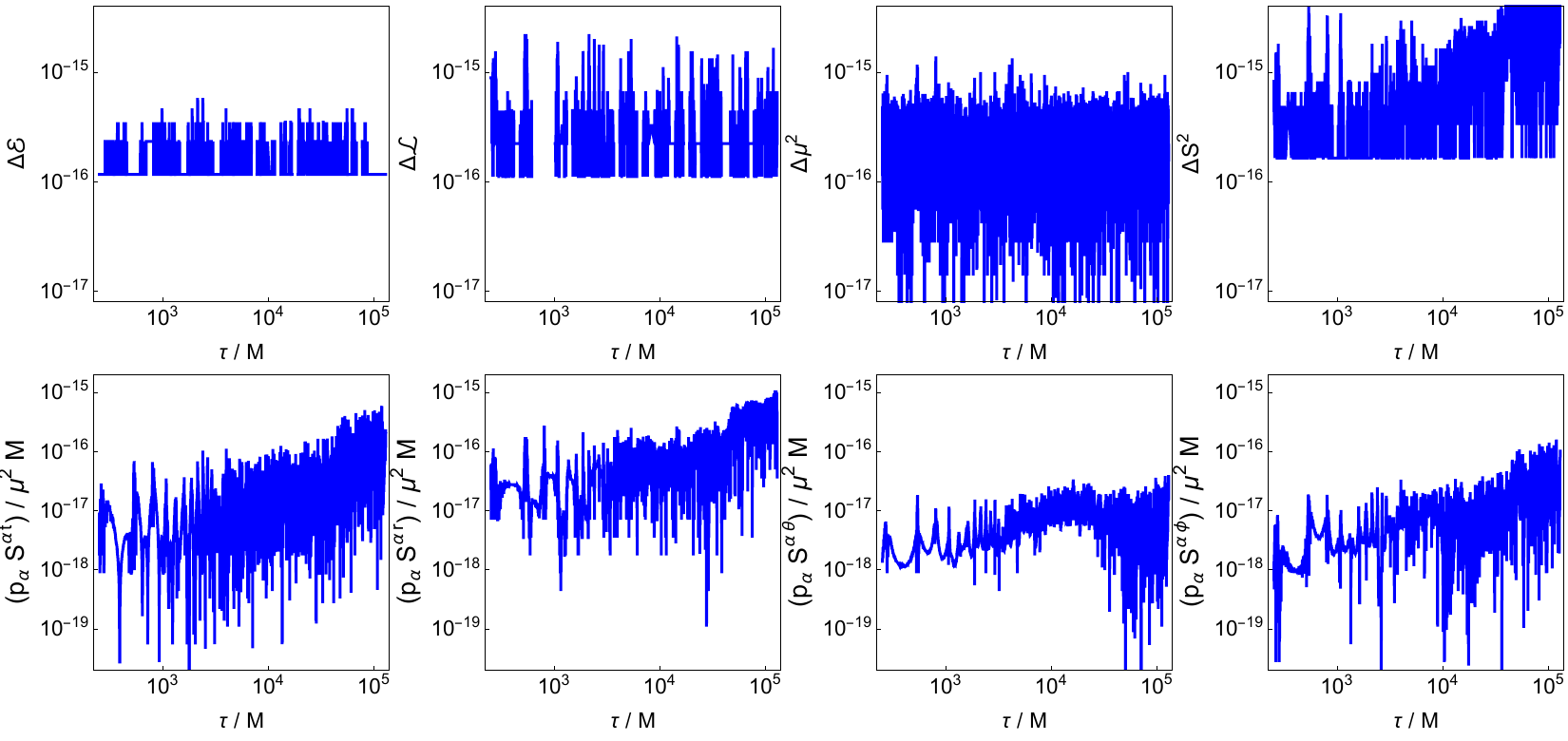}
\end{center}
\caption{Logarithmic plots showing the relative errors in the energy $\EE$, angular momentum $\LL$, mass $\mu$, and spin magnitude $\mathcal{S}$ of the spinning charged test body (first row), corresponding to chaotic off-equatorial orbit ($\hat r_{0}=10.75 $) presented in Fig.~\ref{fig_PS}. The bottom row represents the numerical evolution for the TD SSC~\eqref{eq:TD_spin-cond}. 
\label{fig_Err_chaotic}}
\end{figure*}

The performance of numerical integrators plays a crucial role in accurately solving mathematical problems and simulating dynamical systems. We use a fourth-order Gauss-Runge-Kutta integrator to numerically integrate the equations of motion, and check the accuracy of our numerical integration by computing the relative errors in the conserved quantities, namely, the energy $\EE$, angular momentum $\LL$, mass $\mu$, and spin $\mathcal{S}$, defined by
\bea\label{errors}
\Delta \EE(\tau) &=& \left| 1 - \frac{\EE(\tau)}{\EE(0) } \right|,\\
\Delta \LL(\tau) &=& \left| 1 - \frac{\LL(\tau)}{\LL(0) } \right|,\\
\Delta \mathcal{S}(\tau) &=& \left| 1 - \frac{\mathcal{S}(\tau)}{\mathcal{S}(0) } \right|,\\
\Delta \mu (\tau) &=& \left| 1 - \frac{\mu(\tau)}{\mu(0) } \right|.
\eea
In addition to the conserved quantities, we also assess the preservation of the TD SSC \eqref{eq:TD_spin-cond} throughout the system's evolution. The relative errors in the conserved quantities and the numerical evolution in the TD SSC for regular and chaotic trajectories shown in Fig.~\ref{fig_PS} are presented in Figs.~\ref{fig_Err_regular} and \ref{fig_Err_chaotic}, respectively. We observe that the relative errors in the conserved quantities $\EE$, $\LL$, $\mu$, and $\mathcal{S}$ exhibit small oscillations around the double-digit numerical accuracy, and the amplitude of the oscillations remains bounded throughout the evolution. Similarly, the numerical evolution of TD SSC remains very close to zero and bounded during the evolution, indicating that the constraint is preserved to high accuracy over the entire evolution.

\section{Lengthy expressions} \label{sec:anal_exp}

In this section, we present lengthy expressions for the quantities used in the main text. The expressions for $\mathcal{F}_0$, $\mathcal{F}_2$, and $\mathcal{H}$ in Eqs.~\eqref{time-Eq},~\eqref{radial-Eq},~\eqref{eq:Veffap} are given by
\begin{widetext}
\bea\non
\mathcal{F}_0 &=& \left(4 \BB \LL+ \EE^2 + 2 \right) \left(2 \BB^4 \hat r^7 + 2 \BB^2 \LL^2 \hat r^3\right) + \LL^2 \left(3 + 4 \BB^2 \LL^2 \right) + \hat r^2 \left(1 - 2 \BB \LL \left(4 \BB \LL \left(1 + 2 \BB \LL \right)+7 \right) \right) - 8 \BB^4 \hat r^6 \left(1 + 2 \BB \LL \right) \\ &+&
\BB^2 \hat r^4 \left(11 + 8 \BB \LL \left(2 + 3 \BB \LL \right) \right) - 4 \BB^2 \hat r^5 \left(1 + \BB \LL \left(2 + \EE^2 + 3 \BB \LL \right) \right)  - 2 \BB \hat r \left(\BB^5 (\hat r-2) \hat r^7 + \BB \LL^4 - 2 \LL \hat r^2 \right),\\\non
\mathcal{F}_2 &=& - \frac{6}{\hat r^3} + \frac{1}{\hat r} \left(4 \BB^2 \left(2 + \left(\EE^2 - 1 \right) \hat r \right) \right) + \frac{1}{\hat r^5} \left(2 \BB^2 \hat r^4 \left(4 \EE^2 \hat r+1\right) - 8 \BB \LL \hat r^2 + 6 \LL^2 + 6 \hat r^2 \right) - \mathcal{F}_3 \left(\mathcal{F}_3 - 4 \BB \right) \EE^2\\ 
&+& \frac{1}{\hat r^3 \mathcal{F}_4} \Big[\hat r^3 \left(\BB^2 \hat r \left(2 \hat r - 5 \right) + \EE^2 \left(\hat r - 4 \right) \right) + 2 \BB \LL \left(5-2 \hat r \right) \hat r^2 + \LL^2 \left(2 \hat r-5 \right) \Big],\\
\mathcal{F}_3 &=& \frac{1}{\mathcal{F}_4} (\hat r-3) \left(\BB \hat r^2 - \LL \right), \qquad \mathcal{F}_4 = \EE^2 \hat r^3 -(\hat r-2) \left(\LL- \BB \, \hat r^2\right)^2  + (2 - \hat r) \hat r^2,
\eea
\begin{align}
\mathcal{H} &= 2 \left(\BB^2 \hat r^4 \left(\left(8 \EE^2 - 2\right) \hat r+5\right)-4 \BB \LL \hat r^2 + 3 \LL^2 + \hat r^2\right) \left(\hat r^6 \left(\hat r \left(\BB^2 (\hat r-2) \hat r-\EE^2+1\right)-2\right) - 2 \BB \LL (\hat r-2) \hat r^6 + \LL^2 (\hat r-2) \hat r^4\right) \nonumber \\
&+ \hat r^8 \left(\hat r \left(\EE^2 (\hat r-2) - 2 -\BB^2 \hat r \right)+4\right)+8 \BB \EE^2 (\hat r-3) \hat r^9 \left(\LL - \BB \hat r^2\right) + 2 \BB \LL \hat r^8 - \LL^2 \hat r^6.
\end{align}
The quadratic spin corrections in the energy~\eqref{eq:EE} and the angular momentum~\eqref{eq:LL} are given by
\bea\non
\EE_{2} &=& \frac{1}{2 \EE_0 \hat r^6 \left(\BB \hat r^2 + \LL_0 (\hat r-3)\right)} \left[\LL_{0} \hat r^4 \left(\BB^2 ((24-5 \hat r) \hat r-25) + 2 \BB \EE_{1} \hat r^2 + \EE_{0}^2 - \EE_{0} \LL_{1} - \EE_{1}^2 (\hat r-3) \hat r^2 \right) \right. \\\non  &+& \left. \BB \hat r^5 \left(\BB^2 ((\hat r-5) \hat r+5) \hat r -\BB \EE_{1} \hat r^3 + \EE_{0}^2 ((\hat r-7) \hat r + 8) - \EE_{0} \LL_{1} (\hat r-4) (2 \hat r-3)-\EE_{1}^2 \hat r^3 + \LL_{1}^2 (\hat r-2)^2\right) \right. \\ &+& \left. \LL_{0}^2 r^2 \left(\BB (\hat r (7 \hat r-33) + 35) - \EE_{1} \hat r^2\right) -\LL_{0}^3 (\hat r-3) (3 \hat r-5) \right], \\
\LL_{2} &=& \frac{1}{2} \left[3 \EE_{1} - \BB + \frac{5 \BB}{\hat r} - \frac{5 \LL_0}{\hat r^3} + \frac{ 6 \EE_0^2 -2 \BB \LL_0 - 9 \EE_0 \LL_1 + 3 \LL_1^2 - \hat r \left(\LL_{0} (\EE_1 - \BB ) + (\EE_0 - \LL_1)^2\right)}{ \left(\hat r-3 \right) \LL_{0} + \BB \hat r^2} \right].
\eea
The expression for the coefficient $\mathcal{Y}$ of the quadratic spin term of radial frequency $\omega_{r}$~\eqref{Omega_R} is given by
\bea\non
\mathcal{Y} &=& \LL^2 \hat r^2 \left(2 \hat r \left(3 \EE^2 \left(8 \BB^2 (\hat r-4) \hat r^2 - 15\right) - 2 \BB^2 \hat r (3 (\hat r-20) \hat r+160)+60\right)-357\right) \\\non &+& 4 \BB \LL \hat r^4 \left(\hat r \left(2 \EE^2 \left(\hat r \left(8 \BB^2 \hat r+3\right) - 12 \right) + \BB^2 \hat r (42-11 \hat r)-18\right)+65\right) + 24 \LL^4 (7 \hat r-18) \\ &+& \hat r^5 \left(\BB^2 \hat r \left(-4 \BB^2 \hat r^4 + 20 \hat r-75\right) + \EE^2 \left(\hat r \left(16 \BB^4 \hat r^4-18 \BB^2 \hat r+3\right)-24\right)\right)+60 \BB \LL^3 (14-5 \hat r) \hat r^2.
\eea
The expression for the coefficient $\mathcal{Z}$ of the quadratic spin term of the radial frequency $\Omega_{r}$~\eqref{distant_Omega_R} is given by
\bea\non
\mathcal{Z} &=& \left(\hat r - 2 \right)^4  \Bigl[12 \BB^4 \hat r^4 \left(\BB^2 \LL^2 (9 \hat r-28) \hat r^4 + \BB \LL r^4 \left(9-4 \BB^2 (\hat r-1) \hat r^2\right) + \BB^2 \hat r^6 \left(\BB^2 \hat r^3 - 2 \right) + \LL^4 \left(19 \hat r - 88 \right) \right. \\\non &+& \left. 8 \BB \LL^3 (9-2 \hat r) \hat r^2  \right) \Bigl] + 2 \BB \left(\hat r - 2 \right)^3  \Bigl[ 2 \LL^5 \left(\hat r \left(9 - 4 \BB^2 \hat r (3 \hat r - 13) \left(\left(\EE^2+3\right) \hat r-6\right)\right)-36\right)  -66 \BB \LL^4 (\hat r-4) \hat r^2 \\\non &+& 2 \BB \LL^4 \hat r^3 \EE^2 \left( 70 + \hat r \left(\BB^2 \hat r (19 \hat r - 88) - 18 \right) \right) - 4 \BB^2 \LL^3 \hat r^4 \left(\hat r \left(2 \EE^2 \left(\hat r \left(2 \BB^2 \hat r (2 \hat r - 9) - 9\right) + 40 \right) - 27 \right) + 96 \right) \\\non &+& \BB^2 \LL \hat r^6 \left(4 \BB^2 \hat r^2 \left(\hat r \left(\EE^2 \left(\hat r \left(- 2 \BB^2 (\hat r-1) \hat r + \EE^2+6\right)-8\right)+3\right)-3\right)-39\right) + 36 \BB^3 \LL^2 \hat r^6 \left(8 + \left(6 \EE^2 - 3\right) \hat r\right) \\\non &+& 2 \BB^3 \EE^2 \LL^2 \hat r^8 \left(\BB^2 \hat r (9 \hat r-28) - 24\right) + 2 \BB^3 \hat r^8 \left(\BB^4 \EE^2 \hat r^6 - 3 \BB^2 \hat r^3 + 6\right) + 6 \BB \LL^6 (\hat r-4) \left(\left(\EE^2 + 3 \right) \hat r - 6 \right) \Bigl] \\\non &+& \left(\hat r - 2 \right)^2 \Bigl[ 3 \LL^4 \left(\hat r \left(-4 \BB^2 \EE^4 (\hat r-4) \hat r^3+10 \EE^2 (5 \hat r-12)+3\right)-12\right) + 4 \BB \LL \hat r^4 \left(\EE^2 \hat r \left(\hat r \left(6 \EE^2 \left(2 \BB^2 \hat r^2 + \hat r-4\right) \right. \right. \right. \\\non &-& \left. \left. \left. 2 \BB^2 \hat r (\hat r (4 \hat r-19) + 10) - 11 \right) + 32\right) + 3\right) + \LL^2 \hat r^2 \left(\hat r \left(-2 \EE^4 \hat r \left(4 \BB^2 \hat r^2 \left(2 \hat r \left(\BB^2 (\hat r-5) \hat r-3\right)+23\right)+45\right) \right. \right. \\\non &+&  \left. \left. \EE^2 \left(2 \hat r \left(2 \BB^2 \hat r (\hat r (15 \hat r-74)+34)+51\right) - 249\right) + 12 \BB^2 \hat r (7 \hat r-17) \right) - 6 \right) + 3 \BB^4 \hat r^9 - 6 \hat r^6 \left(\BB^2 + 4 \EE^4 \right) \\ &+& 2 \BB \LL^3 r^2 \left(66 + \hat r \left(4 \EE^2 \left(36 + \hat r \left(6 + \hat r \left(-6 + \BB^2 \EE^2 \hat r (3 \hat r-14)\right)\right)\right)-21\right)\right)
\Bigl].
\eea
The quadratic spin coefficient $\hat{\Omega}_{2} (\hat r, \BB)$ in the expression of orbital frequency $\Omega_{\phi}$ \eqref{eq:Anyl_Omega_Phi} takes the form
\bea\non
\hat{\Omega}_{2} &=& \frac{-\left(\hat{r}-2 -\hat{r}^3 \hat{\Omega}_0^2\right)^{\frac{3}{2}} \left(\hat{r}^6 \left(\hat{\Omega}_0^4+3 \hat{\Omega}_0^2 \hat{\Omega}_1 + \hat{\Omega}_1^2\right)+\hat{r}^3 \left(\hat{\Omega}_0^2+2 \hat{\Omega}_1\right)-2 \right) - \BB \left(\hat{r}-2\right) \left(3 \hat{r}^3 \hat{\Omega}_1-4\right)\hat{\Omega}_0 \hat{\Omega}_1 \hat{r}^{11/2}}{2 \hat{r}^{11/2} \left(\hat{r}^3 \hat{\Omega}_0^2-\hat{r}+2\right) \left(\BB \left(2 - \hat{r} + 2 \hat{r}^3 \hat{\Omega}_0^2\right)-\hat{\Omega}_0 \sqrt{\hat{r} \left(\hat{r}-2 -\hat{r}^3 \hat{\Omega}_0^2\right)}\right)}\\\ &+&
\frac{\BB \hat{r}^{5/2} \hat{\Omega}_0 \left(2 \hat{r}^9 \hat{\Omega}_0^6+2 \hat{r}^3 \hat{\Omega}_0^2 \left(\hat{r}^3 \hat{\Omega}_1 \left(\hat{r}^3 \hat{\Omega}_1 -4 \hat{r}+7\right)+2 \hat{r}-4\right)+\hat{r}^6 \hat{\Omega}_0^4 \left(6 \hat{r}^3 \hat{\Omega}_1 -3 \hat{r}+4\right)-\hat{r}+2\right)}{2 \hat{r}^{11/2} \left(\hat{r}^3 \hat{\Omega}_0^2-\hat{r}+2\right) \left(\BB \left(2 - \hat{r} + 2 \hat{r}^3 \hat{\Omega}_0^2\right)-\hat{\Omega}_0 \sqrt{\hat{r} \left(\hat{r}-2 -\hat{r}^3 \hat{\Omega}_0^2\right)}\right)}.
\eea

\end{widetext}



\def\prc{Phys. Rev. C}
\def\pre{Phys. Rev. E}
\def\prd{Phys. Rev. D}
\def\jcap{Journal of Cosmology and Astroparticle Physics}
\def\apss{Astrophysics and Space Science}
\def\mnras{Monthly Notices of the Royal Astronomical Society}
\def\apj{The Astrophysical Journal}
\def\aap{Astronomy and Astrophysics}
\def\actaa{Acta Astronomica}
\def\pasj{Publications of the Astronomical Society of Japan}
\def\apjl{Astrophysical Journal Letters}
\def\pasa{Publications Astronomical Society of Australia}
\def\nat{Nature}
\def\physrep{Physics Reports}
\def\araa{Annual Review of Astronomy and Astrophysics}
\def\apjs{The Astrophysical Journal Supplement}
\def\aapr{The Astronomy and Astrophysics Review}
\def\procspie{Proceedings of the SPIE}

\bibliographystyle{unsrt}
\bibliography{reference}

\end{document}